\documentclass[twocolumn]{aastex62}
\graphicspath{{./}{figures/}}

\usepackage{graphicx}
\usepackage{hyperref}
\usepackage{mhchem}
\usepackage{xspace}
\usepackage{lineno}

\newcommand{\htwo}{H$_2$\xspace}
\newcommand{\hii}{\ion{H}{2}\xspace}
\newcommand{\av}{A$_V$\xspace}
\newcommand{\avtonh}{$A_{\rm V}/N_{\rm H}$\xspace}

\newcommand{\htohtwo}{H/H$_2$\xspace}

\newcommand{\seppc}{$0.043 \pm 0.013(\rm{stat.}) \pm 0.0036 (\rm{syst.})$ pc}
\newcommand{\separc}{$0\farcs146 \pm 0\farcs042 (\rm{stat.}) \pm 0\farcs012 (\rm{syst.})$ }
\received{January 27, 2025}
\accepted{July 7, 2025}
\submitjournal{ApJ}

\begin{document}

\shortauthors{Clark et al.}
\shorttitle{N13 PDR}

\title{The Resolved Structure of a Low Metallicity Photodissociation Region}

\author[0000-0003-3334-4267]{Ilyse Y. Clark}
\affiliation{Department of Astronomy \& Astrophysics, University of California, San Diego, 9500 Gilman Drive, La Jolla, CA 92093}

\author[0000-0002-4378-8534]{Karin Sandstrom}
\affiliation{Department of Astronomy \& Astrophysics, University of California, San Diego, 9500 Gilman Drive, La Jolla, CA 92093}

\author[0000-0003-0030-9510]{Mark Wolfire}
\affiliation{Department of Astronomy, University of Maryland, College Park, MD 20742, USA}

\author[0000-0002-5480-5686]{Alberto D. Bolatto}
\affiliation{Department of Astronomy, University of Maryland, College Park, MD 20742, USA}

\author[0000-0002-5235-5589]{J\'er\'emy Chastenet}
\affiliation{Sterrenkundig Observatorium, Universiteit Gent, Krijgslaan 281-S9, 9000 Gent, Belgium}

\author[0000-0002-5782-9093]{Daniel~A.~Dale}
\affiliation{Department of Physics and Astronomy, University of Wyoming, Laramie, WY 82071, USA}

\author[0000-0003-4224-6829]{Brandt A. L. Gaches}
\affiliation{Department of Space, Earth and Environment, Chalmers University of Technology, Gothenburg SE-412 96, Sweden}
\affiliation{Faculty of Physics, University of Duisburg-Essen, Lotharstraße 1, 47057 Duisburg, Germany}

\author[0000-0001-6708-1317]{Simon C.~O.\ Glover}
\affiliation{Universit\"{a}t Heidelberg, Zentrum f\"{u}r Astronomie, Institut f\"{u}r Theoretische Astrophysik, Albert-Ueberle-Str.\ 2, 69120 Heidelberg, Germany}

 \author[0000-0001-7046-4319]{Javier R. Goicoechea}
 \affiliation{Instituto de F\'{\i}sica Fundamental   (CSIC). Calle Serrano 121-123, 28006, Madrid, Spain.}

\author[0000-0001-5340-6774]{Karl D.\ Gordon}
\affiliation{Space Telescope Science Institute, 3700 San Martin
  Drive, Baltimore, MD, 21218, USA}
\affiliation{Sterrenkundig Observatorium, Universiteit Gent, Krijgslaan 281-S9, 9000 Gent, Belgium}

\author[0000-0002-9768-0246]{Brent Groves}
\affiliation{International Centre for Radio Astronomy Research, University of Western Australia, 7 Fairway, Crawley, 6009 WA, Australia}

\author[0009-0005-0750-2956]{Lindsey Hands}
\affiliation{Department of Astronomy \& Astrophysics, University of California, San Diego, 9500 Gilman Drive, La Jolla, CA 92093}

\author[0000-0002-0560-3172]{Ralf Klessen}
\affiliation{Universit\"{a}t Heidelberg, Zentrum f\"{u}r Astronomie, Institut f\"{u}r Theoretische Astrophysik, Albert-Ueberle-Str.\ 2, 69120 Heidelberg, Germany}
\affiliation{Universit\"{a}t Heidelberg, Interdisziplin\"{a}res Zentrum f\"{u}r Wissenschaftliches Rechnen, Im Neuenheimer Feld 225, 69120 Heidelberg, Germany}
\affiliation{Harvard-Smithsonian Center for Astrophysics, 60 Garden Street, Cambridge, MA 02138, U.S.A.}
\affiliation{Elizabeth S. and Richard M. Cashin Fellow at the Radcliffe Institute for Advanced Studies at Harvard University, 10 Garden Street, Cambridge, MA 02138, U.S.A.}

\author[0000-0001-9419-6355]{Ilse De Looze}
\affiliation{Sterrenkundig Observatorium, Universiteit Gent, Krijgslaan 281-S9, 9000 Gent, Belgium}

\author[0000-0003-1545-5078]{J. D. T. Smith}
\affiliation{Ritter Astrophysical Observatory, University of Toledo, Toledo, OH 43606, USA}

\author[0000-0002-5895-8268]{Dries Van De Putte}
\affiliation{Department of Physics \& Astronomy, The University of Western Ontario, London ON N6A 3K7, Canada}

\author[0000-0001-6941-7638]{Stefanie K. Walch}
\affiliation{I. Physics Institute, University of Cologne. Zülpicher Str. 77, 50937, Cologne, Germany.}

\correspondingauthor{Ilyse Clark} 
\email{iyclark@ucsd.edu}

%----------------------------------------------------------------------------------------------------------------------
%----------------------------------------------------------------------------------------------------------------------

\begin{abstract}
Photodissociation Regions (PDRs) are key to understanding the feedback processes that shape interstellar matter in galaxies. One important type of PDR is the interface between \hii regions and molecular clouds, where far-ultraviolet (FUV) radiation from massive stars heats gas and dissociates molecules. Photochemical models predict that as metallicity decreases, the C/CO transition occurs at greater depths in the PDR compared to the \htohtwo transition, increasing the extent of CO-dark \htwo gas in low-metallicity environments. This prediction has been difficult to test outside the Milky Way due to the lack of high spatial resolution observations tracing \htwo and CO. This study examines a low-metallicity PDR in the N13 region of the Small Magellanic Cloud (SMC) where we spatially resolve the ionization front, the \htwo dissociation front, and the C/CO transition using $^{12}$CO J=2$-$1, 3$-$2 and [CI] 1-0 observations from the Atacama Large Millimeter/sub-mm Array (ALMA) and near-infrared spectroscopy of the \htwo 2.12 1-0 S(1) vibrational line, and H recombination lines from the James Webb Space Telescope (JWST). Our analysis shows that the separation between the H/\htwo and C/CO boundaries is approximately \seppc\ (equivalent to \separc at the SMC's distance of 62 kpc), defining the spatial extent of the CO-dark \htwo region. Compared to our plane-parallel PDR models, we find that a constant pressure model matches the observed structure better than a constant density one. Overall, we find that the PDR model does well at predicting the extent of the CO-dark \htwo layer in N13. This study represents the first resolved benchmark for low metallicity PDRs.
\end{abstract}

,

\section{Introduction} \label{sec:intro}
Photodissociation Regions (PDRs) occur where far-ultraviolet (FUV; 6 eV $< h\nu <$ 13.6 eV) photons drive the chemistry and thermal balance of the interstellar medium. A common type of PDR is created when massive O and B stars ionize their surroundings inside or near a molecular cloud, leading to distinct layers of ionized, atomic, and molecular gas. The PDR extends from the cloud surface, where the radiation emerging from the H~II region photoionizes atoms with ionization potential less than 13.6 eV, to deeper layers in the molecular gas where photo-processes can still be important.
A classic example of such a region is the Orion Bar PDR \citep{PDR1985,Berné_2022}. Since PDRs occur wherever FUV photons govern the properties of the interstellar medium (ISM), they represent a significant portion of the atomic and molecular gas in a galaxy \citep{PDR1985, Hollenbach97, Hollenbach99, wolfire2022}. Understanding their characteristics and evolution is crucial, as a large part of the molecular gas reservoir potentially fueling future star formation resides in PDRs.

Early PDR studies, including seminal work by \cite{PDR1985}, focused on the Orion Bar and emphasized the penetration depth of the FUV radiation into the cloud, set by the ratio of extinction to column density ($A_{\rm V}/N_{\rm H}$), and its crucial role in determining the chemical and thermal structure. 
Recent observations with ALMA \citep{Goicoechea16, Goicoechea17, Goicoechea25}, and JWST \citep{Peeters24, Habart24, Chown24, VanDePutte24, Fuente24} have pushed the spatial resolution of Orion Bar measurements to 0.0002 pc. This comprehensive multi-wavelength high-resolution dataset of the Orion Bar revealed unexpected small-scale filaments and globules ($\sim 10^{-3}$ pc) along with ridges that follow the boundaries of the PDR. In general, the large-scale PDR structure follows plane-parallel geometry, but with many complex embedded small-scale features, which are not well understood \citep{Goldsmith08, Joblin18}. 

A PDR's structure is expected to be highly dependent on metallicity \citep{Rollig06} due to effects from changes in heating and cooling rates and decreased dust extinction (e.g., lower \avtonh). In higher metallicity regions, abundant species, such as C$^+$ and O, play crucial roles in cooling and regulating the thermal balance. However, with fewer metals, the gas cooling efficiency decreases \citep{Tielens10, Draine11} due to reductions of important coolants such as [CII]. This effect may be offset by a lower grain photoelectric heating \citep[e.g.,][]{Jameson2018}, caused by a lower abundance of polycyclic aromatic hydrocarbons \citep[PAHs;][]{Chastenet2019}, which dominate the photoelectric heating \citep{Bakes1994,Wolfire1995}. The resulting thermal balance determines the distribution of gas temperature in the PDR which may affect the chemistry and abundance of atoms and molecules. 

PDRs are also critical for understanding the cold molecular gas content of the ISM, because they encompass the transition from H to \htwo and ionized carbon (C$^+$) to neutral carbon (C) to carbon monoxide (CO).
Cold \htwo ($T_{\rm gas}\lesssim 100$~K) is hard to directly observe due to the required excitation energy of its rotational levels, \textit{E(\htwo) $>>$ kT$_{\rm gas}$}. Because of this observational limitation, CO is often used to trace the bulk cold molecular gas, as it is highly abundant and easily detectable at typical molecular cloud densities and temperatures \citep{Bolatto13}. 

This makes understanding the transitions from C$^+$/C/CO and H$^+$/H/\htwo crucial for defining where we can trace \htwo using CO. These two transitions are not fundamentally at the same spatial location in a PDR, due to differences in shielding mechanisms. H$_2$ is able to self-shield via the Lyman Werner bands \citep{Wolfire10, Gnedin&Draine2014}, while the C/CO transition is primarily governed by dust shielding and occurs at higher \av, deeper in the PDR. Processes like CO self shielding and mutual H$_2$/CO shielding can also contribute to the survival of CO \citep{vanDishoeck&Black88}.

In low metallicity environments, the separation between the H/\htwo and C$^+$/C/CO transitions is expected to increase. This is a consequence of the dust-to-gas ratio dropping with metallicity \citep{Remy-Ruyer2014, Roman-Duval2022} resulting in lower dust extinction relative to the column density of hydrogen ($A_{\rm V}/N_{\rm H}$), allowing FUV photons to penetrate deeper into the cloud \citep{Wolfire10, Jameson2018}, photodissociating CO, while H$_2$ is protected by self-shielding. As a result, low metallicity PDRs tend to have larger extents for the same A$_V$ \citep{Bolatto13, Leroy11}, causing a larger separation in the chemical transitions. Notably, the shielding of CO by dust is expected to only occur for \av $\gtrsim 2$ \citep{vanDishoeck&Black88, Sternberg&Dalgarno95, Smith14, Glover16}.

Due to the larger separations in the locations of the H/H$_2$ and the C$^+$/C/CO transition at low metallicity, a significant portion of the H$_2$ mass resides in the ``CO-dark'' region (where the carbon is either C$^+$ or C) relative to the CO bright region \citep{Wolfire10, Bell07, Madden20, Bisbas24}. Studies of the SMC point to almost 80\% of the H$_2$ mass being in the CO-dark phase \citep{Israel97, Leroy11, Bolatto11, Pineda17, Jameson2018}, compared to only around 30\% in the Milky Way \citep{Grenier05, Pineda13}. This results in a metallicity-dependent $X_{\rm CO}$ conversion factor for unresolved clouds, which accounts for the large amounts of CO-dark H$_2$ observed in low metallicity environments \citep{Bolatto13, Madden20, Gong2020}. Given that star formation earlier in the history of the universe occurred in low metallicity gas, this metallicity dependence could greatly impact our understanding of high-z observations. Though crucial to studies of ISM physics, resolved predictions of low metallicity PDR models have never been directly tested due to a lack of observations that can resolve each of the individual boundaries of a PDR. Resolving a low metallicity PDR will therefore shed light on the CO-to-H$_2$ conversion factor metallicity dependence in other low metallicity environments such as those in the high-redshift universe. 

Prior to JWST, it was not possible to resolve PDR structures in extragalactic PDRs, particularly the H/H$_2$ transition (as traced by ro-vibrational H$_2$ emission), due to limits of angular resolution in the near- and mid-infrared. In the Milky Way, to resolve the layers of a PDR, it is essential to reach typical scales of a few $10^{-3}$ pc  \citep{Joblin18, Berné_2022}.  While this resolution is still out of reach with JWST anywhere but the Milky Way, the predicted larger spatial extent of PDRs at low metallicity means it is now possible to resolve PDRs in the SMC, at a distance of 62 kpc \citep[1$''=0.3$ pc;][]{Scowcroft16} and metallicity of 1/5 Z$_{\odot}$ \citep{Toribio17}. The capabilities of JWST and ALMA therefore enable, for the first time, resolving low-metallicity extragalactic PDR structures.

This paper employs JWST and ALMA observations to spatially resolve key PDR transitions in the N13 PDR in the SMC. Using these results, we compare to steady-state plane-parallel PDR models, previously applied to SMC observations \citep{Jameson2018}, with a set of reasonable assumptions for SMC conditions. In Section \ref{sec:obs&reduction}, we introduce our target and data products. In Section \ref{sec:Methods}, we discuss the creation and analysis of intensity maps and the comparison to PDR models. In Section \ref{sec: Results}, we analyze the observed PDR structure. In Section \ref{sec:Disscussion}, we evaluate PDR models and potential influences on the PDR and conclude that the constant pressure model for N13 aligns best with our observations. In Section \ref{sec: conclusions}, we discuss the implications of these results for CO-dark \htwo.

\section{Observations \& Data Reduction} \label{sec:obs&reduction}

\subsection{Target} \label{sec: Target}
\begin{figure*}[ht!]
    \begin{center}
        \includegraphics[width=\textwidth,trim=0mm 0mm 0mm 0mm,clip]{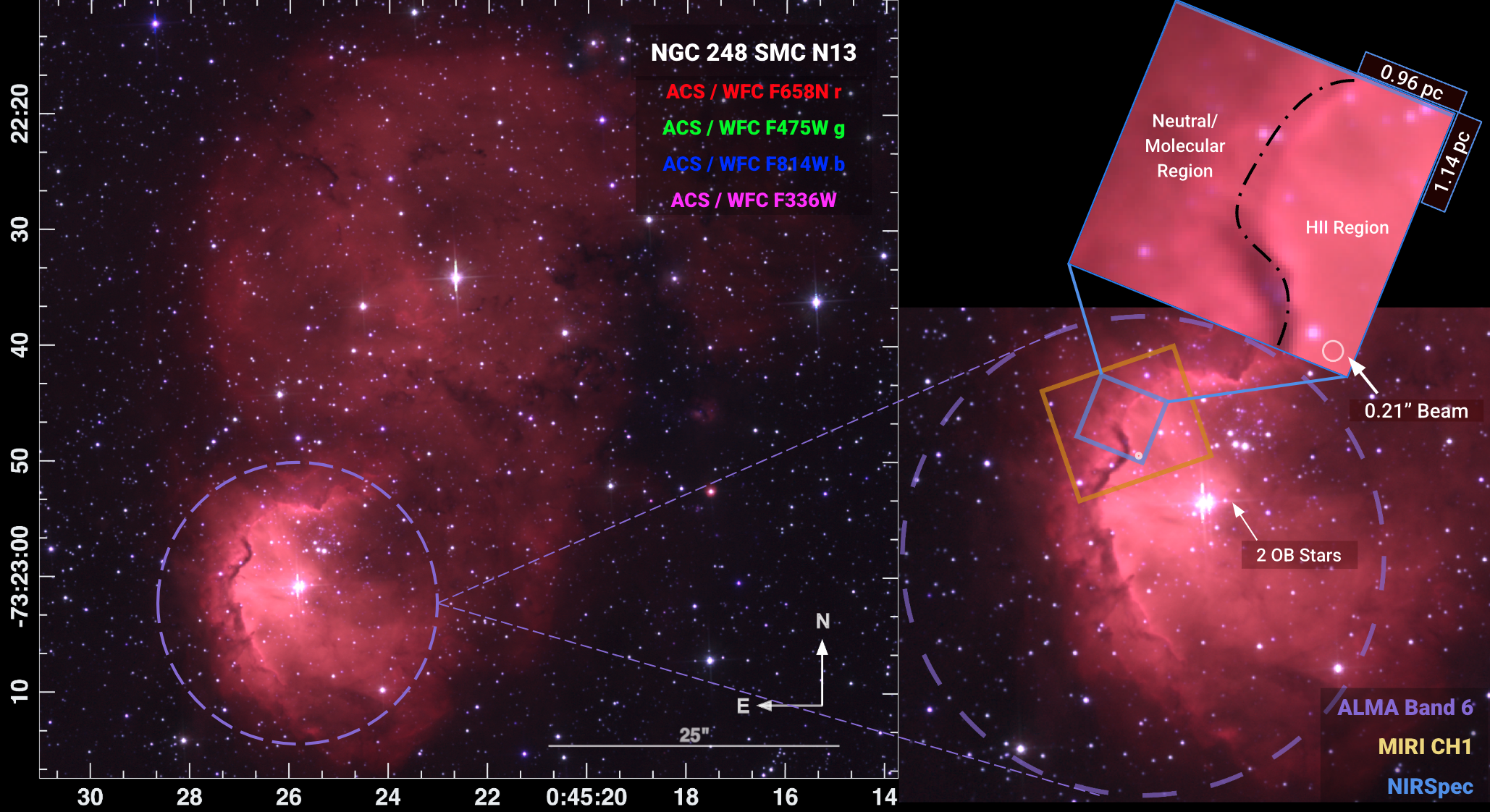}
    \end{center}
    \caption{In the left panel, we show an HST image of NGC 248, the star-forming region that contains the N13 \hii\ region as seen inside of the purple dashed circle. In the right panel, we show the N13 PDR with the apertures of the different telescope instruments overlaid. This panel also shows a zoom-in of the dark edge of the PDR that shows where the molecular gas and dust reside, with a dot-dashed line giving the approximate by-eye location of the PDR boundary. The two ionizing OB stars that power the N13 PDR are labeled below the apertures. The zoom-in on the location of the NIRSpec aperture} shows the sub-parsec scale spatial resolution we achieve with JWST and ALMA.
    \label{fig:2}
\end{figure*}

We investigate a PDR in the N13 region of the Small Magellanic Cloud (SMC) located at R.A.: $00^{\rm h} 45^{\rm m} 26\fs 760$, Decl.:$- 73\degr 22' 55\farcs66$. The SMC sits at a distance of 62 kpc \citep{Scowcroft16} and a sub-solar metallicity of $Z=0.2 Z_\odot$ with no systematic gradients across the galaxy \citep{Toribio17}. N13 is a useful laboratory to explore the effects of a low metallicity environment on the gas and dust properties due to a simple stellar population of two OB stars, similar to the Orion Bar. We selected the region based on the appearance of an edge-on geometry for the PDR in narrowband H$\alpha$ observations from Hubble \citep[HST;][]{Smidge}. Edge-on geometry maximizes the angular separation of the layers to avoid blending and allows for precise spatial identifications of each layer. The geometry appears simple from HST imaging, but we investigate potential inclination effects in Section \ref{sec:inclination}. At the distance of the SMC, for an edge-on PDR, we can resolve each PDR layer at spatial resolutions between $0.03-0.21$ pc with the JWST NIRSpec and MIRI-MRS integral field unit (IFU) resolution between 0\farcs1--0\farcs7. On the left side of Figure \ref{fig:2}, we show N13 in three HST filters, described in the top right corner \citep{Smidge}. We present a zoom-in of the N13 PDR on the right side of Figure \ref{fig:2}, which is located within the purple dashed circle. The right side of Figure \ref{fig:2} also shows the field of view of the JWST and ALMA observations.

\subsection{JWST NIRSpec Integral Field Spectroscopy}
We observed the N13 PDR using JWST NIRSpec \citep{Jakobsen2022,Boker2022} and MIRI-MRS \citep{Argyriou2023} integral field units (IFUs) as part of program GO 2521 in JWST Cycle 1. NIRSpec observations were conducted on July 29, 2023, and MIRI-MRS on July 21, 2023. Both NIRSpec and MIRI-MRS used one pointing with four dithers to sample the point spread function (PSF). For NIRSpec, a ``leakcal'' was taken for each dither to mitigate MSA slit leakage, while MIRI-MRS included ``off'' observations to remove foreground contamination and pixel-based residuals. We used three NIRSpec medium-resolution gratings (G140M/F070LP, G235M/F170LP, G395M/F290LP) and all MIRI-MRS channels and gratings spanning 5–28 \micron. 
In the following, we present results using the higher spatial resolution observations from NIRSpec to dissect the PDR.  The description of MIRI-MRS observations and their analysis will be presented in a future paper.

Data were downloaded from MAST and reduced using the JWST pipeline. For NIRSpec, we used a development version of the pipeline that allowed for $1/f$ noise removal from both the ``on'' and ``leakcal'' observations (JWST pipeline version 1.16.1.dev14) using CRDS \texttt{jwst\_1293.pmap}. We processed the raw \texttt{uncal} files through the Detector1 pipeline, including $1/f$ noise correction with the \texttt{clean\_flicker\_noise} step. We then ran the Spec2 and Spec3 pipeline stages and created a drizzled cube with 0\farcs05 pixels for each grating.  Typical uncertainties per spaxel in the cubes range from $0.3-0.5$ MJy sr$^{-1}$ for G140M and G235M and $0.3-1$ MJy sr$^{-1}$ for G395M.

\subsubsection{Astrometric Alignment}
Given the sub-parsec scale separations we aim to measure in the PDR, ensuring accurate astrometry is critical. The astrometry of the ALMA observations is well understood due to the nature of interferometric measurements with extragalactic radio sources as phase calibrators. The astrometry of the JWST observations, however, can have offsets related to uncertainty in the positions of guide stars. 

To correct the JWST astrometry, we compared to archival data from HST \citep{Smidge} in the ACS/WFC F475W filter which we aligned to Gaia DR3 (HST has similar astrometric uncertainties related to guide stars, but a much larger field of view than the JWST IFUs). We found 1000 Gaia DR3 catalog stars within the HST field to use as our astrometric reference. Using the \textit{Photutils Centroids} Python package \citep{photutils}, we measured the centroid positions of these stars in the HST F475W image and computed average offsets in RA and Dec to find the overall astrometric shift in the HST data. We find a Gaia-HST offset in RA of $-0\farcs174 \pm 0\farcs039$ and in Dec of $0\farcs158 \pm 0\farcs010$ where the errors are the standard deviation over the 1000 Gaia DR3 stars. 

We then identify four stars within the NIRSpec G140M cube that are also evident in the HST imaging and use these to correct the JWST astrometry, using the same centroid and averaging method as described above. We find an HST-JWST offset in RA of $-0\farcs444 \pm 0\farcs022$ and in Dec of $-0\farcs197 \pm 0\farcs011$, where the error listed here is the standard deviation of the offsets. To check our astrometric correction, we compared Gaia astrometry-corrected HST data with the ALMA $^{12}$CO J=2$-$1 moment zero map. The alignment between the CO emission and the dust lane in the HST image matched well, giving us confidence that the JWST-HST-ALMA alignment was robust.

\subsubsection{Integrated Line Maps from JWST}\label{sec:linemaps}
To measure the integrated intensity of spectral lines in the JWST spectral cubes, we used two different approaches based on the line's intensity, the complexity of decomposing its emission from surrounding PAH features, and potentially blended spectral lines. For H recombination lines, and the 3.3 $\micron$ PAH emission feature, we used the python implementation of the PAHFIT package\footnote{\url{https://github.com/PAHFIT/pahfit}} \citep{Smith2007}. This model works well for PAH features and for bright and/or blended emission lines. For fainter lines, like the H$_2$ 1$-$0 S(1) 2.12 \micron\ vibrational line, errors in the local PAHFIT continuum fitting can be significant, so we instead do a local continuum fit and integrate under the line. In this case, we defined continuum regions around each line, fit a 1-d polynomial, and then subtracted the fitted continuum before integrating under the line. 

\begin{figure*}[h]
    \centering
    \includegraphics[width=0.43\textwidth]{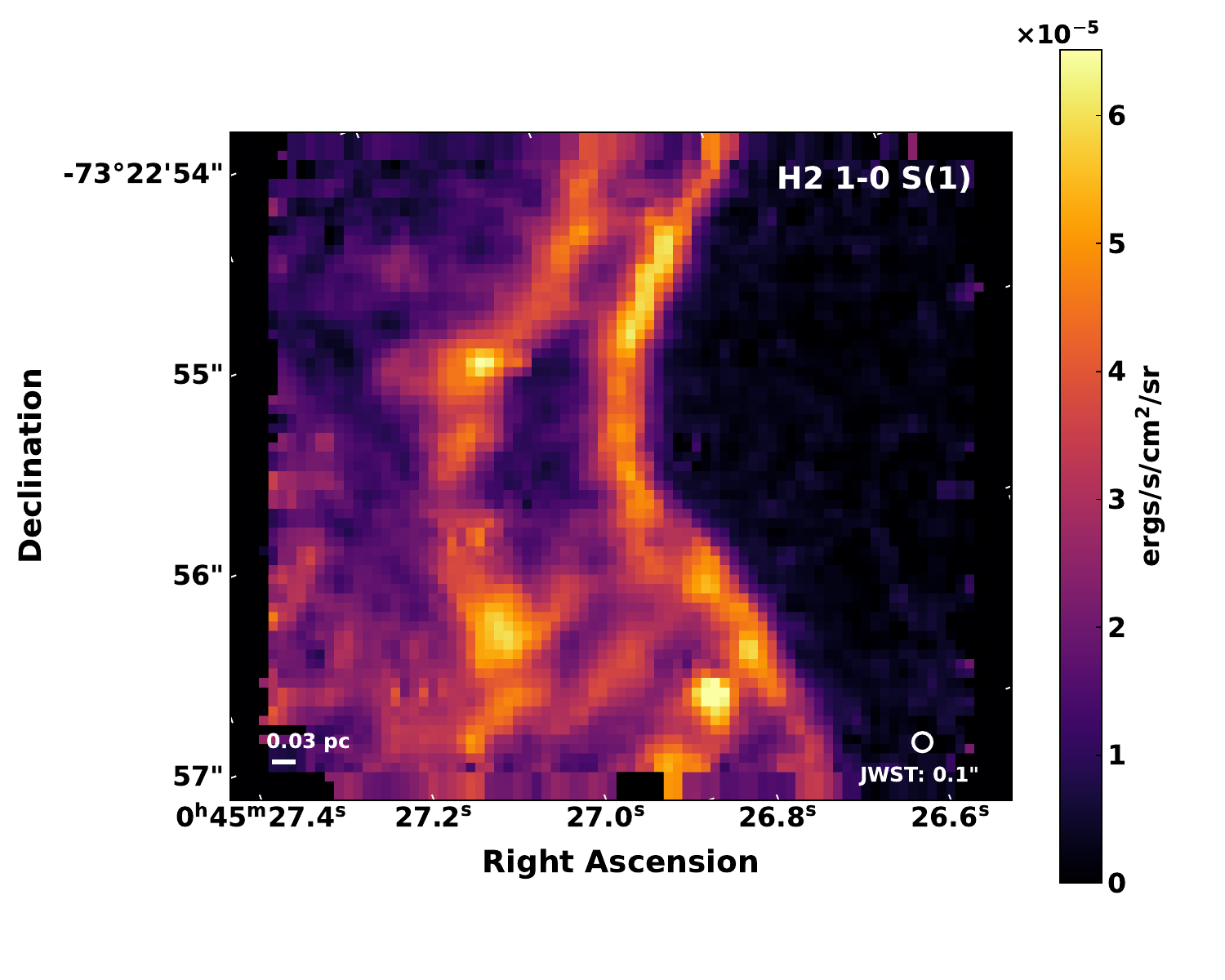}
    \includegraphics[width=0.43\textwidth]{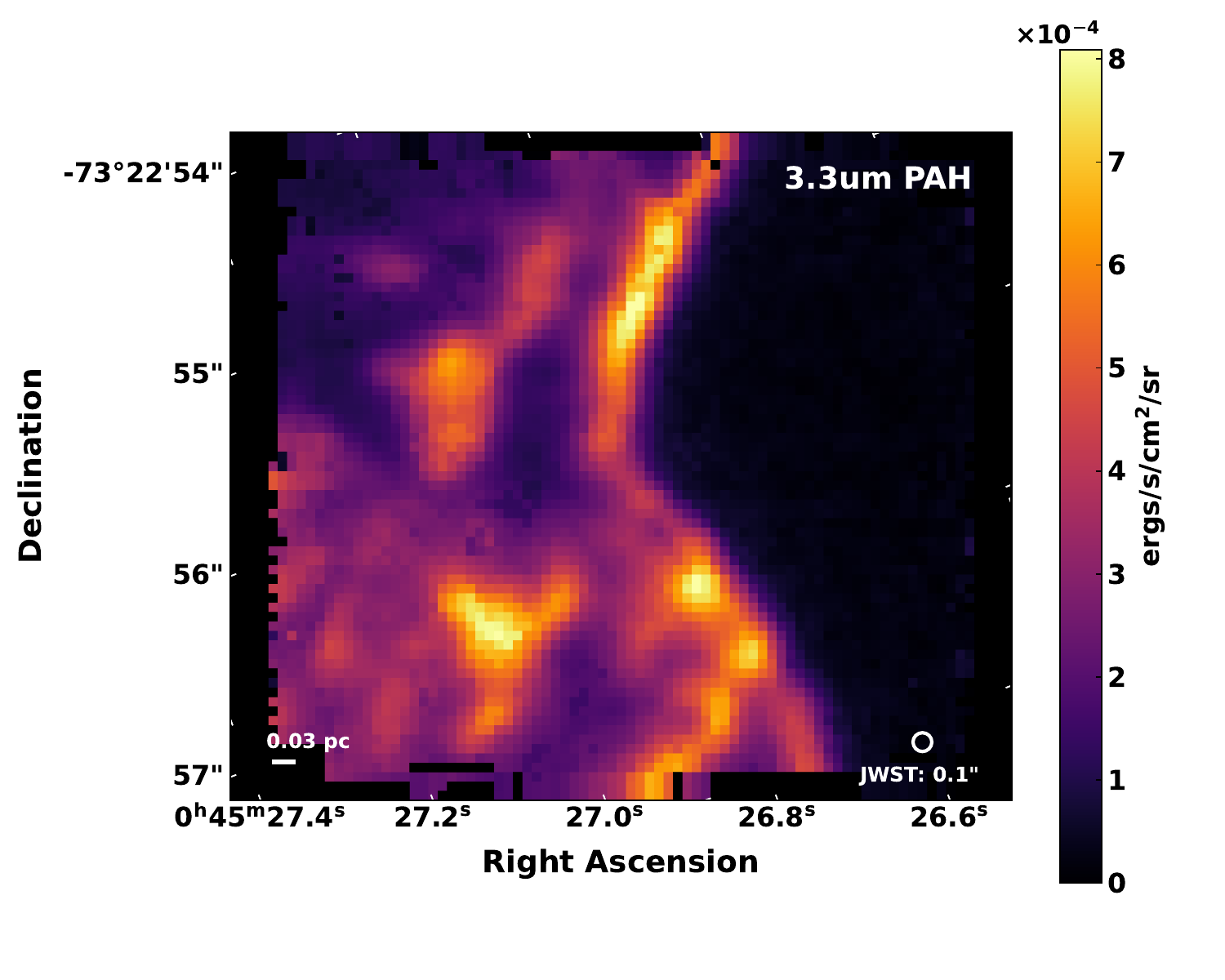} \\ \vspace{-0.2in}
    \includegraphics[width=0.43\textwidth]{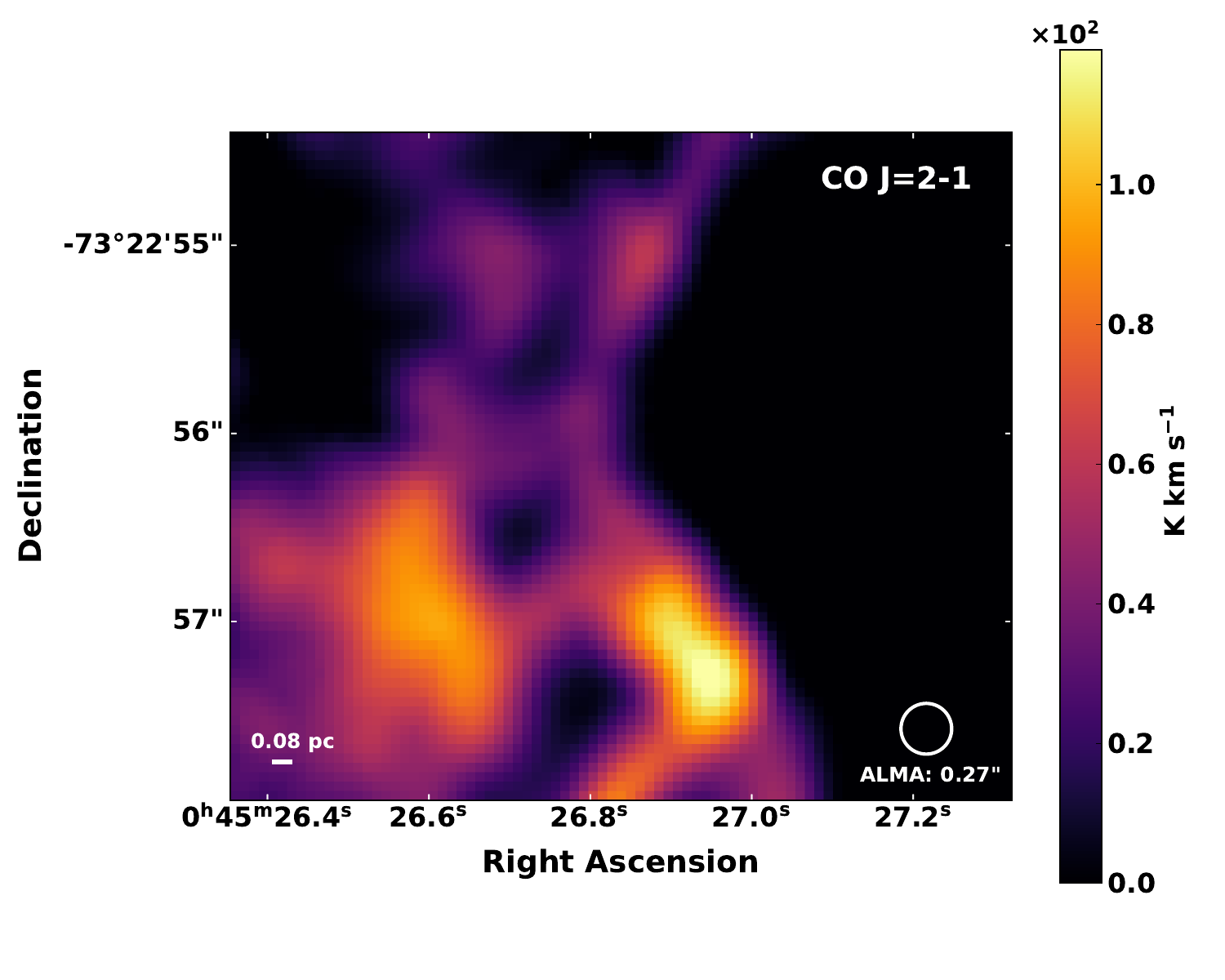}
    \includegraphics[width=0.43\textwidth]{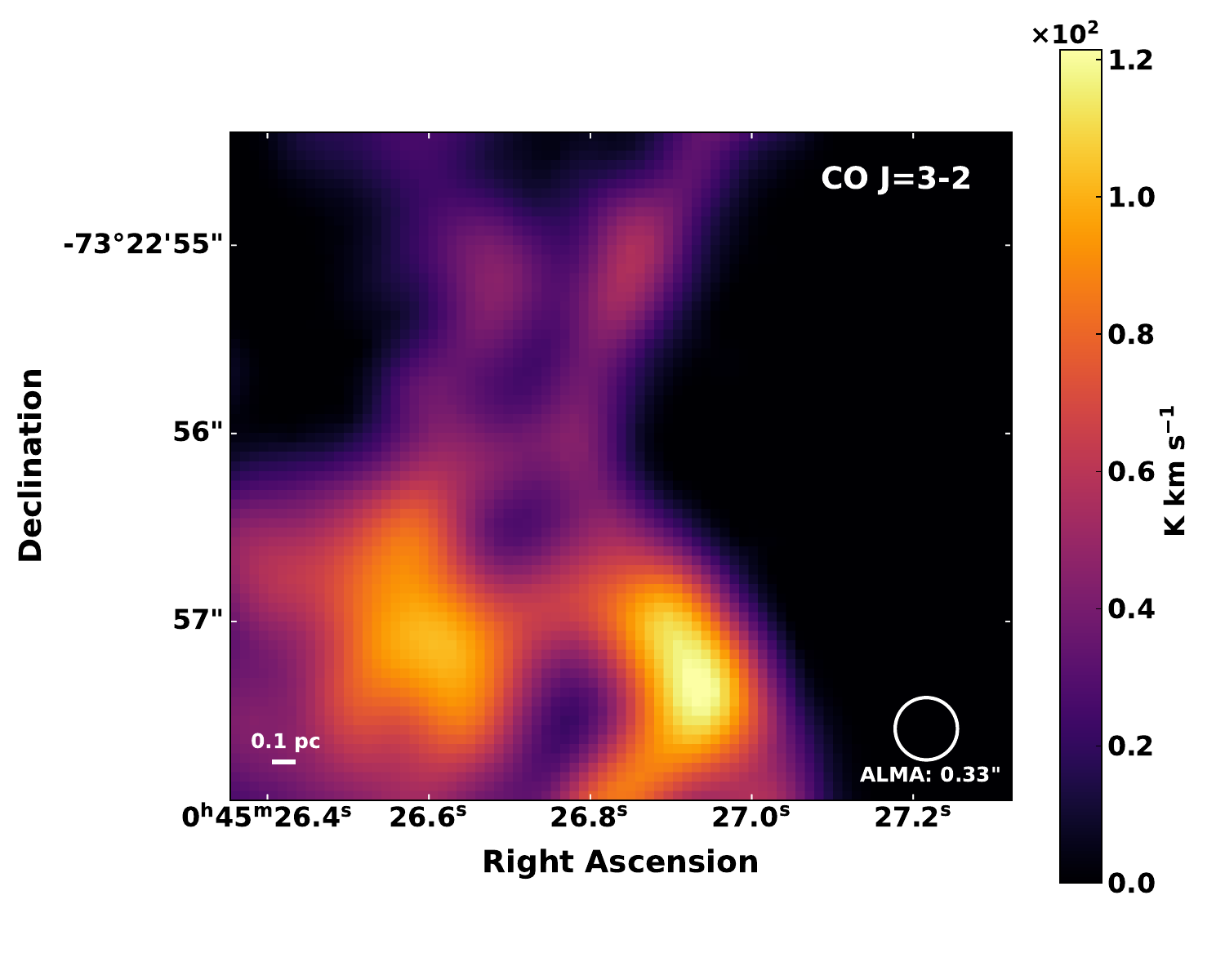} \\ \vspace{-0.2in}
    \includegraphics[width=0.43\textwidth]{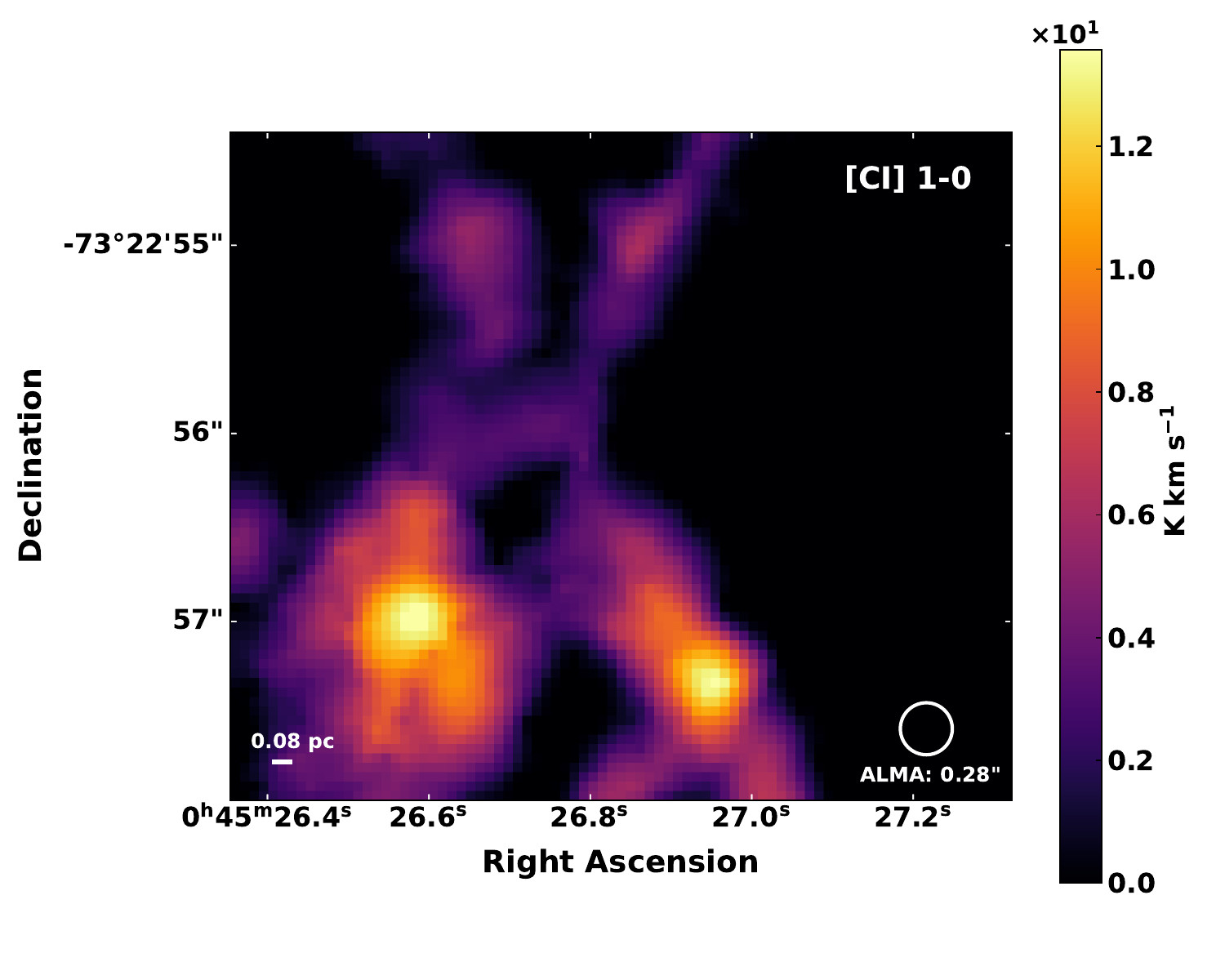}  
    \includegraphics[width=0.43\textwidth]{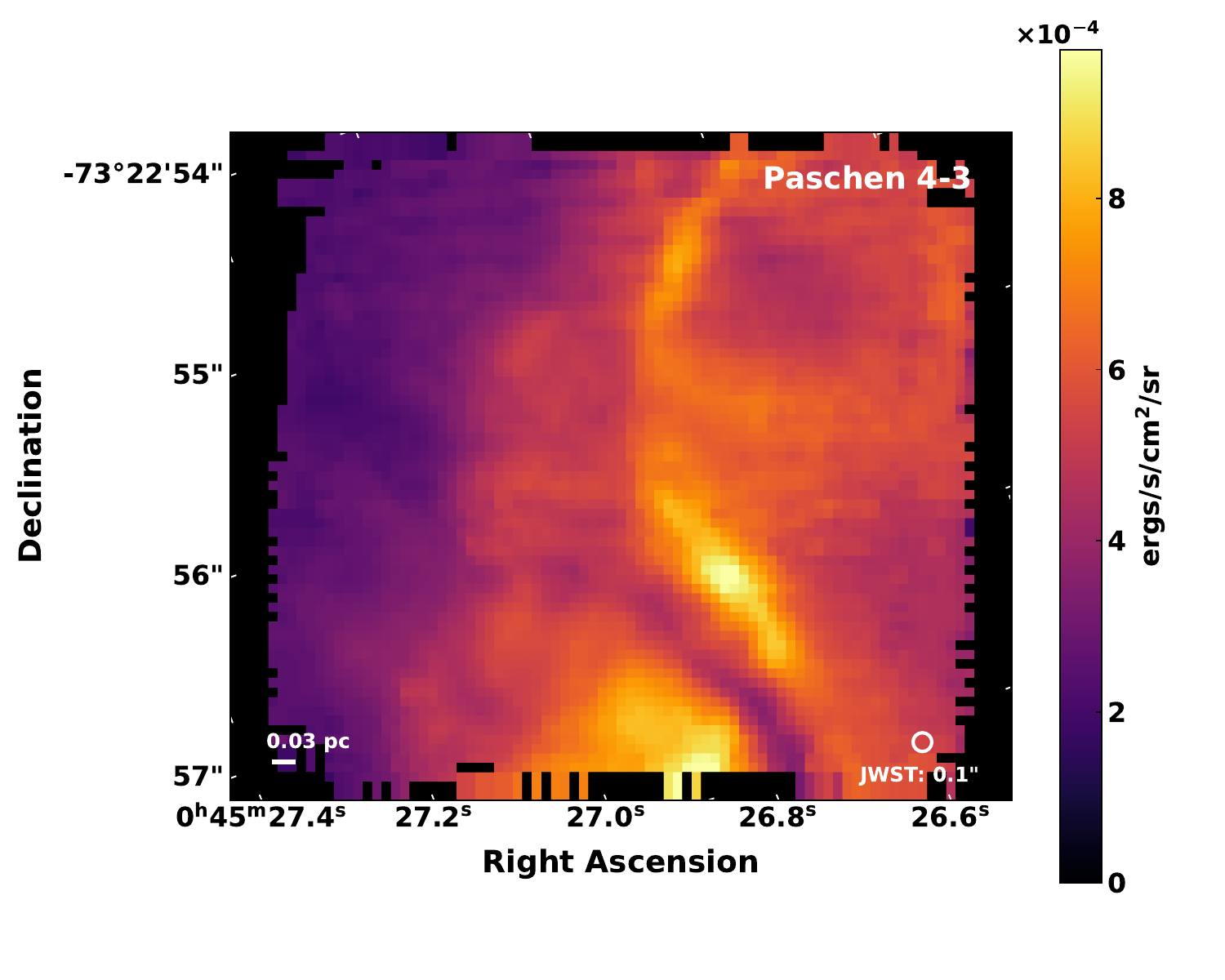}\\ \vspace{-0.2in}
    \includegraphics[width=0.43\textwidth]{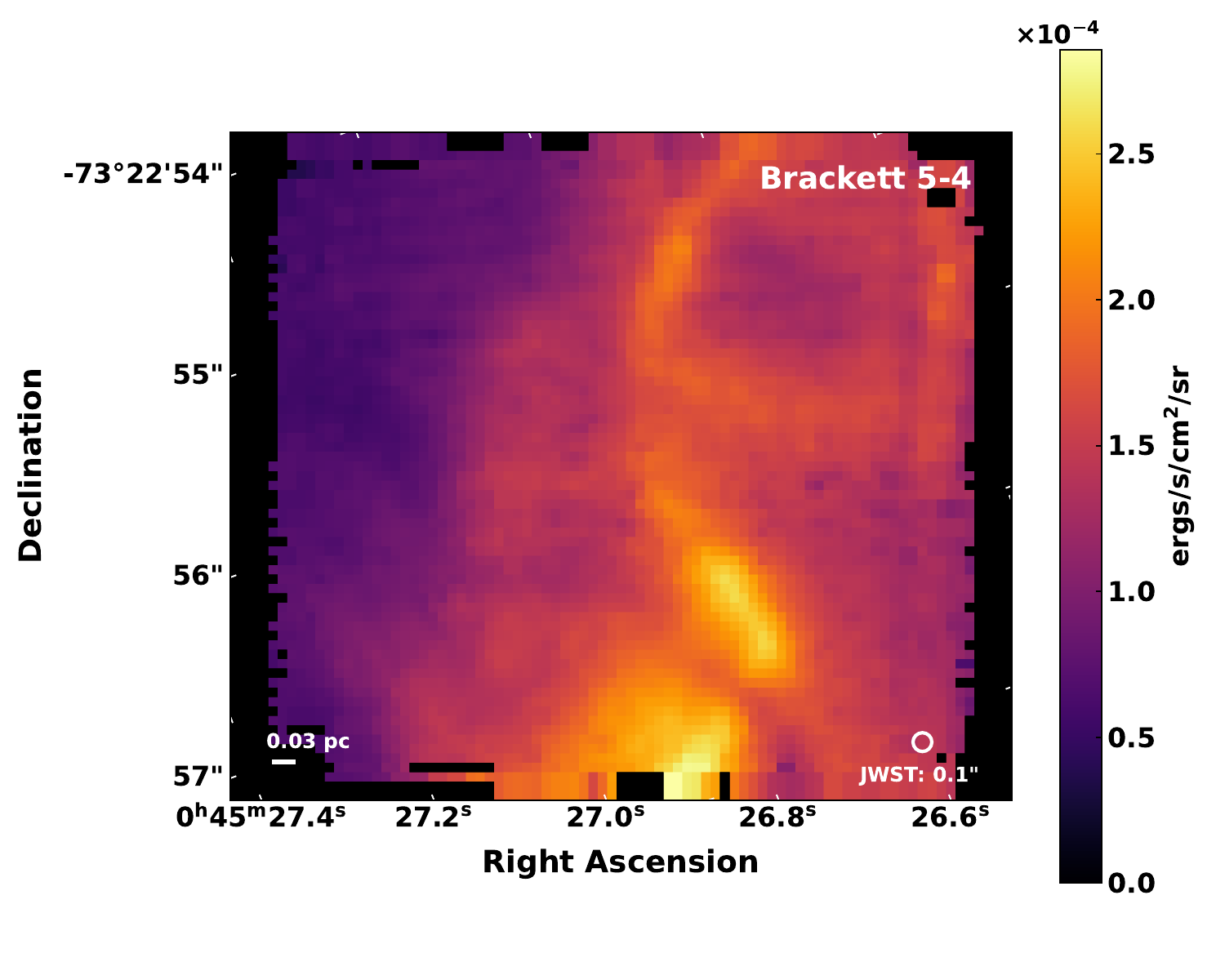}
    \includegraphics[width=0.43\textwidth]{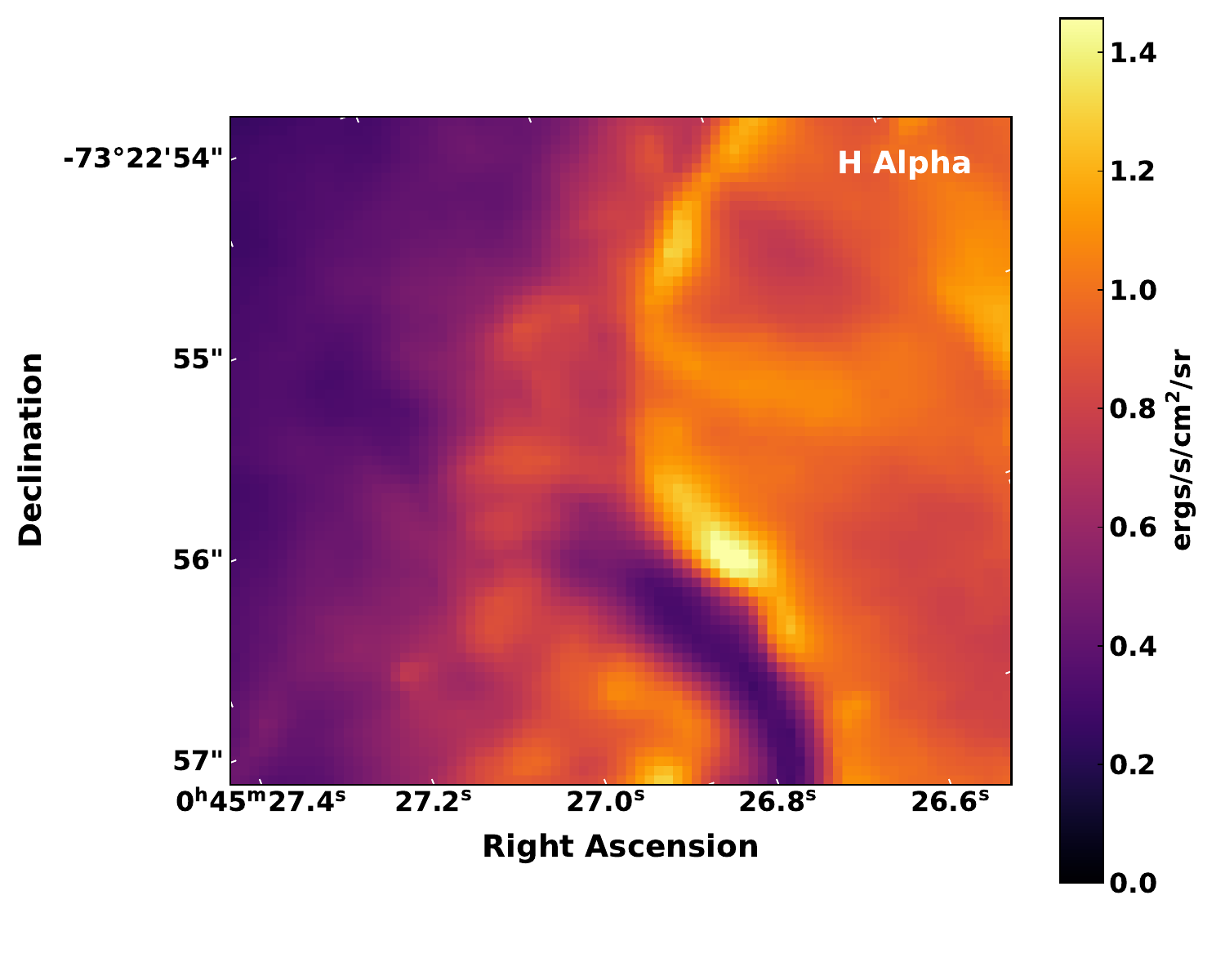} \vspace{-0.2in}
    \caption{Integrated intensity maps for the H$_2$ 2.12 \micron\ line, 3.3 \micron\ PAH feature, H~I 4-3 Paschen $\alpha$, H~I 5-4 Brackett $\alpha$, and the moment zero maps of $^{12}$CO J=2$-$1, $^{12}$CO J=3$-$2, and [CI] 1-0. We also show the F658N HST filter for reference, as we use this map for the H-$\alpha$ radial profile discussed in Sec.~\ref{sec: linear profiles}. The structure of the PDR is well resolved in all of the tracers, showing two dissociation fronts (DFs) and a filamentary structure. }
    \label{fig:maps}
\end{figure*}

We fit emission lines and PAH features in each spaxel and created maps of the integrated feature strengths.
We applied the fitting to all spaxels in the NIRSpec cubes.  We created resolved integrated intensity maps for key lines such as the 2.12 \micron\ H$_2$ 1-0 S(1), 4.05 \micron\ H~I 5-4 Brackett $\alpha$, 1.87 \micron\ H~I 4-3 Paschen $\alpha$, and the 3.3 \micron\ PAH feature.  We calculate S/N values at the first peak of the radial profiles (discussed in Sec.\ref{sec: linear profiles}) to be $\sim 58$ for H$_2$ 2.12 \micron, $\sim 179$ for H~I 4-3 Paschen $\alpha$, $\sim 87$ for H~I 5-4 Brackett $\alpha$, and $\sim 29$ for the 3.3 \micron\ PAH feature. We also use archival data from the HST F658N narrowband photometry, obtained from \cite{Smidge}, to trace H-$\alpha$. Figure~\ref{fig:maps} shows the resulting line and PAH maps for N13. Further analysis of these maps is provided in Section~\ref{sec:dfproject}. 
 
\begin{figure*}[]
    \centering
    \includegraphics[width=\textwidth, trim={2.2cm, 0cm, 0cm, 0cm}]{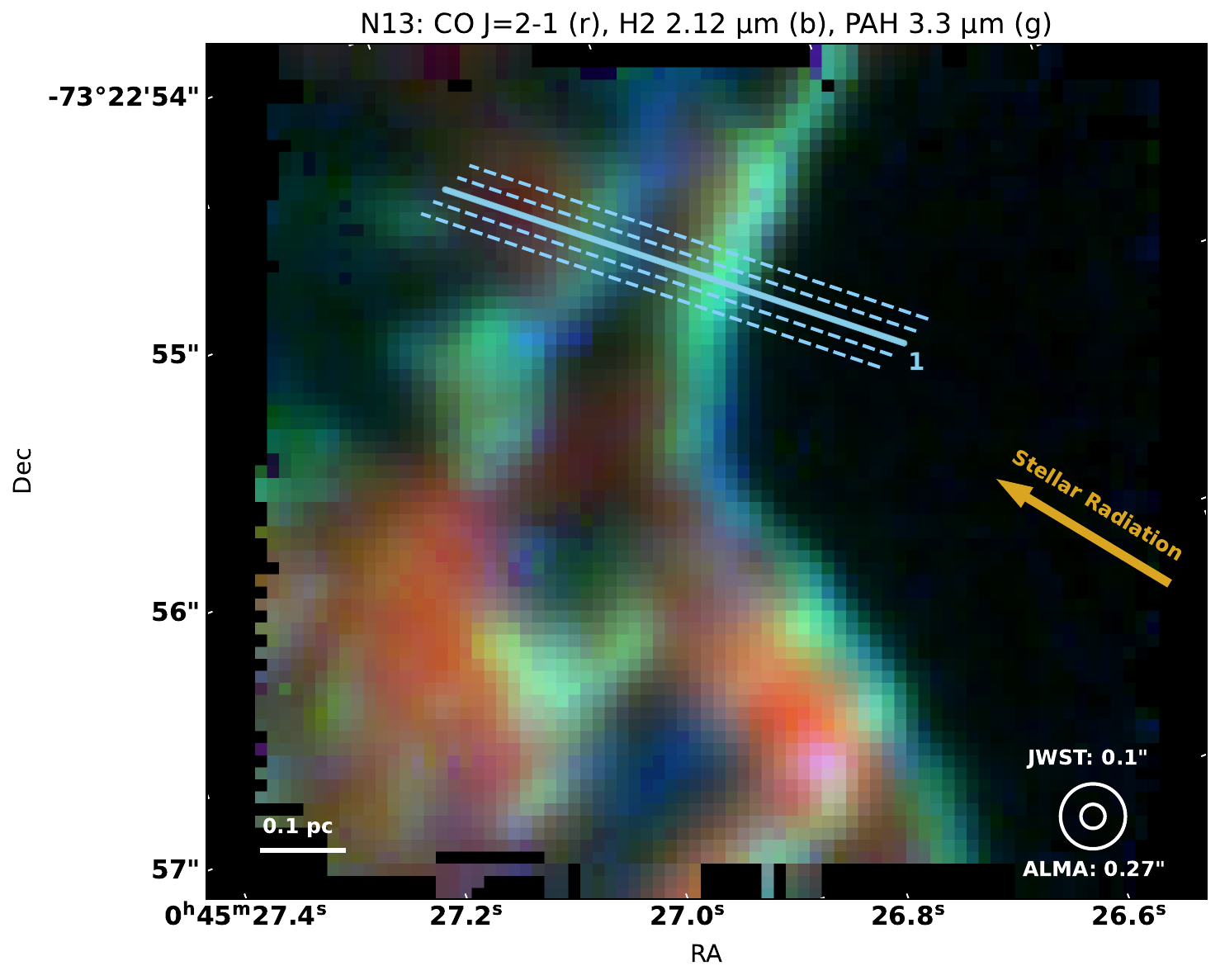}
    \caption{Three-color image of the N13 PDR, with H$_2$ 2.12 \micron\ in blue, CO J=2-1 in red, and the 3.3 \micron\ PAH feature in green. We show our main radial line profile in blue, labeled as ``1'', interpolated at 0\farcs01 spacing. The dashed blue lines indicate the additional slices used to calculate uncertainty, separated by the pixel scale. The starting coordinates for each slice are in the \hii region, closer to the illuminating stars located to the right of the map. The radial profiles are discussed further in Sec. \ref{sec:radialprofile}}
    \label{fig:3-color image}
\end{figure*}

\subsection{ALMA}\label{alma-phangs-pipe}
We obtained data for $^{12}$CO J=2$-$1 in Band 6, $^{12}$CO J=3$-$2 in Band 7, and [CI] $^3{\rm P}_1$-${^3}{\rm P}_0$ 1$-$0 in Band 8 using the ALMA 12-m array and Atacama Compact Array (ACA) 7-m array in Project ID 2021.1.01065.S. The target angular resolution of 0\farcs25 (0.075 pc) was set to resolve the PDR layers given predictions from PDR models described in Section~\ref{sec:models}. We observed a single pointing for all ALMA observations, as the PDR is smaller than the field of view in all Bands. The 12m and 7m configurations included in each observation were set to recover angular scales up to at least 15$''$, which covers the angular extent of the molecular cloud in N13 detected in previous observations \citep{Saldano2024} and is larger than the JWST field of view\footnote{At present, the second 12m configuration for the Band 8 [CI] observations has not been observed. Using the full Band 8 dataset for CO J$=$(3$-$2) which has the same set of configurations, we tested the effect of the missing 12m configuration on the location of the peak in the PDR and did not see any significant differences. The lack of 12m data yields lower than planned S/N, but does not affect the observed peak location which is the focus of this paper.}. We did not observe the $^{12}$CO J=1$-$0 line due to low surface brightness with the extended configuration necessary to reach 0\farcs25 resolution. The Band 6, 7, and 8 observations used 0.09, 0.12, and 0.09 km ${\rm s}^{-1}$ velocity resolution, significantly higher resolution than the line widths of $\sim$0.5 km s$^{-1}$. The observed bandwidths for Band 6, 7, and 8 each cover $>140$ km s$^{-1}$, encompassing the velocity extent of the emission in this portion of the SMC. 
 
We used version 1.0 of the PHANGS-ALMA pipeline\footnote{The PHANGS-ALMA pipeline can be found at \url{https://github.com/akleroy/phangs_imaging_scripts}.} \citep{Leroy21b} to image the calibrated data from the 12m array and ACA and generate cubes and moment maps.  We convolve all cubes to have circular Gaussian beams, but do not convolve to matched spatial resolution.  For our goal of identifying the layers of the PDR, the highest resolution version of the data is ideal.  The final resolution of the cubes are:  0.270$''$ for $^{12}$CO J=2$-$1, 0.331$''$ for CO 3$-$2, and 0.276$''$ for [CI] 1$-$0. The moment map generation includes a step of signal masking to create high confidence moment maps, following the ``broad'' mask procedure in the PHANGS pipeline.  
In our $^{12}$CO J=2$-$1, $^{12}$CO J=3$-$2, and [CI] 1$-$0 moment zero maps we find an RMS value of $1.464 {\rm ~K~km~s^{-1}}$, $1.098 {\rm ~K~km~s^{-1}}$, and $1.112 {\rm ~K~km~s^{-1}}$ respectively.  We calculate S/N values at the peak of the radial profiles (discussed in Sec.~\ref{sec: linear profiles}) to be $\sim 39$ for $^{12}$CO J=2$-$1, $\sim 56$ for $^{12}$CO J=3$-$2, and $\sim 10$ for [CI] 1$-$0

\section{Methods}\label{sec:Methods}
We aim to map the spatial structure of the N13 PDR and compare to PDR models in order to determine the separations between the IF, DF, and C/CO transition. Observationally, the available tracers are emission lines that emerge from gas at different depths in the PDR. While it is possible to select tracers that should have distinct spatial profiles across PDR boundaries (e.g. peaks or drops), we are limited by the angular resolution of our observational dataset and by the lack of direct observables for the abundances of the relevant species. In order to compare to models, we use the volume emissivity of the relevant emission lines, which translates the abundance and temperature profiles of the PDR models into ``observable'' space. We further convolve each model emissivity profile to match the angular resolution of the corresponding observed emission line. This allows us to locate the emission peaks in both models and observations for each relevant tracer to identify the PDR boundaries.

The emission lines we use to compare the observations and models include HI Paschen-$\alpha$ $1.87$\micron, which should show a decrease in emission after the IF; \htwo 1-0\,S(1) $2.12$\micron, which should peak near the DF; and [CI] $^3{\rm P}_1$-${^3}{\rm P}_0$ $609$ \micron, $^{12}$CO J=2$-$1, and $^{12}$CO J=3$-$2 which should trace the transition from C to CO. The correspondence between emission profiles and the expected location of the PDR boundaries is discussed in Section~\ref{sec:dfproject}. We create radial profiles to characterize each key tracer map, as described in Section~\ref{sec:radialprofile}.

Our comparison PDR models are tailored for SMC conditions and anchored to the density and/or pressure observed in the nearby H~II region, as described in Section~\ref{sec:models}. In identifying the best-matched model to our observations we allow for small changes in density and pressure in the models, within their uncertainties, and select the model that provides the closest match to the observed peak spacings between the DF and the C/CO transition. We align model predicted peak locations with our observations rather than doing a formal fit of the models. A formal fitting procedure is not warranted given the limited comparison (DF to C and DF to CO peak locations), and the large number of additional model parameters. Translating PDR models to observables via the volume emissivities is a standard way to compare the location of emission peaks from edge-on models with observations \cite[e.g.,][]{Joblin18, Goicoechea19}. The calculation of model intensities would depend on additional considerations such as the angle of the line-of-sight and optical depth effects in the line. Our approach enables a matched resolution comparison of the spatial separation between the DF and the C/CO transition, allowing us to evaluate how well the models reproduce the observed low metallicity structure. In doing this comparison between modeled and observed boundary separations between peaks we characterize N13's PDR structure and the extent of CO-dark \htwo content between the DF and the C/CO boundary. We step through this process in more detail below.

\subsection{Radial Profile Analysis}\label{sec:radialprofile}
To analyze the spatial separation of the PDR boundaries of N13, we generated radial profiles along a slice perpendicular to the PDR.  We selected the end coordinates of a perpendicular slice by visual inspection using the CARTA software package \citep{CARTA}. This slice was selected to be as perpendicular as possible to the PDR H$_2$ emission to yield a clean, simple radial profile. We chose this particular placement of the slice to avoid complex structure from an embedded YSO in the south and to have enough coverage for the profile to not extend past the edge of the cube in the north. We find this location of the slice to produce profiles that have a clear peak for PDR boundary analysis for all of the maps analyzed.

We measured the integrated intensity of H$_2$ 2.12 $\micron$, CO J=2$-$1, CO J=3$-$2, and [CI] 1$-$0 emission along the slice with a step size of 0\farcs01, which oversamples the resolution element of JWST and ALMA. We used the \texttt{griddata} cubic interpolation method to measure the intensities at each point of the slice.
In Figure \ref{fig:3-color image}, we show the main perpendicular slice in blue. We provide the radial profile measurements for each map in Table \ref{tab:radial slice h2 co} and Table \ref{tab:radial slice hi pah}. The parallel dashed blue lines represent additional slices, offset by the pixel scale, used to estimate uncertainties in the peak spatial placement. It is important to note that the OB stars powering N13 are not the starting point of the slice. The primary goal of the slice is to locate the different peaks from each PDR tracer and measure their separations relative to each other. In Figure \ref{fig:models_conv}, the $x$-axis of the profiles increases away from the OB stars, where zero marks the point closest to the stars (5.992$''$ or 1.81 pc pc away from the OB stars).

\subsection{PDR Models}\label{sec:models}
The density distribution of the gas is one of the main parameters in PDR modeling. Many PDR models use constant density, plane-parallel, semi-infinite slabs of gas and dust to model observations \citep[e.g.][]{PDR1985}. While constant density plane parallel models can successfully describe line intensities and the spatial separation of layers in some PDRs, there are several processes that can modify this picture. Constant thermal pressure models tend to increase the density in the deeper (cooler) layers and lead to a convergence of layers \citep{Joblin18}. Density inhomogeneities like high-density clumps, produced by compression from turbulence \citep{Glover2010} or photoevaporation \citep{Gorti02}, can lead to spatially unresolved H/H$_2$ and C$^+$/C/CO transitions surrounding the denser clumps. Photoevaporation from the PDR surface into the H~II region can lead to an advection flow that draws the H$_2$ and C/CO layers towards the IF leading to a convergence of layers \citep{Storzer1998,Bron18, Mailard2021}.  Endoergic carbon chemistry, overcome by FUV-pumped, excited H$_2$ can produce carbon species such as CO and HCO$^+$ coincident with the H$_2$ \citep{Sternberg&Dalgarno95,Goicoechea16}, also leading to overlapping H$_2$ and C/CO layers. The plane-parallel (steady-state) model is a more simplistic geometry for PDR structure, but may still provide a reasonably good fit for some PDRs.
 
Differences between the plane-parallel and clumpy models mainly hinge on the surface-to-volume ratio of the model PDR. Future efforts may be able to model 3D PDR structures from simulated molecular clouds 
\citep{Bisbas12} or use models 
that directly couple the hydrodynamics and chemistry \citep[][also see the review by \citealt{wolfire2022}]{Glover12, Grassi14, Bisbas15, Seifried17, Lupi18, Haid19, Seifried20, Bellomi20,Hu21,Gaches23,Gurman2024}.

Currently, it is not clear that low metallicity PDRs would necessarily be preferentially isobaric (constant pressure), isochoric (constant density), or clumpy. However, we can now test different model predictions for the C/CO transition in the low metallicity environment of the SMC. It is important to note that in models that have a fixed level of turbulence, along with a fixed radiation field spectrum and intensity and cosmic-ray rate, a low-metallicity cloud tends to be less clumpy than a high-metallicity one \citep{Glover2010}. This difference is primarily due to higher temperatures and lower turbulent Mach numbers at low metallicity. However, because there are still density substructures that form at low metallicities, we cannot necessarily conclude that clumpy models are not appropriate for low metallicity clouds.

To analyze our observations, we use our PDR model based on that of \cite{PDR1985} with updates to the dominant chemistry and thermal processes given in \cite{Kaufman06}, \cite{Wolfire10},
\cite{Hollenbach2012}, and \cite{Neufeld16}, and tailored for the SMC as in \cite{Jameson2018}.
These are plane-parallel models of a layer of gas and dust exposed to a far-ultraviolet radiation field with a fixed spectral shape and a cosmic-ray flux\footnote{The use of a fixed spectral shape is justified because \htwo and CO share the same narrow photodissociation wavelength band \citep{vanDishoeck&Black88}, implying that their photo processes scale similarly with an increase in FUV flux.} The abundances of the atomic and molecular species, as a function of depth into the cloud, are found from steady-state chemical balance, and the gas temperature from thermal equilibrium \citep[see e.g.,][for an exploration of non-steady state models at low metallicity]{Mailard2021}. We use a primary cosmic-ray ionization per H of $3.3\times 10^{-17}$ s$^{-1}$ H$^{-1}$ estimated from scaling the local Galactic value from \cite{Neufeld2017} by a factor 0.15 for the reduced density of cosmic rays in the SMC measured by {\em Fermi} \citep{Abdo2010}, assumed to be homogeneous along the line of sight through the PDR. The assumed dust and metal abundances are customized for the N13 PDR in the SMC. We use gas phase abundances of metals that are 1/5 of the local Galactic values \citep{Toribio17}, $A_V/N_{\rm H}= 5.35\times10^{-23}$ cm$^{2}$ from \cite{Gordon2024}, a small grain abundance of $1/7.7$ of the Galactic value from \cite{Sandstrom10}, and an appropriate FUV extinction curve \citep{Gordon03,Gordon2024} resulting in a factor of two higher FUV opacity in the photo rates compared to those listed in \cite{Heays17} for the Galactic case \citep[see also][and the Appendix for additional model details]{Jameson2018}. Additional studies of PDRs at low metallicity include \citet{Kaufman06,Rollig06, Bialy19, Bisbas21, Bisbas25}.

We estimated $G_0$ from the massive star that dominates the ionizing photon production rate in N13, which is equivalent to an O7 star of $T_{\rm eff}{\sim} 38$ kK \citep{Ramachandran2019}, and use the FUV luminosity of a Galactic star of the same $T_{\rm eff}$, $L_{\rm FUV}=1.6\times 10^5$ $L_\odot$ \citep{Conti2008,Parravano2003} and a distance of 1.8 pc from the star to PDR boundary. This yields $G_0 \sim 10^3$ in units of the Habing field  \citep[$1.6\times 10^{-3}$ erg cm$^{-2}$ s$^{-1}$;][]{Habing68}.\footnote{This assumes the minimum distance between the PDR and star and gives an upper limit to $G_0$. The field can be lower if the star is substantially in the foreground or background, however, the spherical appearance of the N13 region suggests this distance is reasonable.} We test models of constant density, $n$, and constant thermal pressure $P_{\rm th}/k$, where $n$ is the density of hydrogen nuclei and $k$ is the Boltzmann constant. These correspond to the limiting cases of a cloud completely dominated by magnetic pressure so that $T$ drops without changing the density, and a cloud in which the magnetic pressure is negligible. We use a thermal pressure $P_{\rm th}/k \sim 7.6 \times 10^{6}$~K cm$^{-3}$, which provides the best match to the observed separation between the DF and CI/CO peaks by minimizing the distance between the model and the observed peaks (see Sec.\ \ref{sec:PDR_Comps}).

With this pressure, our PDR model gives a density of $n\sim 3.9\times 10^4$ cm$^{-3}$ at the cloud edge, which we use as our constant density model since it ensures pressure balance between the HII region and the PDR \citep[see e.g., the analysis in][]{Seo2019}. In contrast to the constant pressure case, we do not force the constant density model to match the observed peak separation. Matching the constant density peak separations would require a pressure 7 times higher than what is observed in the HII region, making it physically unrealistic. We note that future, more detailed studies with the JWST observations will refine these numbers. For the present effort, the models are used for comparison with the observed locations of the IF, DF, and C/CO transitions as shown at the bottom of Figure \ref{fig:models_conv}, colored by emission line. In Figure \ref{fig:models_conv}, we plot both the constant density (dotted) and the constant pressure (dash-dotted) models along with each corresponding radial profile for the emission as discussed in Section \ref{sec: Results}.

\section{Results} \label{sec: Results}
\subsection{Observed Structure of N13}\label{sec:dfproject}
To quantify the structure of the PDR, we use the radial profiles to find the locations of the peak intensity of various emission lines, related to the ionization front (IF), dissociation front (DF), and C/CO transition. We then compare these measurements to line intensities from PDR models.

To locate the dissociation front in the N13 PDR, we used the H$_2$ 2.12 $\micron$ vibrational line, which is a good indicator of the H~I to H$_2$ transition (see Appendix). This line is the 1-0 vibrational transition and is, therefore, the last vibrational transition in the H$_2$ fluorescent cascade, making it one of the strongest H$_2$ lines in the NIRSpec wavelength range, and a key marker for identifying the DF \citep[see][for the same measurement in the Orion Bar PDR]{Peeters24}. 

\begin{figure*}[tbhp]
    \centering
    \includegraphics[width=0.769\textwidth, trim=0mm 4mm 0mm 0mm,clip]{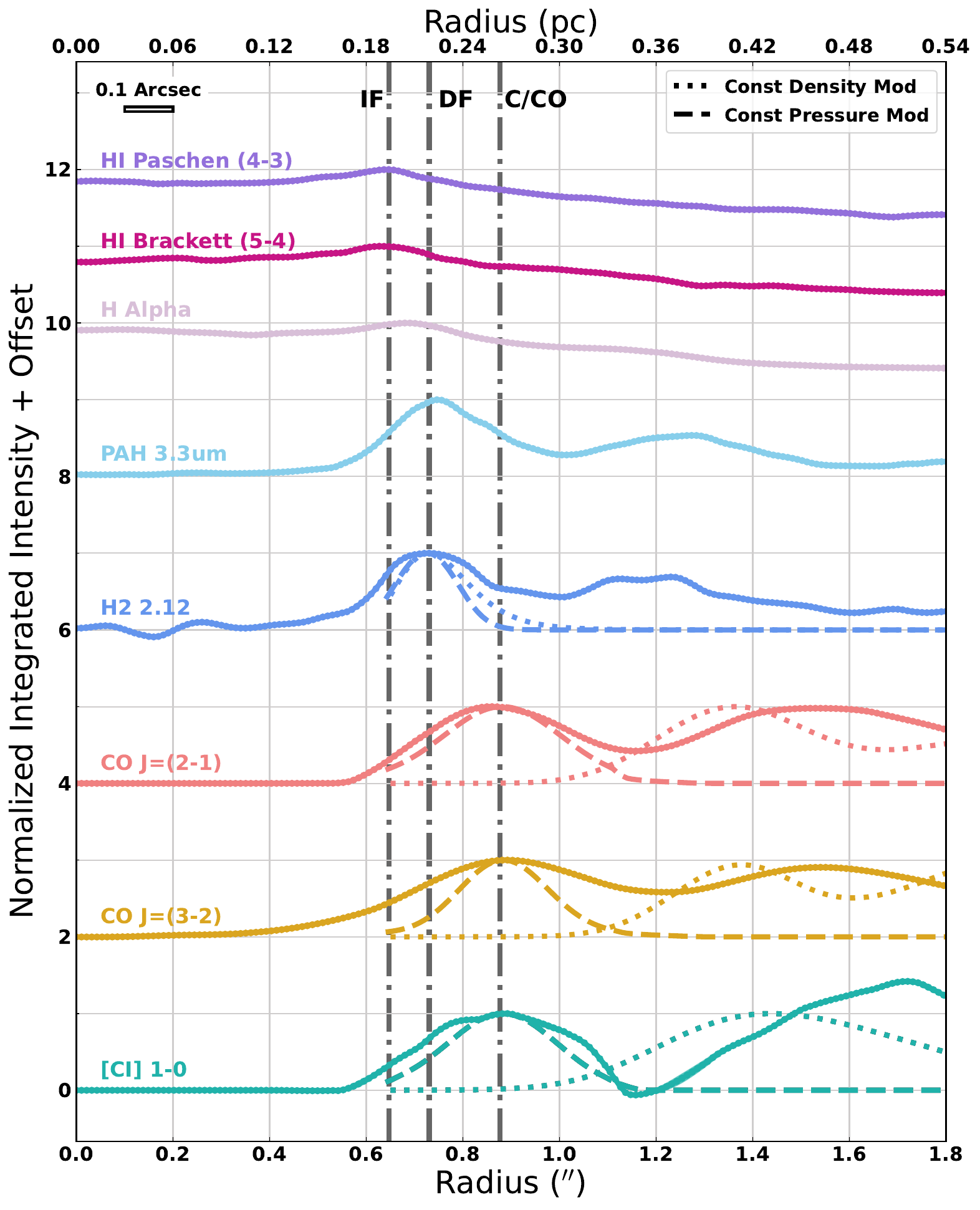}\\
    \caption{We present the stacked and normalized linear radial profiles of the integrated intensity maps of HI 4-3 Paschen $\alpha$, HI 5-4 Brackett $\alpha$, H$\alpha$ from HST F658N photometry, 3.3 $\micron$ PAH feature, H$_2$ 2.12 $\micron$, $^{12}$CO J=2$-$1, $^{12}$CO J=3$-$2, and [CI] 1-0. We observe a peak in the HI Paschen 4-3 profile before the dissociation front, followed by a subsequent decline, which we interpret as the ionization front. We identify the ionization front (IF), the dissociation front (DF), and the C/CO transition from left to right as seen in the vertical dashed-dotted gray lines in each panel. We also present the stacked and normalized linear radial emissivity profiles, convovled for each beam, of the $P_{\rm th}/k=7.6 \times 10^{6}$ K cm$^{-3}$ constant pressure (dashed) and $n=3.9 \times 10^{4}$ cm$^{-3}$ constant density (dotted) models which are color-coded to match the observed profiles.}
    \label{fig:models_conv}
\end{figure*}

Our radial profiles of the H$_2$ 2.12 $\micron$ (1-0 S(1) line, shown in Figure~\ref{fig:models_conv}, reveal two distinct intensity peaks that we interpret as two separate DFs. Similar multiple DF structures have been observed in the Orion Bar \citep{Peeters24}. Our analysis will focus on the first ``primary'' DF in all subsequent discussions. The second DF also exhibits a similar structure to the first DF. To understand the nature of the H2 emission generating the observed peak, we create a map of H$_2$ 2.24 $\micron$ (2-1 S(1)) line, and generate a corresponding radial profile for comparison against the H$_2$ 2.12 $\micron$ emission. We analyze the ratio of the 1-0 S(1) to 2-1 S(1) intensities at the first peak of the H$_2$ 1-0 S(1) radial profile, located at 0\farcs.73. At this position, we measure a 1-0 S(1)/2-1 S(1) ratio of 2.17, characteristic of a primarily H$_2$ FUV pumped region. In the Orion Bar, \citet{Peeters24} find similar \htwo v=1–0 S(1)/v=2–1 S(1) line intensity ratios of $\sim$ 3–5 in each of the three different DFs. We also find that the structure of the H$_2$ 2.12 \micron~and 3.3 $\micron$ PAH radial profiles look strikingly similar, in contrast to the Orion Bar PDR, where the PAH emission peaks at the IF and is bright in the atomic gas region \citep{Peeters24}. We also observed two peaks in the PAH feature integrated intensity radial profile. 

With the Paschen-$\alpha$ and Brackett-$\alpha$ line maps, we explore the location of the IF. We look for a decline in HI recombination line intensities that corresponds to the edge of the H~II region. This location marks the transition from H$^+$ to H. We compare the HI recombination lines to the H$_2$ 2.12 \micron, CO, and [CI] emission in Figure \ref{fig:models_conv}. We observe the ionized gas to be more extended but do observe a peak, and a subsequent dip, right before the DF which we interpret as the IF. We measure the IF as the peak in the HI Paschen 4-3 line to sit at $0\farcs647 \pm 0\farcs060$, or $0.195 \pm 0.013$ pc, from the arbitrary zero point of the radial profile.

We attempt to measure the location of the C/CO transition using the ALMA maps (${}^{12}$CO J=2$-$1, ${}^{12}$CO J=3$-$2, and [CI] 1$-$0). However, as is evident in Figure~\ref{fig:maps}, the [CI] and CO exhibit similar structures. The small separations between CO and [CI] emission could be taken as a signature of clumpiness in the PDR, where in unresolved cases, clumps make it appear that both CO and [CI] are co-spatial \citep{Bolatto99, Rollig06, Glover16, Izumi21}. However, this small overlap could also be consistent with a constant pressure model where density increases in the cooler, shielded gas leading to a convergence of the [CI] and CO layers below our resolution, a topic we discuss further in Sections~\ref{sec:PDR_Comps} and~\ref{sec:clumpy}. Because the separations between the ${}^{12}$CO J=2$-$1, ${}^{12}$CO J=3$-$2, and [CI] 1$-$0 layers are unresolved, we quote an upper limit on their spacing and use the average position of the peak in ${}^{12}$CO J=2$-$1 and ${}^{12}$CO J=3$-$2 to define the boundary of the C/CO transition in Section~\ref{sec: linear profiles}. We note that the location of the peak in CO emission is very close, but is not exactly located, at the C/CO boundary determined from C and CO modeled abundances, as shown in the Appendix.

We note that we also observe CO ice absorption near the first DF (R.A.\ 00:45:26.794, Dec. $-$73:22:57.697) indicating the presence of an embedded young stellar object. This position is not near our radial profile so does not affect our measurements of the PDR layer spacings.

\subsection{Distances Between PDR Boundaries} \label{sec: linear profiles}
A key goal of this study is to resolve the PDR boundaries in a low-metallicity environment. In Figure~\ref{fig:models_conv}, we overplot and normalize the radial profiles to the peak value of each emission line over the whole profile. We also show the convolved constant pressure (dashed) and constant density (dotted) models for the \htwo 2.12 \micron, [CI] 1$-$0, $^{12}$CO J=2$-$1, $^{12}$CO J=3$-$2 line. We label the ionization front (IF), dissociation front (DF), and the C/CO boundary on each of the plots in vertical gray dash-dotted lines. We also present the ionized gas tracers HI Paschen $\alpha$, Brackett $\alpha$, and H$\alpha$, which are not included in the models. To compare to the PDR models, we make the same measurements on the modeled line emissivity to characterize the separations, convolving each model tracer to match the corresponding resolution of each observed emission line. We note that comparing model emissivities with observables is typical for edge-on PDRs \citep{Joblin18, Goicoechea19}, but the absolute line intensities depend on the viewing angle. The emissivity and intensity profiles could be different in the case of opacity effects, especially for the CO lines. However,  we find that the emissivity peaks close to the edge of where the CO abundance starts to rise, where the gas temperatures and densities are sufficiently high to excited the lines.

Figure \ref{fig:models_conv} indicates that we have measured the separation between the DF and the peaks in [CI] and CO emission, indicating we have resolved the PDR structure. Table \ref{table:slice} lists the locations and separations for the H$_2$ 2.12 \micron, [CI] 1$-$0, $^{12}$CO J=2$-$1, $^{12}$CO J=3$-$2, HI Paschen $\alpha$, HI Brackett $\alpha$, and H$\alpha$ lines, with distances converted to parsecs using the SMC distance of 62 kpc \citep{Scowcroft16}. We also present the separations measured from the constant density/pressure models. We take location of the C/CO transition to be the average of the ${}^{12}$CO J=2$-$1 and ${}^{12}$CO J=3$-$2. We give an upper limit on the separation between $^{12}$CO J=2$-$1 and [CI] 1$-$0 as they are not distinguished within their respective uncertainties. The separations between the DF, [CI], and CO emission suggest a compact PDR structure, which we compare to models in the following sections.

\begin{deluxetable*}{c|c|c}
\caption{Key Locations and Separations of PDR Layers} 
\label{table:slice}
\tabletypesize{} 
\startdata  
\\
% & & \\
\multicolumn{3}{c}{\textbf{Radial Slice Locations}} \\
\hline  
 & RA & Dec \\
\hline
Start Coordinates & $0:45:26.6275$ & $-73:22:56.1209$ \\   
End Coordinates & $0:45:26.9911$ & $-73:22:54.8568$ \\   
\hline  
\hline  
\multicolumn{3}{c}{\textbf{Locations of First Peak}} \\
\hline  
Species & Arcsecond ($''$) & Parsec (pc)\\  
\hline  
H$_2$ $2.12$ \micron & $0\farcs730 \pm 0\farcs040$ & $0.220 \pm 0.013$\\  
3.3$\micron$ PAH  & $0\farcs746 \pm 0\farcs032$  & $0.225 \pm 0.009$\\ 
$^{12}$CO J=2$-$1 & $0\farcs860 \pm 0\farcs011$ & $0.260 \pm 0.003$\\   
${\rm [CI]}$ 1-0 & $0\farcs892 \pm 0\farcs013$ & $0.270 \pm 0.004$\\ 
$^{12}$CO J=3$-$2 & $0\farcs892 \pm 0\farcs011$  & $0.270 \pm 0.003$\\ 
HI Paschen $\alpha$ & $0\farcs647 \pm 0\farcs060$ & $0.195 \pm 0.018$\\ 
HI Brackett $\alpha$  & $0\farcs627 \pm 0\farcs070$ & $0.189 \pm 0.021$\\ 
H$\alpha$  & $0\farcs680 \pm 0\farcs145$ & $0.205 \pm 0.044$\\ 
\hline   
\hline  
\multicolumn{3}{c}{\textbf{PDR Layer Separations}} \\
\hline   
Separation Type & Arcsecond ($''$) & Parsec (pc) \\
\hline 
IF $\rightarrow$ DF & $0\farcs083 \pm 0\farcs033$ & $0.025 \pm 0.009$ \\
DF $\rightarrow$ C/CO$_{\rm avg}$ & $0\farcs146 \pm 0\farcs042$ & $0.043 \pm 0.013$ \\
$^{12}$CO J=2$-$1 $\rightarrow$ $^{12}$CO J=3$-$2 & $< 0\farcs032$ & $< 0.009 $\\  
$^{12}$CO J=2$-$1$\rightarrow$ ${\rm [CI]}$ 1-0  &  $< 0\farcs032$ & $< 0.009$ \\   
Const. Density Model IF $\rightarrow$ DF & $0\farcs080$ & $0.024$ \\ 
Const. Density Model DF $\rightarrow$ $^{12}$CO J=2$-$1 & $0\farcs650$ & $0.195$ \\ 
Const. Density Model DF $\rightarrow$ [CI] 1$-$0 & $0\farcs710$ & $0.213$ \\
Const. Pressure Model IF $\rightarrow$ DF & $0\farcs090$ & $0.027$ \\ 
Const. Pressure Model DF $\rightarrow$ $^{12}$CO J=2$-$1 & $0\farcs150$ & $0.045$ \\
Const. Pressure Model DF $\rightarrow$ [CI] 1$-$0 & $0\farcs150$ & $0.045$ \\
\enddata  
\tablecomments{R.A.\ and Dec.\ of profile cuts perpendicular to the PDR as seen on the H$_2$ 2.12 \micron\ line integrated intensity map (see Fig.\ \ref{fig:maps}). The beginning points (0.0$''$) are offset from the star by 5.992$''$ (1.81 pc). Profiles are sampled at 0.01$''$ spacing. We also present the separations in the peak intensities of the radial profiles. We show the separation of the layers in angular arcsecond units ($''$) and parsec (pc) scales. The errors on the peak measurements come from the standard deviation of peak calculations on each of the 5 slices, as seen in Fig. \ref{fig:3-color image}. We use the average of the ${}^{12}$CO J=2$-$1 and ${}^{12}$CO J=3$-$2 peaks to define the C/CO transition at $0\farcs876 \pm 0\farcs016$, since $^{12}$CO J=2$-$1 and [CI] 1$-$0 are not distinguished within their respective uncertainties. We note that the models of a single, constant density PDR have two peaks in CO emission, the first of which is used to calculate the model seperations.} 
\end{deluxetable*}    

\section{Discussion}\label{sec:Disscussion}
Our radial profile analysis reveals clear separations between key species in the DF and the C/CO transition. We find the separation between the DF and C/CO to be \seppc. The statistical (stat) error come from the four adjacent slices around slice 1. The systematic (syst) error is determined from the astrometric alignment adjustments between Gaia to HST with 1000 stars and HST to JWST with four stars, where we take the error on the mean for each and add them in quadrature. This work marks the first time an extragalactic low metallicity PDR has ever been resolved.

\subsection{Comparison to PDR Models} \label{sec:PDR_Comps}
We compared our results to a constant pressure, $P_{\rm th}/k \sim 7.6\times 10^6$ K cm$^{-3}$ SMC PDR model, convolving the model line emissivities to match the H$_2$ 2.12 \micron, [CI], and CO resolutions using a Gaussian kernel at the spatial resolution of each individual line. The convolved models were then overlaid on our radial profiles shown in Figure \ref{fig:models_conv}. We also show a constant density ($n=3.9\times 10^4$ cm$^{-3}$) PDR model for comparison.

Since the observed position of the DF from the H$_2$ 1-0 S(1) line is well defined and the models have an arbitrary $x$ location, we shift the peak in the models to match the observed peak to compare the spacing of the boundaries. 
This adds one constant spatial shift to the models and does not change the spacing between the model peaks.

Our results show that the constant density models overestimate the H$_2$ to C/CO separation by $\sim $1$''$ as presented in Table \ref{table:slice}. In contrast, the constant pressure models fit the observed spacings well, reproducing the separations between the H/H$_2$ and the C/CO transition as well as the coincidence of [CI] and CO emission at our resolution.
The best-fit pressure is only $\sim 35\%$ higher than  (and consistent within the uncertainties of) an estimate of the thermal pressure in the adjacent ionized gas $(P_{\rm th}/k  = 5.6\times 10^6\, \rm{K\, cm^{-3}})$from electron density measurements using low angular resolution Spitzer spectroscopy of [SIII] \citep{Sandstrom12} and a temperature of $T_{\rm e} \sim 12500$ K \citep[e.g.,][]{Dufour77}. The separation of peaks scales as $1/P_{\rm th}$ for pressures within a factor of 2 of the best fit with a similar dependence for $A_{\rm V}/N_{\rm H}$. Changes in the cosmic-ray ionization rate by a factor of 2 have a negligible effect on the peak separations. 
 
We also note that our fitted $P_{\rm th}$ at $G_0{\sim} 10^3$ is somewhat lower than that in \cite{Joblin18} based on high-J CO lines measured in Milky Way PDRs but is close to the fit in \cite{Wu2018} and \cite{Seo2019} for an \hii region in thermal pressure equilibrium with a surrounding PDR at the Str\"omgren radius. The pressure is higher than that shown in \cite{wolfire2022} (Fig.\ 13) for a compilation of extragalactic observations, possibly due to our much higher spatial resolution, which avoids beam averaging over environments.

\subsection{Other Resolved PDRs in the Milky Way}
Compared to Milky Way PDRs, we observe notable differences in the separation between the H/\htwo and C/CO boundaries. In the Orion Bar, \citet{Goicoechea16,Goicoechea25} find an H/\htwo to C/CO boundary separation of $\sim 0.002$ pc which is around 25 times smaller than in N13. Similarly in the Horsehead Nebula, \cite{Hernandez-Vera23} finds a separation of $\lesssim 0.004$ pc or 830 au with a $G_0$ of $\sim$100. In both of these PDRs, a constant pressure model best describes the observed structure.  When compared to N13, with a separation of $\sim 0.043$ pc, our analysis highlights that PDRs are much more extended and CO-dark \htwo gas plays a more prominent role in lower metallicity environments. We confirm the long-standing theory that the extent of the CO-dark \htwo layer increases at low metallicities \citep{Bolatto13, Glover16, Madden20}. In Orion, \citet{Habart24} also find the spatial extent from the IF to DF to be $0.02-0.04$ pc. This separation is similar to N13 at $0.025 \pm 0.009$ pc.

Interestingly, the 3.3 $\micron$ feature in N13 shows a fundamental difference to the Orion Bar PDR. In N13, this PAH feature peaks close to the DF at $0\farcs016 \pm 0\farcs008$ behind the \htwo 2.12 $\micron$ peak. In the Orion Bar, \cite{Peeters24} finds clear bright peaks of the 3.3 \micron\ feature towards the IF and fainter peaks slightly behind the DF. The differences between Orion and N13 could be caused by a metallicity effect, potentially due to higher penetration of FUV photons from the lack of dust shielding. However, further work is needed to confirm whether metallicity is the primary driver of this difference.

\subsection{Inclination and Geometry Effects} \label{sec:inclination}
We examined N13 for inclination effects to see whether the PDR orientation impacts the measured separations between the DF and the C/CO transition. Ideally, the PDR should be at near edge-on inclination (i.e. $\sim0\degr$), where a well-defined DF and maximally separated boundaries are expected.

We tested the constant density and constant pressure models, scaling the separations of different layers by $\sin(i)$ for 0--90\degr\ inclinations $i$ between the line-of-sight and the PDR surface. The $n=3.9 \times 10^{4}$ cm$^{-3}$ constant density model requires an unrealistic inclination of $\sim$ 80$\degr$ to match the observed separations (i.e., a nearly face-on PDR), inconsistent with the visual appearance in HST data showing an approximately edge-on geometry. In contrast, the constant pressure models match the observations better with an inclination of $\lesssim$ 30$\degr$. 

In addition to the inclination of the individual PDR front, the overall geometry of the region is also of interest in explaining the existence of multiple DFs. In the Orion Bar, \cite{Habart24}, similarly finds  multiple DFs which they attribute to terraced structure with three steps to explain the succession of H$_2$ ridges across the bar, along with an inclination of the bar at 1 to 8 degrees \citep{Salgado16}. This could indicate that N13 has a geometry similar to Orion with possibly two terraced surfaces along with a slight tilt. Large-scale geometry effects likely explain the presence of two DFs, as our analysis indicates a small inclination for the constant pressure model.

\subsection{Clumpy PDRs}\label{sec:clumpy}
Unresolved clumpy structures have been used to explain the overlap of CO and [CI] emission in some PDRs \citep{Cubick08}.
Physical drivers, like turbulence, can create a multi-phase clumpy medium with uneven radiation penetration, increasing the temperature deeper in the PDR, and enhancing chemical processes \citep{Glover2015}. If the PDR was clumpy, CO clumps could remain unresolved in our observations (at $\lesssim 0.075$ pc). 
Our results do show nearly co-spatial $^{12}$CO J=2$-$1, J=3$-$2, and [CI] 1$-$0. This implies a potentially clumpy gas distribution, although this type of model may not accurately match the separation of the \htwo and C/CO layers. Due to uncertainty in whether a clumpy model would match all the observed spacings, we cannot dismiss the possibility of a clumpy structure based solely on the observed spacing of the [CI] and CO species. 

Another way to constrain the presence of sub-resolution clumps is to use the peak temperature from optically thick CO emission. This peak temperature may not represent the actual gas temperature if the clumps are still unresolved, as the expected peak $T_{\rm pk}$ for optically thick CO is much lower than the actual gas temperature in these cases\footnote{Additionally, $T_{\rm pk}$ can also be lower than T$_{\rm gas}$ if n $\lesssim$ n$_{\rm critical}$. We found that for the CO 2-1 transition the densities in the constant pressure model always exceeded the critical density for collisions with \htwo.}. If we observe a much lower $T_{\rm pk}$, it may be consistent with clumpiness as an explanation for the almost co-spatial overlap of CO/CI. 

\begin{figure*}[]
    \centering
    \includegraphics[width=0.49\textwidth, trim=0mm 0mm 0mm 0mm,clip]{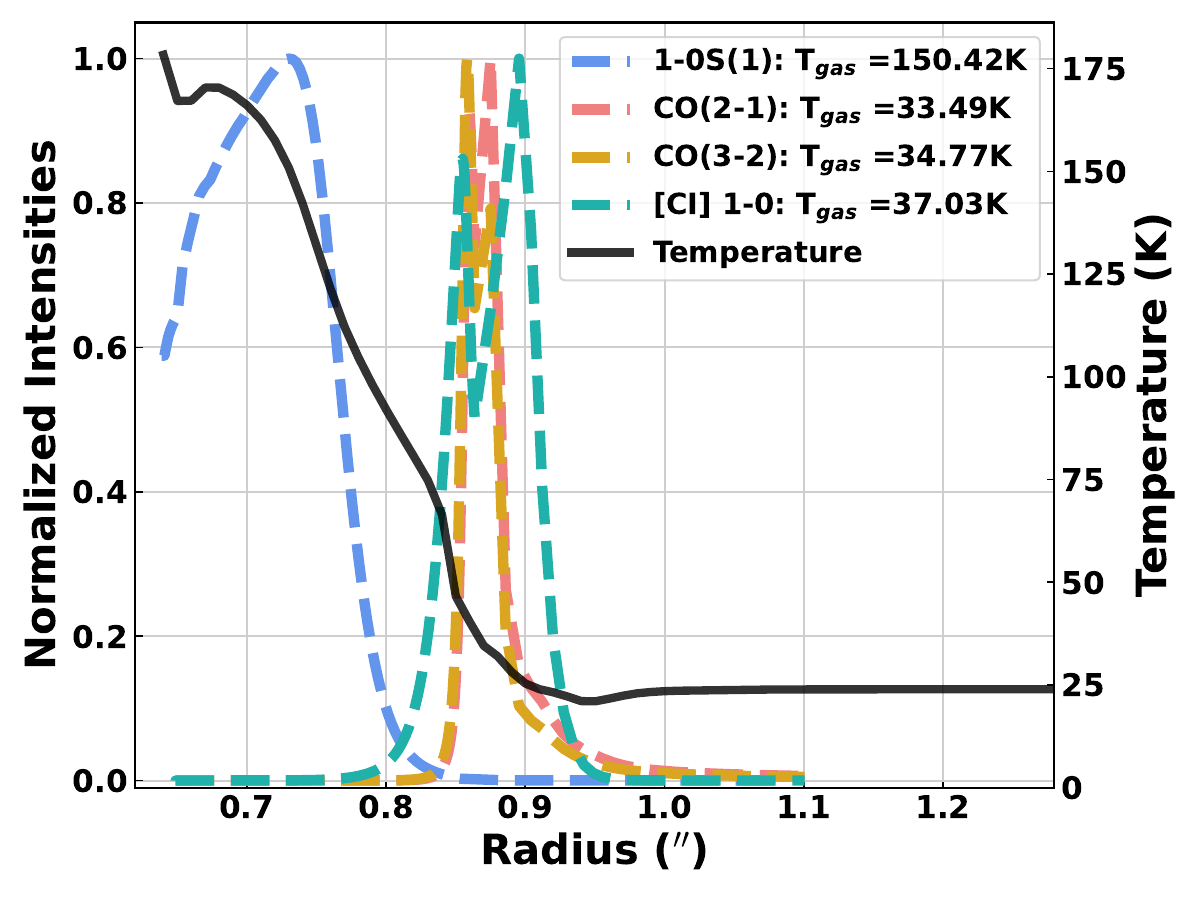}
    \includegraphics[width=0.49\textwidth, trim=0mm 4mm 0mm 0mm,clip]{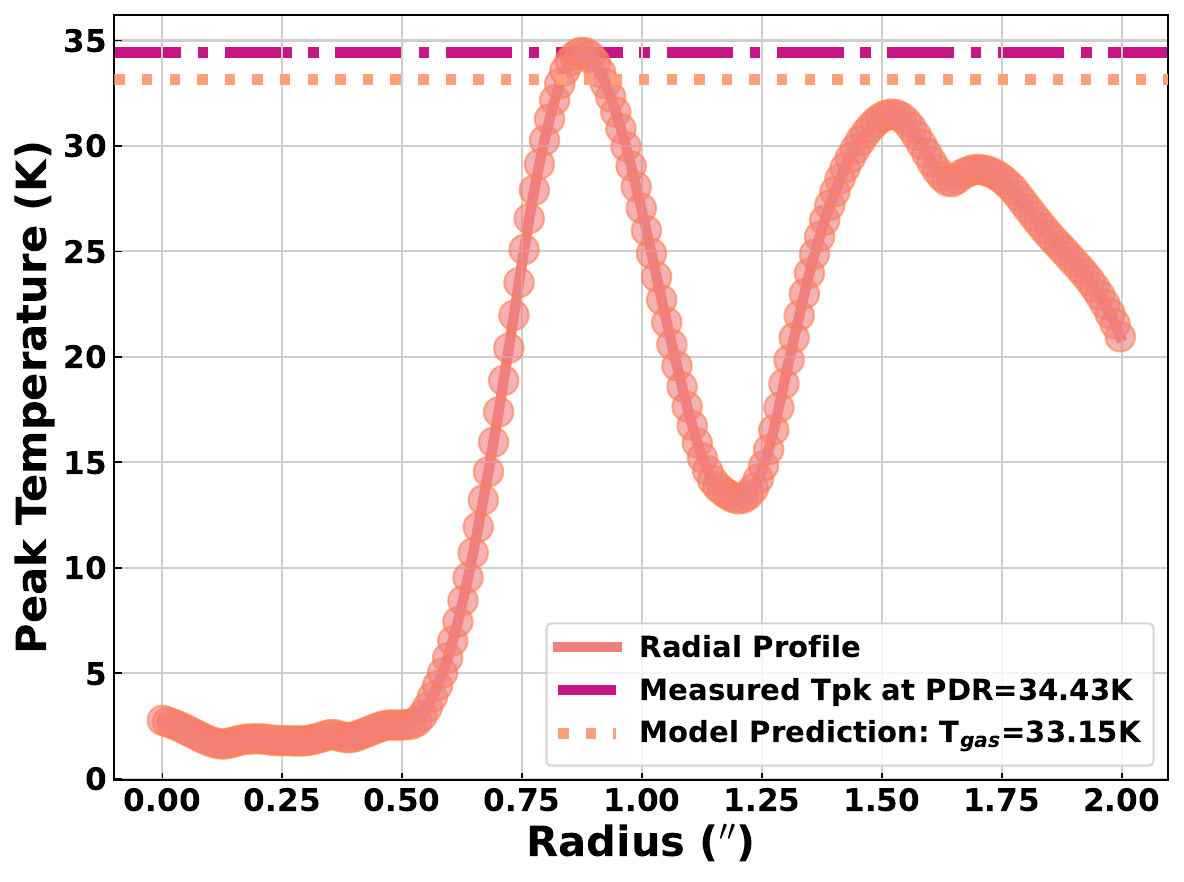}
    \caption{In the left panel we plot the models, the theoretical unconvolved temperature profile across the PDR in a solid black line, and the emissivity of each relevant species in colored dashed lines. The model emission-weighted gas temperature ($T_{\rm gas}$) for each species is listed in the legend. In the right panel, we show the model peak temperature($T_{\rm gas}$) and the observed optically thick $^{12}$CO J=2$-$1 peak temperature profile $T_{\rm pk}$}
    \label{fig:Temp&Tpk}
\end{figure*}

To test if clumps play a role in the PDR structure of N13, we create a linear radial profile of the CO peak temperature maps produced from the PHANGS-ALMA Pipeline (discussed in Sec. \ref{alma-phangs-pipe}). We compare the peak temperature radial slice to the emission-weighted gas temperature ($T_{\rm gas}$) from the constant pressure PDR models for the optically thick $^{12}$CO J=2$-$1 in Figure \ref{fig:Temp&Tpk}.

We find that the peak temperature of the $^{12}$CO J=2$-$1 emission is $\sim$34 K, only $\sim 1 K$ more than the model predictions for the line. There is strong agreement between the modeled and observed peaks indicative of a high filling factor and a well-matched model for N13. Additionally, the similar observed peak temperature to the model suggests the absence of sub-resolution clumps. Alternatively, in regimes with a lower observed peak temperature this may indicate that the CO emission is not filling the entire beam, likely due to a low filling factor caused by unresolved clumps.
We note that this offset in the modeled peak temperature could come from PDR model uncertainty, since the temperature structure is most sensitive to potential metallicity-driven variations in the heating and cooling.

The lack of strong evidence for the presence of sub-resolution clumps from Figure \ref{fig:Temp&Tpk}, and the sufficient constant pressure model match to the observed separations of the DF and C/CO transitions, point to the plane-parallel models being adequate to explain the structure of N13. However, models like Kosma-$\tau$ \citep{Rollig2022}, highlight the need to account for small-scale structures that may be influencing the observed emission patterns. Further exploration of clumpy PDR and constant pressure models is necessary to disentangle these effects.

\subsection{CO-dark H$_2$} 

In low-metallicity environments, where dust-to-gas is lower (i.e.,\ $A_{\rm V}/N_{\rm H}$ is lower), there is expected to be a higher proportion of CO-dark H$_2$ \citep{Glover2010, Schruba12, Bolatto13, Nordon16, Madden20, Hu21, Bisbas21}, increasing the uncertainty in calibrating the $X_{\rm CO}$ factor. These lower metallicity environments in particular lead to deeper UV penetration, typically decreasing the amount of CO that can survive close to the dissociation front. This metallicity effect also leads to a more extended molecular zone \citep{Bolatto99, Rollig06, Schneider2021}, adding further uncertainty on the constraint for the $X_{\rm CO}$ conversion factor. It is important to note, that there is also a geometric aspect to the problem where complex filamentary geometry can dramatically increase the expected fraction of CO-dark \htwo gas compared to simple plane-parallel or spherical shell models \citep{Smith14}. 

The separation between the DF and the C/CO transition is observed to be \seppc, while the predicted separation for the constant pressure model is 0.045 pc. This corresponds to a modeled CO-dark gas column density of $N_{\rm H_2}\sim 1.1\times 10^{22}$ cm$^{-2}$. We find that a Galactic PDR with the same incident FUV field and thermal pressure would have a CO-dark gas column density of $N_{\rm H_2}\sim 2.1\times 10^{21}$ cm$^{-2}$, equivalent to a DF to C/CO separation of $0\farcs04$ or 0.012pc. This difference indicates that the SMC N13 PDR has a CO-dark gas column density 5 times greater than that of a Milky Way PDR. The plane-parallel model depth extends past the observed emission peaks but the total depth is not tied 
to a specific molecular cloud model, and thus we are unable to obtain a unique H$_2$ column density. Therefore we cannot calculate a precise value of $X_{\rm{CO}}$. The typical Galactic $X_{\rm{CO}}$ factor is $2\times10^{20}$ cm$^{-2}$/(K km s$^{-1})$.

It is important to note that, in this case, there are no significant improvements in using [CI] 1$-$0 over CO to trace CO-dark H$_2$.
In N13 the [CI] 1$-$0 emission is not particularly bright in the higher  A$_{\rm{V}}$ molecular material traced by CO, as it would be expected if [CI] mostly arises from photodissociation in a thin PDR layer. Therefore it does not do a good job at capturing the bulk of the CO emitting molecular gas. Some studies propose [CI] as an alternative to trace molecular gas in regions where CO emission is weak or absent, or even find [CI] a better tracer of H$_2$ than CO in general \citep{Gerin00, Papadopoulos04, Kramer08, Glover2015, Bisbas2025}. Our results suggest that using [CI] to trace H$_2$, particularly in low-metallicity environments, has limitations. Further exploration of the conditions under which neutral carbon can be a reliable tracer of molecular gas is needed in order to establish its usefulness.

\section{Conclusions $\&$ Implications for Low Metallicity PDRs} \label{sec: conclusions}

For the first time, we have spatially resolved an extragalactic, low metallicity, photodissociation region showcasing the capabilities of the JWST in conjunction with ALMA for studying PDRs in the SMC. We measure our separation for the transition from the DF to the average of the $^{12}$CO J=2$-$1 and $^{12}$CO J=3$-$2 peaks to be \separc or \seppc. Our findings reveal that the N13 PDR has separation between the H/H$_2$ and C/CO transitions are consistent with the plane-parallel constant pressure model at $0.045$ pc ($0\farcs150$), while the constant density model at $0.195$ pc ($0\farcs650$), overestimates the separations. A reduced $A_{\rm V}/N_{\rm H}$ for SMC conditions is also necessary in the models to match the observed spacings. This reasonable match between the constant pressure models and our observations suggests that traditional plane-parallel PDR models do a good job describing the spatial extent of the CO-dark H$_2$ in low metallicity environments. Understanding the spatial extent of CO-dark gas is crucial for refining the $X_{\rm CO}$ conversion factor, highlighting the role CO-dark \htwo plays in the ISM of low-metallicity galaxies, which resemble conditions in the early universe. The PDR model that describes N13 has a CO-dark H$_2$ column density 5 times higher than the comparable Milky Way model.

Understanding the mechanisms driving PDR structure at low metallicity is critical for tracing molecular gas and understanding the evolution of the ISM throughout the early universe. Future efforts for N13 will explore the impact of the spectrum of the ionizing OB stars on the PDR; the temperature structure of the neutral gas using the H$_2$ rotational ladder and CO spectral line energy distribution; and the nature of the small dust grain population; in addition to improving models to better describe metallicity-driven changes. 
We present the first resolved extragalactic and  low-metallicity PDR. However, a larger sample of low metallicity PDRs with sub-mm and infrared data is essential for robust constraints on state-of-the-art PDR models and improving our understanding of low-metallicity astrochemistry.

\section*{Acknowledgments}

We thank the anonymous referee for helpful and constructive feedback which improved the manuscript. This work is based on observations made with the NASA/ESA/CSA JWST. The data were obtained from the Mikulski Archive for Space Telescopes at the Space Telescope Science Institute, which is operated by the Association of Universities for Research in Astronomy, Inc., under NASA contract NAS 5-03127 for JWST. These observations are associated with program \#2521. The specific observations analyzed can be accessed via \dataset[https://doi.org/10.17909/ppkp-5h63]{https://doi.org/10.17909/ppkp-5h63}. Support for program \#2521 was provided by NASA through a grant from the Space Telescope Science Institute, which is operated by the Association of Universities for Research in Astronomy, Inc., under NASA contract NAS 5-03127.

This research is based on observations made with the NASA/ESA Hubble Space Telescope obtained from the Space Telescope Science Institute, which is operated by the Association of Universities for Research in Astronomy, Inc., under NASA contract NAS 5–26555. These observations are associated with program GO 13659.

This paper makes use of the following ALMA data: ADS/JAO.ALMA\#2021.1.01065.S. ALMA is a partnership of ESO (representing its member states), NSF (USA) and NINS (Japan), together with NRC (Canada), NSTC and ASIAA (Taiwan), and KASI (Republic of Korea), in cooperation with the Republic of Chile. The Joint ALMA Observatory is operated by ESO, AUI/NRAO and NAOJ. The National Radio Astronomy Observatory is a facility of the National Science Foundation operated under cooperative agreement by Associated Universities, Inc.

This research has made use of NASA's Astrophysics Data System. This research made use of Astropy, a community-developed core Python package for Astronomy \citep{2018AJ....156..123A, 2013A&A...558A..33A} This research made use of SciPy \citep{Virtanen_2020}. This research made use of matplotlib, a Python library for publication quality graphics \citep{Hunter:2007}. This research made use of ds9, a tool for data visualization supported by the Chandra X-ray Science Center (CXC) and the High Energy Astrophysics Science Archive Center (HEASARC) with support from the JWST Mission office at the Space Telescope Science Institute for 3D visualization. This research made use of NumPy \citep{harris2020array} 

I.C.\ acknowledges funding support from NRAO Student Observing Grant SOSPA8-009. This material is based upon work supported by the National Science Foundation under Grant No.\ 2308274. I.C.\ and K.S.\ thank Michael Busch, Aida Wofford, and Hannah Koziol for helpful conversations and suggestions that improved the paper.

S.C.O.G.\ and R.S.K.\ acknowledge financial support from the European Research Council via the ERC Synergy Grant ``ECOGAL'' (project ID 855130) and from the German Excellence Strategy via the Heidelberg Cluster of Excellence (EXC 2181 - 390900948) ``STRUCTURES''. R.S.K.\ furthermore thanks the German Ministry for Economic Affairs and Climate Action for funding in project ``MAINN'' (funding ID 50OO2206). R.S.K.\ is also grateful to the 2024/25 Class of Radcliffe Fellows for their company and for highly interesting and stimulating discussions.
M.G.W. was supported in part by NASA grants JWST-GO-02521.002-A and JWST-AR-01557.001-A.
SW acknowledges the support of the Deutsche Forschungsgemeinschaft (DFG) for funding through the SFB~1601 ``Habitats of massive stars across cosmic time’' (sub-project B6) and of the project ''NRW-Cluster for data-intensive radio astronomy: Big Bang to Big Data (B3D)'' funded through the programme ''Profilbildung 2020'', an initiative of the Ministry of Culture and Science of the State of North Rhine-Westphalia.
J.R.G. thanks the Spanish MCINN for funding support under grant PID2023-146667NB-I00.
JC acknowledges funding from the Belgian Science Policy Office (BELSPO) through the PRODEX project ``JWST/MIRI Science exploitation'' (C4000142239). BALG is supported by the German Deutsche Forschungsgemeinschaft, DFG, in the form of an Emmy Noether Research Group - Project-ID 542802847.
 
\bibliography{bib} 

\begin{thebibliography}{}
\expandafter\ifx\csname natexlab\endcsname\relax\def\natexlab#1{#1}\fi
\providecommand{\url}[1]{\href{#1}{#1}}
\providecommand{\dodoi}[1]{doi:~\href{http://doi.org/#1}{\nolinkurl{#1}}}
\providecommand{\doeprint}[1]{\href{http://ascl.net/#1}{\nolinkurl{http://ascl.net/#1}}}
\providecommand{\doarXiv}[1]{\href{https://arxiv.org/abs/#1}{\nolinkurl{https://arxiv.org/abs/#1}}}

\bibitem[{{Abdo} {et~al.}(2010){Abdo}, {Ackermann}, {Ajello}, {Baldini}, {Ballet}, {Barbiellini}, {Bastieri}, {Bechtol}, {Bellazzini}, {Berenji}, {Blandford}, {Bloom}, {Bonamente}, {Borgland}, {Bouvier}, {Brandt}, {Bregeon}, {Brez}, {Brigida}, {Bruel}, {Buehler}, {Buson}, {Caliandro}, {Cameron}, {Caraveo}, {Carrigan}, {Casandjian}, {Cecchi}, {{\c{C}}elik}, {Charles}, {Chekhtman}, {Cheung}, {Chiang}, {Ciprini}, {Claus}, {Cohen-Tanugi}, {Conrad}, {Dermer}, {de Palma}, {Digel}, {Silva}, {Drell}, {Dubois}, {Dumora}, {Favuzzi}, {Fegan}, {Fukazawa}, {Funk}, {Fusco}, {Gargano}, {Gasparrini}, {Gehrels}, {Germani}, {Giglietto}, {Giordano}, {Giroletti}, {Glanzman}, {Godfrey}, {Grenier}, {Grondin}, {Grove}, {Guiriec}, {Hadasch}, {Harding}, {Hayashida}, {Hays}, {Horan}, {Hughes}, {Jean}, {J{\'o}hannesson}, {Johnson}, {Johnson}, {Kamae}, {Katagiri}, {Kataoka}, {Kerr}, {Kn{\"o}dlseder}, {Kuss}, {Lande}, {Latronico}, {Lee}, {Lemoine-Goumard}, {Llena Garde}, {Longo}, {Loparco}, {Lovellette}, {Lubrano}, {Makeev}, {Martin},
  {Mazziotta}, {McEnery}, {Michelson}, {Mitthumsiri}, {Mizuno}, {Monte}, {Monzani}, {Morselli}, {Moskalenko}, {Murgia}, {Nakamori}, {Naumann-Godo}, {Nolan}, {Norris}, {Nuss}, {Ohsugi}, {Okumura}, {Omodei}, {Orlando}, {Ormes}, {Panetta}, {Parent}, {Pelassa}, {Pepe}, {Pesce-Rollins}, {Piron}, {Porter}, {Rain{\`o}}, {Rando}, {Razzano}, {Reimer}, {Reimer}, {Reposeur}, {Ripken}, {Ritz}, {Romani}, {Sadrozinski}, {Sander}, {Saz Parkinson}, {Scargle}, {Sgr{\`o}}, {Siskind}, {Smith}, {Smith}, {Spandre}, {Spinelli}, {Strickman}, {Strong}, {Suson}, {Takahashi}, {Takahashi}, {Tanaka}, {Thayer}, {Thayer}, {Thompson}, {Tibaldo}, {Torres}, {Tosti}, {Tramacere}, {Uchiyama}, {Usher}, {Vandenbroucke}, {Vasileiou}, {Vilchez}, {Vitale}, {Waite}, {Wang}, {Winer}, {Wood}, {Yang}, {Ylinen}, \& {Ziegler}}]{Abdo2010}
{Abdo}, A.~A., {Ackermann}, M., {Ajello}, M., {et~al.} 2010, \aap, 523, A46, \dodoi{10.1051/0004-6361/201014855}

\bibitem[{{Allers} {et~al.}(2005){Allers}, {Jaffe}, {Lacy}, {Draine}, \& {Richter}}]{Allers05}
{Allers}, K.~N., {Jaffe}, D.~T., {Lacy}, J.~H., {Draine}, B.~T., \& {Richter}, M.~J. 2005, \apj, 630, 368, \dodoi{10.1086/431919}

\bibitem[{{Argyriou} {et~al.}(2023){Argyriou}, {Glasse}, {Law}, {Labiano}, {{\'A}lvarez-M{\'a}rquez}, {Patapis}, {Kavanagh}, {Gasman}, {Mueller}, {Larson}, {Vandenbussche}, {Glauser}, {Royer}, {Dicken}, {Harkett}, {Sargent}, {Engesser}, {Jones}, {Kendrew}, {Noriega-Crespo}, {Brandl}, {Rieke}, {Wright}, {Lee}, \& {Wells}}]{Argyriou2023}
{Argyriou}, I., {Glasse}, A., {Law}, D.~R., {et~al.} 2023, \aap, 675, A111, \dodoi{10.1051/0004-6361/202346489}

\bibitem[{{Astropy Collaboration} {et~al.}(2013){Astropy Collaboration}, {Robitaille}, {Tollerud}, {Greenfield}, {Droettboom}, {Bray}, {Aldcroft}, {Davis}, {Ginsburg}, {Price-Whelan}, {Kerzendorf}, {Conley}, {Crighton}, {Barbary}, {Muna}, {Ferguson}, {Grollier}, {Parikh}, {Nair}, {Unther}, {Deil}, {Woillez}, {Conseil}, {Kramer}, {Turner}, {Singer}, {Fox}, {Weaver}, {Zabalza}, {Edwards}, {Azalee Bostroem}, {Burke}, {Casey}, {Crawford}, {Dencheva}, {Ely}, {Jenness}, {Labrie}, {Lim}, {Pierfederici}, {Pontzen}, {Ptak}, {Refsdal}, {Servillat}, \& {Streicher}}]{2013A&A...558A..33A}
{Astropy Collaboration}, {Robitaille}, T.~P., {Tollerud}, E.~J., {et~al.} 2013, \aap, 558, A33, \dodoi{10.1051/0004-6361/201322068}

\bibitem[{{Astropy Collaboration} {et~al.}(2018){Astropy Collaboration}, {Price-Whelan}, {Sip{\H o}cz}, {G{\"u}nther}, {Lim}, {Crawford}, {Conseil}, {Shupe}, {Craig}, {Dencheva}, {Ginsburg}, {VanderPlas}, {Bradley}, {P{\'e}rez-Su{\'a}rez}, {de Val-Borro}, {Aldcroft}, {Cruz}, {Robitaille}, {Tollerud}, {Ardelean}, {Babej}, {Bach}, {Bachetti}, {Bakanov}, {Bamford}, {Barentsen}, {Barmby}, {Baumbach}, {Berry}, {Biscani}, {Boquien}, {Bostroem}, {Bouma}, {Brammer}, {Bray}, {Breytenbach}, {Buddelmeijer}, {Burke}, {Calderone}, {Cano Rodr{\'{\i}}guez}, {Cara}, {Cardoso}, {Cheedella}, {Copin}, {Corrales}, {Crichton}, {D'Avella}, {Deil}, {Depagne}, {Dietrich}, {Donath}, {Droettboom}, {Earl}, {Erben}, {Fabbro}, {Ferreira}, {Finethy}, {Fox}, {Garrison}, {Gibbons}, {Goldstein}, {Gommers}, {Greco}, {Greenfield}, {Groener}, {Grollier}, {Hagen}, {Hirst}, {Homeier}, {Horton}, {Hosseinzadeh}, {Hu}, {Hunkeler}, {Ivezi{\'c}}, {Jain}, {Jenness}, {Kanarek}, {Kendrew}, {Kern}, {Kerzendorf}, {Khvalko}, {King}, {Kirkby}, {Kulkarni},
  {Kumar}, {Lee}, {Lenz}, {Littlefair}, {Ma}, {Macleod}, {Mastropietro}, {McCully}, {Montagnac}, {Morris}, {Mueller}, {Mumford}, {Muna}, {Murphy}, {Nelson}, {Nguyen}, {Ninan}, {N{\"o}the}, {Ogaz}, {Oh}, {Parejko}, {Parley}, {Pascual}, {Patil}, {Patil}, {Plunkett}, {Prochaska}, {Rastogi}, {Reddy Janga}, {Sabater}, {Sakurikar}, {Seifert}, {Sherbert}, {Sherwood-Taylor}, {Shih}, {Sick}, {Silbiger}, {Singanamalla}, {Singer}, {Sladen}, {Sooley}, {Sornarajah}, {Streicher}, {Teuben}, {Thomas}, {Tremblay}, {Turner}, {Terr{\'o}n}, {van Kerkwijk}, {de la Vega}, {Watkins}, {Weaver}, {Whitmore}, {Woillez}, {Zabalza}, \& {Astropy Contributors}}]{2018AJ....156..123A}
{Astropy Collaboration}, {Price-Whelan}, A.~M., {Sip{\H o}cz}, B.~M., {et~al.} 2018, \aj, 156, 123, \dodoi{10.3847/1538-3881/aabc4f}

\bibitem[{{Bakes} \& {Tielens}(1994)}]{Bakes1994}
{Bakes}, E.~L.~O., \& {Tielens}, A.~G.~G.~M. 1994, \apj, 427, 822, \dodoi{10.1086/174188}

\bibitem[{{Bell} {et~al.}(2007){Bell}, {Viti}, \& {Williams}}]{Bell07}
{Bell}, T.~A., {Viti}, S., \& {Williams}, D.~A. 2007, \mnras, 378, 983, \dodoi{10.1111/j.1365-2966.2007.11830.x}

\bibitem[{{Bellomi} {et~al.}(2020){Bellomi}, {Godard}, {Hennebelle}, {Valdivia}, {Pineau des For{\^e}ts}, {Lesaffre}, \& {P{\'e}rault}}]{Bellomi20}
{Bellomi}, E., {Godard}, B., {Hennebelle}, P., {et~al.} 2020, \aap, 643, A36, \dodoi{10.1051/0004-6361/202038593}

\bibitem[{{Bialy} \& {Sternberg}(2019)}]{Bialy19}
{Bialy}, S., \& {Sternberg}, A. 2019, \apj, 881, 160, \dodoi{10.3847/1538-4357/ab2fd1}

\bibitem[{{Bisbas} {et~al.}(2012){Bisbas}, {Bell}, {Viti}, {Yates}, \& {Barlow}}]{Bisbas12}
{Bisbas}, T.~G., {Bell}, T.~A., {Viti}, S., {Yates}, J., \& {Barlow}, M.~J. 2012, \mnras, 427, 2100, \dodoi{10.1111/j.1365-2966.2012.22077.x}

\bibitem[{{Bisbas} {et~al.}(2015){Bisbas}, {Haworth}, {Barlow}, {Viti}, {Harries}, {Bell}, \& {Yates}}]{Bisbas15}
{Bisbas}, T.~G., {Haworth}, T.~J., {Barlow}, M.~J., {et~al.} 2015, \mnras, 454, 2828, \dodoi{10.1093/mnras/stv2156}

\bibitem[{{Bisbas} {et~al.}(2021){Bisbas}, {Tan}, \& {Tanaka}}]{Bisbas21}
{Bisbas}, T.~G., {Tan}, J.~C., \& {Tanaka}, K. E.~I. 2021, \mnras, 502, 2701, \dodoi{10.1093/mnras/stab121}

\bibitem[{{Bisbas} {et~al.}(2024){Bisbas}, {Zhang}, {Gjergo}, {Zhao}, {Luo}, {Quan}, {Jiang}, {Sun}, {Topkaras}, {Li}, \& {Guo}}]{Bisbas24}
{Bisbas}, T.~G., {Zhang}, Z.-Y., {Gjergo}, E., {et~al.} 2024, \mnras, 527, 8886, \dodoi{10.1093/mnras/stad3782}

\bibitem[{{Bisbas} {et~al.}(2025{\natexlab{a}}){Bisbas}, {Zhang}, {Kyrmanidou}, {Luo}, {Zhao}, {Topkaras}, {Jiang}, {Quan}, \& {Li}}]{Bisbas25}
{Bisbas}, T.~G., {Zhang}, Z.-Y., {Kyrmanidou}, M.-C., {et~al.} 2025{\natexlab{a}}, \aap, 697, A115, \dodoi{10.1051/0004-6361/202554067}

\bibitem[{{Bisbas} {et~al.}(2025{\natexlab{b}}){Bisbas}, {Zhang}, {Kyrmanidou}, {Luo}, {Zhao}, {Topkaras}, {Jiang}, {Quan}, \& {Li}}]{Bisbas2025}
---. 2025{\natexlab{b}}, arXiv, arXiv:2503.12073, \dodoi{10.48550/arXiv.2503.12073}

\bibitem[{{B{\"o}ker} {et~al.}(2022){B{\"o}ker}, {Arribas}, {L{\"u}tzgendorf}, {Alves de Oliveira}, {Beck}, {Birkmann}, {Bunker}, {Charlot}, {de Marchi}, {Ferruit}, {Giardino}, {Jakobsen}, {Kumari}, {L{\'o}pez-Caniego}, {Maiolino}, {Manjavacas}, {Marston}, {Moseley}, {Muzerolle}, {Ogle}, {Pirzkal}, {Rauscher}, {Rawle}, {Rix}, {Sabbi}, {Sargent}, {Sirianni}, {te Plate}, {Valenti}, {Willott}, \& {Zeidler}}]{Boker2022}
{B{\"o}ker}, T., {Arribas}, S., {L{\"u}tzgendorf}, N., {et~al.} 2022, \aap, 661, A82, \dodoi{10.1051/0004-6361/202142589}

\bibitem[{{Bolatto} {et~al.}(1999){Bolatto}, {Jackson}, \& {Ingalls}}]{Bolatto99}
{Bolatto}, A.~D., {Jackson}, J.~M., \& {Ingalls}, J.~G. 1999, \apj, 513, 275, \dodoi{10.1086/306849}

\bibitem[{{Bolatto} {et~al.}(2013){Bolatto}, {Wolfire}, \& {Leroy}}]{Bolatto13}
{Bolatto}, A.~D., {Wolfire}, M., \& {Leroy}, A.~K. 2013, \araa, 51, 207, \dodoi{10.1146/annurev-astro-082812-140944}

\bibitem[{{Bolatto} {et~al.}(2011){Bolatto}, {Leroy}, {Jameson}, {Ostriker}, {Gordon}, {Lawton}, {Stanimirovi{\'c}}, {Israel}, {Madden}, {Hony}, {Sandstrom}, {Bot}, {Rubio}, {Winkler}, {Roman-Duval}, {van Loon}, {Oliveira}, \& {Indebetouw}}]{Bolatto11}
{Bolatto}, A.~D., {Leroy}, A.~K., {Jameson}, K., {et~al.} 2011, \apj, 741, 12, \dodoi{10.1088/0004-637X/741/1/12}

\bibitem[{Bradley {et~al.}(2023)Bradley, Sip{\H o}cz, Robitaille, Tollerud, Vin{\'{\i}}cius, Deil, Barbary, Wilson, Busko, Donath, G{\"u}nther, Cara, Lim, Me{\ss}linger, Conseil, Bostroem, Droettboom, Bray, Bratholm, Barentsen, Craig, Rathi, Pascual, Perren, Georgiev, de~Val-Borro, Kerzendorf, Bach, Quint, \& Souchereau}]{photutils}
Bradley, L., Sip{\H o}cz, B., Robitaille, T., {et~al.} 2023, astropy/photutils: 1.8.0, 1.8.0,  Zenodo, \dodoi{10.5281/zenodo.7946442}

\bibitem[{{Bron} {et~al.}(2018){Bron}, {Daudon}, {Pety}, {Levrier}, {Gerin}, {Gratier}, {Orkisz}, {Guzman}, {Bardeau}, {Goicoechea}, {Liszt}, {{\"O}berg}, {Peretto}, {Sievers}, \& {Tremblin}}]{Bron18}
{Bron}, E., {Daudon}, C., {Pety}, J., {et~al.} 2018, \aap, 610, A12, \dodoi{10.1051/0004-6361/201731833}

\bibitem[{{Burton} {et~al.}(1990){Burton}, {Hollenbach}, \& {Tielens}}]{BurtonHollenbachTielens90}
{Burton}, M.~G., {Hollenbach}, D.~J., \& {Tielens}, A.~G.~G.~M. 1990, \apj, 365, 620, \dodoi{10.1086/169516}

\bibitem[{{Chandra} {et~al.}(1996){Chandra}, {Maheshwari}, \& {Sharma}}]{Chandra1996}
{Chandra}, S., {Maheshwari}, V.~U., \& {Sharma}, A.~K. 1996, \aaps, 117, 557

\bibitem[{{Chastenet} {et~al.}(2019){Chastenet}, {Sandstrom}, {Chiang}, {Leroy}, {Utomo}, {Bot}, {Gordon}, {Draine}, {Fukui}, {Onishi}, \& {Tsuge}}]{Chastenet2019}
{Chastenet}, J., {Sandstrom}, K., {Chiang}, I.-D., {et~al.} 2019, \apj, 876, 62, \dodoi{10.3847/1538-4357/ab16cf}

\bibitem[{{Chown} {et~al.}(2024){Chown}, {Sidhu}, {Peeters}, {Tielens}, {Cami}, {Bern{\'e}}, {Habart}, {Alarc{\'o}n}, {Canin}, {Schroetter}, {Trahin}, {Van De Putte}, {Abergel}, {Bergin}, {Bernard-Salas}, {Boersma}, {Bron}, {Cuadrado}, {Dartois}, {Dicken}, {El-Yajouri}, {Fuente}, {Goicoechea}, {Gordon}, {Issa}, {Joblin}, {Kannavou}, {Khan}, {Lacinbala}, {Languignon}, {Le Gal}, {Maragkoudakis}, {Meshaka}, {Okada}, {Onaka}, {Pasquini}, {Pound}, {Robberto}, {R{\"o}llig}, {Schefter}, {Schirmer}, {Vicente}, {Wolfire}, {Zannese}, {Aleman}, {Allamandola}, {Auchettl}, {Baratta}, {Bejaoui}, {Bera}, {Black}, {Boulanger}, {Bouwman}, {Brandl}, {Brechignac}, {Br{\"u}nken}, {Buragohain}, {Burkhardt}, {Candian}, {Cazaux}, {Cernicharo}, {Chabot}, {Chakraborty}, {Champion}, {Colgan}, {Cooke}, {Coutens}, {Cox}, {Demyk}, {Meyer}, {Foschino}, {Garc{\'\i}a-Lario}, {Gavilan}, {Gerin}, {Gottlieb}, {Guillard}, {Gusdorf}, {Hartigan}, {He}, {Herbst}, {Hornekaer}, {J{\"a}ger}, {Janot-Pacheco}, {Kaufman}, {Kemper}, {Kendrew},
  {Kirsanova}, {Klaassen}, {Kwok}, {Labiano}, {Lai}, {Lee}, {Lefloch}, {Le Petit}, {Li}, {Linz}, {Mackie}, {Madden}, {Mascetti}, {McGuire}, {Merino}, {Micelotta}, {Misselt}, {Morse}, {Mulas}, {Neelamkodan}, {Ohsawa}, {Omont}, {Paladini}, {Palumbo}, {Pathak}, {Pendleton}, {Petrignani}, {Pino}, {Puga}, {Rangwala}, {Rapacioli}, {Ricca}, {Roman-Duval}, {Roser}, {Roueff}, {Rouill{\'e}}, {Salama}, {Sales}, {Sandstrom}, {Sarre}, {Sciamma-O'Brien}, {Sellgren}, {Shenoy}, {Teyssier}, {Thomas}, {Togi}, {Verstraete}, {Witt}, {Wootten}, {Zettergren}, {Zhang}, {Zhang}, \& {Zhen}}]{Chown24}
{Chown}, R., {Sidhu}, A., {Peeters}, E., {et~al.} 2024, \aap, 685, A75, \dodoi{10.1051/0004-6361/202346662}

\bibitem[{Comrie {et~al.}(2021)Comrie, Wang, Hsu, Moraghan, Harris, Pang, Pi?ska, Chiang, Chang, Hwang, Jan, Lin, \& Simmonds}]{CARTA}
Comrie, A., Wang, K.-S., Hsu, S.-C., {et~al.} 2021, {CARTA: The Cube Analysis and Rendering Tool for Astronomy}, 2.0.0,  Zenodo, \dodoi{10.5281/zenodo.4905459}

\bibitem[{{Conti} {et~al.}(2008){Conti}, {Crowther}, \& {Leitherer}}]{Conti2008}
{Conti}, P.~S., {Crowther}, P.~A., \& {Leitherer}, C. 2008, {From Luminous Hot Stars to Starburst Galaxies}

\bibitem[{{Cubick} {et~al.}(2008){Cubick}, {Stutzki}, {Ossenkopf}, {Kramer}, \& {R{\"o}llig}}]{Cubick08}
{Cubick}, M., {Stutzki}, J., {Ossenkopf}, V., {Kramer}, C., \& {R{\"o}llig}, M. 2008, \aap, 488, 623, \dodoi{10.1051/0004-6361:20079270}

\bibitem[{{Draine}(2011)}]{Draine11}
{Draine}, B.~T. 2011, {Physics of the Interstellar and Intergalactic Medium}

\bibitem[{{Dufour} \& {Harlow}(1977)}]{Dufour77}
{Dufour}, R.~J., \& {Harlow}, W.~V. 1977, \apj, 216, 706, \dodoi{10.1086/155513}

\bibitem[{{Flower} \& {Roueff}(1998)}]{Flower1998}
{Flower}, D.~R., \& {Roueff}, E. 1998, Journal of Physics B Atomic Molecular Physics, 31, 2935, \dodoi{10.1088/0953-4075/31/13/012}

\bibitem[{{Flower} \& {Roueff}(1999)}]{Flower1999}
---. 1999, Journal of Physics B Atomic Molecular Physics, 32, 3399, \dodoi{10.1088/0953-4075/32/14/310}

\bibitem[{{Fuente} {et~al.}(2024){Fuente}, {Roueff}, {Le Petit}, {Le Bourlot}, {Bron}, {Wolfire}, {Babb}, {Yan}, {Onaka}, {Black}, {Schroetter}, {Van De Putte}, {Sidhu}, {Canin}, {Trahin}, {Alarc{\'o}n}, {Chown}, {Kannavou}, {Bern{\'e}}, {Habart}, {Peeters}, {Goicoechea}, {Zannese}, {Meshaka}, {Okada}, {R{\"o}llig}, {Le Gal}, {Sales}, {Palumbo}, {Baratta}, {Madden}, {Neelamkodan}, {Zhang}, \& {Stancil}}]{Fuente24}
{Fuente}, A., {Roueff}, E., {Le Petit}, F., {et~al.} 2024, \aap, 687, A87, \dodoi{10.1051/0004-6361/202449229}

\bibitem[{{Gaches} {et~al.}(2023){Gaches}, {Walch}, {W{\"u}nsch}, \& {Mackey}}]{Gaches23}
{Gaches}, B. A.~L., {Walch}, S., {W{\"u}nsch}, R., \& {Mackey}, J. 2023, \mnras, 522, 4674, \dodoi{10.1093/mnras/stad1206}

\bibitem[{{Gerin} \& {Phillips}(2000)}]{Gerin00}
{Gerin}, M., \& {Phillips}, T.~G. 2000, \apj, 537, 644, \dodoi{10.1086/309072}

\bibitem[{{Glover} \& {Clark}(2012)}]{Glover12}
{Glover}, S. C.~O., \& {Clark}, P.~C. 2012, \mnras, 426, 377, \dodoi{10.1111/j.1365-2966.2012.21737.x}

\bibitem[{{Glover} \& {Clark}(2016)}]{Glover16}
---. 2016, \mnras, 456, 3596, \dodoi{10.1093/mnras/stv2863}

\bibitem[{{Glover} {et~al.}(2015){Glover}, {Clark}, {Micic}, \& {Molina}}]{Glover2015}
{Glover}, S. C.~O., {Clark}, P.~C., {Micic}, M., \& {Molina}, F. 2015, \mnras, 448, 1607, \dodoi{10.1093/mnras/stu2699}

\bibitem[{{Glover} \& {Mac Low}(2011)}]{Glover2010}
{Glover}, S.~C.~O., \& {Mac Low}, M.~M. 2011, \mnras, 412, 337, \dodoi{10.1111/j.1365-2966.2010.17907.x}

\bibitem[{{Gnedin} \& {Draine}(2014)}]{Gnedin&Draine2014}
{Gnedin}, N.~Y., \& {Draine}, B.~T. 2014, \apj, 795, 37, \dodoi{10.1088/0004-637X/795/1/37}

\bibitem[{{Goicoechea} {et~al.}(2019){Goicoechea}, {Santa-Maria}, {Bron}, {Teyssier}, {Marcelino}, {Cernicharo}, \& {Cuadrado}}]{Goicoechea19}
{Goicoechea}, J.~R., {Santa-Maria}, M.~G., {Bron}, E., {et~al.} 2019, \aap, 622, A91, \dodoi{10.1051/0004-6361/201834409}

\bibitem[{{Goicoechea} {et~al.}(2016){Goicoechea}, {Pety}, {Cuadrado}, {Cernicharo}, {Chapillon}, {Fuente}, {Gerin}, {Joblin}, {Marcelino}, \& {Pilleri}}]{Goicoechea16}
{Goicoechea}, J.~R., {Pety}, J., {Cuadrado}, S., {et~al.} 2016, \nat, 537, 207, \dodoi{10.1038/nature18957}

\bibitem[{{Goicoechea} {et~al.}(2017){Goicoechea}, {Cuadrado}, {Pety}, {Bron}, {Black}, {Cernicharo}, {Chapillon}, {Fuente}, \& {Gerin}}]{Goicoechea17}
{Goicoechea}, J.~R., {Cuadrado}, S., {Pety}, J., {et~al.} 2017, \aap, 601, L9, \dodoi{10.1051/0004-6361/201730716}

\bibitem[{{Goicoechea} {et~al.}(2025){Goicoechea}, {Pety}, {Cuadrado}, {Bern{\'e}}, {Dartois}, {Gerin}, {Joblin}, {K{\l}os}, {Lique}, {Onaka}, {Peeters}, {Tielens}, {Alarc{\'o}n}, {Bron}, {Cami}, {Canin}, {Chapillon}, {Chown}, {Fuente}, {Habart}, {Kannavou}, {Le Petit}, {Santa-Maria}, {Schroetter}, {Sidhu}, {Trahin}, {Van De Putte}, \& {Zannese}}]{Goicoechea25}
{Goicoechea}, J.~R., {Pety}, J., {Cuadrado}, S., {et~al.} 2025, \aap, 696, A100, \dodoi{10.1051/0004-6361/202453350}

\bibitem[{{Goldsmith} {et~al.}(2008){Goldsmith}, {Heyer}, {Narayanan}, {Snell}, {Li}, \& {Brunt}}]{Goldsmith08}
{Goldsmith}, P.~F., {Heyer}, M., {Narayanan}, G., {et~al.} 2008, \apj, 680, 428, \dodoi{10.1086/587166}

\bibitem[{{Gong} {et~al.}(2020){Gong}, {Ostriker}, {Kim}, \& {Kim}}]{Gong2020}
{Gong}, M., {Ostriker}, E.~C., {Kim}, C.-G., \& {Kim}, J.-G. 2020, \apj, 903, 142, \dodoi{10.3847/1538-4357/abbdab}

\bibitem[{{Gordon} {et~al.}(2003){Gordon}, {Clayton}, {Misselt}, {Landolt}, \& {Wolff}}]{Gordon03}
{Gordon}, K.~D., {Clayton}, G.~C., {Misselt}, K.~A., {Landolt}, A.~U., \& {Wolff}, M.~J. 2003, \apj, 594, 279, \dodoi{10.1086/376774}

\bibitem[{{Gordon} {et~al.}(2024){Gordon}, {Fitzpatrick}, {Massa}, {Bohlin}, {Chastenet}, {Murray}, {Clayton}, {Lennon}, {Misselt}, \& {Sandstrom}}]{Gordon2024}
{Gordon}, K.~D., {Fitzpatrick}, E.~L., {Massa}, D., {et~al.} 2024, \apj, 970, 51, \dodoi{10.3847/1538-4357/ad4be1}

\bibitem[{{Gorti} \& {Hollenbach}(2002)}]{Gorti02}
{Gorti}, U., \& {Hollenbach}, D. 2002, \apj, 573, 215, \dodoi{10.1086/340556}

\bibitem[{{Grassi} {et~al.}(2014){Grassi}, {Bovino}, {Schleicher}, {Prieto}, {Seifried}, {Simoncini}, \& {Gianturco}}]{Grassi14}
{Grassi}, T., {Bovino}, S., {Schleicher}, D.~R.~G., {et~al.} 2014, \mnras, 439, 2386, \dodoi{10.1093/mnras/stu114}

\bibitem[{{Grenier} {et~al.}(2005){Grenier}, {Casandjian}, \& {Terrier}}]{Grenier05}
{Grenier}, I.~A., {Casandjian}, J.-M., \& {Terrier}, R. 2005, Science, 307, 1292, \dodoi{10.1126/science.1106924}

\bibitem[{{Gurman} {et~al.}(2024){Gurman}, {Hu}, {Sternberg}, \& {van Dishoeck}}]{Gurman2024}
{Gurman}, A., {Hu}, C.-Y., {Sternberg}, A., \& {van Dishoeck}, E.~F. 2024, \apj, 965, 179, \dodoi{10.3847/1538-4357/ad2eac}

\bibitem[{{Habart} {et~al.}(2024){Habart}, {Peeters}, {Bern{\'e}}, {Trahin}, {Canin}, {Chown}, {Sidhu}, {Van De Putte}, {Alarc{\'o}n}, {Schroetter}, {Dartois}, {Vicente}, {Abergel}, {Bergin}, {Bernard-Salas}, {Boersma}, {Bron}, {Cami}, {Cuadrado}, {Dicken}, {Elyajouri}, {Fuente}, {Goicoechea}, {Gordon}, {Issa}, {Joblin}, {Kannavou}, {Khan}, {Lacinbala}, {Languignon}, {Le Gal}, {Maragkoudakis}, {Meshaka}, {Okada}, {Onaka}, {Pasquini}, {Pound}, {Robberto}, {R{\"o}llig}, {Schefter}, {Schirmer}, {Tabone}, {Tielens}, {Wolfire}, {Zannese}, {Ysard}, {Miville-Deschenes}, {Aleman}, {Allamandola}, {Auchettl}, {Baratta}, {Bejaoui}, {Bera}, {Black}, {Boulanger}, {Bouwman}, {Brandl}, {Brechignac}, {Br{\"u}nken}, {Buragohain}, {Burkhardt}, {Candian}, {Cazaux}, {Cernicharo}, {Chabot}, {Chakraborty}, {Champion}, {Colgan}, {Cooke}, {Coutens}, {Cox}, {Demyk}, {Meyer}, {Foschino}, {Garc{\'\i}a-Lario}, {Gavilan}, {Gerin}, {Gottlieb}, {Guillard}, {Gusdorf}, {Hartigan}, {He}, {Herbst}, {Hornekaer}, {J{\"a}ger}, {Janot-Pacheco},
  {Kaufman}, {Kemper}, {Kendrew}, {Kirsanova}, {Klaassen}, {Kwok}, {Labiano}, {Lai}, {Lee}, {Lefloch}, {Le Petit}, {Li}, {Linz}, {Mackie}, {Madden}, {Mascetti}, {McGuire}, {Merino}, {Micelotta}, {Misselt}, {Morse}, {Mulas}, {Neelamkodan}, {Ohsawa}, {Omont}, {Paladini}, {Palumbo}, {Pathak}, {Pendleton}, {Petrignani}, {Pino}, {Puga}, {Rangwala}, {Rapacioli}, {Ricca}, {Roman-Duval}, {Roser}, {Roueff}, {Rouill{\'e}}, {Salama}, {Sales}, {Sandstrom}, {Sarre}, {Sciamma-O'Brien}, {Sellgren}, {Shenoy}, {Teyssier}, {Thomas}, {Togi}, {Verstraete}, {Witt}, {Wootten}, {Zettergren}, {Zhang}, {Zhang}, \& {Zhen}}]{Habart24}
{Habart}, E., {Peeters}, E., {Bern{\'e}}, O., {et~al.} 2024, \aap, 685, A73, \dodoi{10.1051/0004-6361/202346747}

\bibitem[{{Habing}(1968)}]{Habing68}
{Habing}, H.~J. 1968, \bain, 19, 421

\bibitem[{{Haid} {et~al.}(2019){Haid}, {Walch}, {Seifried}, {W{\"u}nsch}, {Dinnbier}, \& {Naab}}]{Haid19}
{Haid}, S., {Walch}, S., {Seifried}, D., {et~al.} 2019, \mnras, 482, 4062, \dodoi{10.1093/mnras/sty2938}

\bibitem[{Harris {et~al.}(2020)Harris, Millman, van~der Walt, Gommers, Virtanen, Cournapeau, Wieser, Taylor, Berg, Smith, Kern, Picus, Hoyer, van Kerkwijk, Brett, Haldane, del R{'{\i}}o, Wiebe, Peterson, G{'{e}}rard-Marchant, Sheppard, Reddy, Weckesser, Abbasi, Gohlke, \& Oliphant}]{harris2020array}
Harris, C.~R., Millman, K.~J., van~der Walt, S.~J., {et~al.} 2020, Nature, 585, 357, \dodoi{10.1038/s41586-020-2649-2}

\bibitem[{{Heays} {et~al.}(2017){Heays}, {Bosman}, \& {van Dishoeck}}]{Heays17}
{Heays}, A.~N., {Bosman}, A.~D., \& {van Dishoeck}, E.~F. 2017, \aap, 602, A105, \dodoi{10.1051/0004-6361/201628742}

\bibitem[{{Hern{\'a}ndez-Vera} {et~al.}(2023){Hern{\'a}ndez-Vera}, {Guzm{\'a}n}, {Goicoechea}, {Maillard}, {Pety}, {Le Petit}, {Gerin}, {Bron}, {Roueff}, {Abergel}, {Schirmer}, {Carpenter}, {Gratier}, {Gordon}, \& {Misselt}}]{Hernandez-Vera23}
{Hern{\'a}ndez-Vera}, C., {Guzm{\'a}n}, V.~V., {Goicoechea}, J.~R., {et~al.} 2023, \aap, 677, A152, \dodoi{10.1051/0004-6361/202347206}

\bibitem[{{Hollenbach} {et~al.}(2012){Hollenbach}, {Kaufman}, {Neufeld}, {Wolfire}, \& {Goicoechea}}]{Hollenbach2012}
{Hollenbach}, D., {Kaufman}, M.~J., {Neufeld}, D., {Wolfire}, M., \& {Goicoechea}, J.~R. 2012, \apj, 754, 105, \dodoi{10.1088/0004-637X/754/2/105}

\bibitem[{{Hollenbach} \& {Tielens}(1997)}]{Hollenbach97}
{Hollenbach}, D.~J., \& {Tielens}, A.~G.~G.~M. 1997, \araa, 35, 179, \dodoi{10.1146/annurev.astro.35.1.179}

\bibitem[{{Hollenbach} \& {Tielens}(1999)}]{Hollenbach99}
---. 1999, Reviews of Modern Physics, 71, 173, \dodoi{10.1103/RevModPhys.71.173}

\bibitem[{{Hu} {et~al.}(2021){Hu}, {Sternberg}, \& {van Dishoeck}}]{Hu21}
{Hu}, C.-Y., {Sternberg}, A., \& {van Dishoeck}, E.~F. 2021, \apj, 920, 44, \dodoi{10.3847/1538-4357/ac0dbd}

\bibitem[{Hunter(2007)}]{Hunter:2007}
Hunter, J.~D. 2007, Computing In Science \& Engineering, 9, 90

\bibitem[{{Israel}(1997)}]{Israel97}
{Israel}, F.~P. 1997, \aap, 328, 471, \dodoi{10.48550/arXiv.astro-ph/9709194}

\bibitem[{{Izumi} {et~al.}(2021){Izumi}, {Fukui}, {Tachihara}, {Fujita}, {Torii}, {Kamazaki}, {Kaneko}, {Silva}, {Iono}, {Momose}, {Sugimoto}, {Nakazato}, {Kosugi}, {Maekawa}, {Takahashi}, {Yoshino}, \& {Asayama}}]{Izumi21}
{Izumi}, N., {Fukui}, Y., {Tachihara}, K., {et~al.} 2021, \pasj, 73, 174, \dodoi{10.1093/pasj/psaa113}

\bibitem[{{Jakobsen} {et~al.}(2022){Jakobsen}, {Ferruit}, {Alves de Oliveira}, {Arribas}, {Bagnasco}, {Barho}, {Beck}, {Birkmann}, {B{\"o}ker}, {Bunker}, {Charlot}, {de Jong}, {de Marchi}, {Ehrenwinkler}, {Falcolini}, {Fels}, {Franx}, {Franz}, {Funke}, {Giardino}, {Gnata}, {Holota}, {Honnen}, {Jensen}, {Jentsch}, {Johnson}, {Jollet}, {Karl}, {Kling}, {K{\"o}hler}, {Kolm}, {Kumari}, {Lander}, {Lemke}, {L{\'o}pez-Caniego}, {L{\"u}tzgendorf}, {Maiolino}, {Manjavacas}, {Marston}, {Maschmann}, {Maurer}, {Messerschmidt}, {Moseley}, {Mosner}, {Mott}, {Muzerolle}, {Pirzkal}, {Pittet}, {Plitzke}, {Posselt}, {Rapp}, {Rauscher}, {Rawle}, {Rix}, {R{\"o}del}, {Rumler}, {Sabbi}, {Salvignol}, {Schmid}, {Sirianni}, {Smith}, {Strada}, {te Plate}, {Valenti}, {Wettemann}, {Wiehe}, {Wiesmayer}, {Willott}, {Wright}, {Zeidler}, \& {Zincke}}]{Jakobsen2022}
{Jakobsen}, P., {Ferruit}, P., {Alves de Oliveira}, C., {et~al.} 2022, \aap, 661, A80, \dodoi{10.1051/0004-6361/202142663}

\bibitem[{{Jameson} {et~al.}(2018){Jameson}, {Bolatto}, {Wolfire}, {Warren}, {Herrera-Camus}, {Croxall}, {Pellegrini}, {Smith}, {Rubio}, {Indebetouw}, {Israel}, {Meixner}, {Roman-Duval}, {van Loon}, {Muller}, {Verdugo}, {Zinnecker}, \& {Okada}}]{Jameson2018}
{Jameson}, K.~E., {Bolatto}, A.~D., {Wolfire}, M., {et~al.} 2018, \apj, 853, 111, \dodoi{10.3847/1538-4357/aaa4bb}

\bibitem[{{Joblin} {et~al.}(2018){Joblin}, {Bron}, {Pinto}, {Pilleri}, {Le Petit}, {Gerin}, {Le Bourlot}, {Fuente}, {Berne}, {Goicoechea}, {Habart}, {K{\"o}hler}, {Teyssier}, {Nagy}, {Montillaud}, {Vastel}, {Cernicharo}, {R{\"o}llig}, {Ossenkopf-Okada}, \& {Bergin}}]{Joblin18}
{Joblin}, C., {Bron}, E., {Pinto}, C., {et~al.} 2018, \aap, 615, A129, \dodoi{10.1051/0004-6361/201832611}

\bibitem[{{K{\'a}losi} {et~al.}(2023){K{\'a}losi}, {Gamer}, {Grieser}, {von Hahn}, {Isberner}, {J{\"a}ger}, {Kreckel}, {Neufeld}, {Paul}, {Savin}, {Schippers}, {Schmidt}, {Wolf}, {Wolfire}, \& {Novotn{\'y}}}]{Kalosi23}
{K{\'a}losi}, {\'A}., {Gamer}, L., {Grieser}, M., {et~al.} 2023, \apjl, 955, L26, \dodoi{10.3847/2041-8213/acf71d}

\bibitem[{{Kaufman} {et~al.}(2006){Kaufman}, {Wolfire}, \& {Hollenbach}}]{Kaufman06}
{Kaufman}, M.~J., {Wolfire}, M.~G., \& {Hollenbach}, D.~J. 2006, \apj, 644, 283, \dodoi{10.1086/503596}

\bibitem[{{K{\l}os} {et~al.}(2021){K{\l}os}, {Dagdigian}, \& {Lique}}]{Klos21}
{K{\l}os}, J., {Dagdigian}, P.~J., \& {Lique}, F. 2021, \mnras, 501, L38, \dodoi{10.1093/mnrasl/slaa192}

\bibitem[{{Kramer} {et~al.}(2008){Kramer}, {Cubick}, {R{\"o}llig}, {Sun}, {Yonekura}, {Aravena}, {Bensch}, {Bertoldi}, {Bronfman}, {Fujishita}, {Fukui}, {Graf}, {Hitschfeld}, {Honingh}, {Ito}, {Jakob}, {Jacobs}, {Klein}, {Koo}, {May}, {Miller}, {Miyamoto}, {Mizuno}, {Onishi}, {Park}, {Pineda}, {Rabanus}, {Sasago}, {Schieder}, {Simon}, {Stutzki}, {Volgenau}, \& {Yamamoto}}]{Kramer08}
{Kramer}, C., {Cubick}, M., {R{\"o}llig}, M., {et~al.} 2008, \aap, 477, 547, \dodoi{10.1051/0004-6361:20077815}

\bibitem[{{Leroy} {et~al.}(2011){Leroy}, {Bolatto}, {Gordon}, {Sandstrom}, {Gratier}, {Rosolowsky}, {Engelbracht}, {Mizuno}, {Corbelli}, {Fukui}, \& {Kawamura}}]{Leroy11}
{Leroy}, A.~K., {Bolatto}, A., {Gordon}, K., {et~al.} 2011, \apj, 737, 12, \dodoi{10.1088/0004-637X/737/1/12}

\bibitem[{{Leroy} {et~al.}(2021){Leroy}, {Hughes}, {Liu}, {Pety}, {Rosolowsky}, {Saito}, {Schinnerer}, {Schruba}, {Usero}, {Faesi}, {Herrera}, {Chevance}, {Hygate}, {Kepley}, {Koch}, {Querejeta}, {Sliwa}, {Will}, {Wilson}, {Anand}, {Barnes}, {Belfiore}, {Be{\v{s}}li{\'c}}, {Bigiel}, {Blanc}, {Bolatto}, {Boquien}, {Cao}, {Chandar}, {Chastenet}, {Chiang}, {Congiu}, {Dale}, {Deger}, {den Brok}, {Eibensteiner}, {Emsellem}, {Garc{\'\i}a-Rodr{\'\i}guez}, {Glover}, {Grasha}, {Groves}, {Henshaw}, {Jim{\'e}nez Donaire}, {Kim}, {Klessen}, {Kreckel}, {Kruijssen}, {Larson}, {Lee}, {Mayker}, {McElroy}, {Meidt}, {Mok}, {Pan}, {Puschnig}, {Razza}, {S{\'a}nchez-Bl'azquez}, {Sandstrom}, {Santoro}, {Sardone}, {Scheuermann}, {Sun}, {Thilker}, {Turner}, {Ubeda}, {Utomo}, {Watkins}, \& {Williams}}]{Leroy21b}
{Leroy}, A.~K., {Hughes}, A., {Liu}, D., {et~al.} 2021, \apjs, 255, 19, \dodoi{10.3847/1538-4365/abec80}

\bibitem[{{Lupi} {et~al.}(2018){Lupi}, {Bovino}, {Capelo}, {Volonteri}, \& {Silk}}]{Lupi18}
{Lupi}, A., {Bovino}, S., {Capelo}, P.~R., {Volonteri}, M., \& {Silk}, J. 2018, \mnras, 474, 2884, \dodoi{10.1093/mnras/stx2874}

\bibitem[{{Madden} {et~al.}(2020){Madden}, {Cormier}, {Hony}, {Lebouteiller}, {Abel}, {Galametz}, {De Looze}, {Chevance}, {Polles}, {Lee}, {Galliano}, {Lambert-Huyghe}, {Hu}, \& {Ramambason}}]{Madden20}
{Madden}, S.~C., {Cormier}, D., {Hony}, S., {et~al.} 2020, \aap, 643, A141, \dodoi{10.1051/0004-6361/202038860}

\bibitem[{{Maillard} {et~al.}(2021){Maillard}, {Bron}, \& {Le Petit}}]{Mailard2021}
{Maillard}, V., {Bron}, E., \& {Le Petit}, F. 2021, \aap, 656, A65, \dodoi{10.1051/0004-6361/202140865}

\bibitem[{{Neufeld} \& {Wolfire}(2016)}]{Neufeld16}
{Neufeld}, D.~A., \& {Wolfire}, M.~G. 2016, \apj, 826, 183, \dodoi{10.3847/0004-637X/826/2/183}

\bibitem[{{Neufeld} \& {Wolfire}(2017)}]{Neufeld2017}
---. 2017, \apj, 845, 163, \dodoi{10.3847/1538-4357/aa6d68}

\bibitem[{{Nordon} \& {Sternberg}(2016)}]{Nordon16}
{Nordon}, R., \& {Sternberg}, A. 2016, \mnras, 462, 2804, \dodoi{10.1093/mnras/stw1791}

\bibitem[{{Papadopoulos} {et~al.}(2004){Papadopoulos}, {Thi}, \& {Viti}}]{Papadopoulos04}
{Papadopoulos}, P.~P., {Thi}, W.~F., \& {Viti}, S. 2004, \mnras, 351, 147, \dodoi{10.1111/j.1365-2966.2004.07762.x}

\bibitem[{{Parravano} {et~al.}(2003){Parravano}, {Hollenbach}, \& {McKee}}]{Parravano2003}
{Parravano}, A., {Hollenbach}, D.~J., \& {McKee}, C.~F. 2003, \apj, 584, 797, \dodoi{10.1086/345807}

\bibitem[{{Paul} {et~al.}(2022){Paul}, {Grieser}, {Grussie}, {von Hahn}, {Isberner}, {K{\'a}losi}, {Krantz}, {Kreckel}, {M{\"u}ll}, {Neufeld}, {Savin}, {Schippers}, {Wilhelm}, {Wolf}, {Wolfire}, \& {Novotn{\'y}}}]{Paul2022}
{Paul}, D., {Grieser}, M., {Grussie}, F., {et~al.} 2022, \apj, 939, 122, \dodoi{10.3847/1538-4357/ac8e02}

\bibitem[{PDRs4AllTeam {et~al.}(2022)PDRs4AllTeam, Berné, Émilie Habart, Peeters, Team:, Abergel, Bergin, Bernard-Salas, Bron, Cami, Dartois, Fuente, Goicoechea, Gordon, Okada, Onaka, Robberto, Röllig, Tielens, Vicente, Wolfire, Team:, Alarcón, Boersma, Canin, Chown, Dicken, Languignon, Gal, Pound, Trahin, Simmer, Sidhu, Putte, time co-authors contributed~to SEPs, Cuadrado, Guilloteau, Maragkoudakis, Schefter, Schirmer, Collaborators:, Cazaux, Aleman, Allamandola, Auchettl, Baratta, Bejaoui, Bera, Bilalbegović, Black, Boulanger, Bouwman, Brandl, Brechignac, Brünken, Burkhardt, Candian, Cernicharo, Chabot, Chakraborty, Champion, Colgan, Cooke, Coutens, Cox, Demyk, Meyer, Engrand, Foschino, García-Lario, Gavilan, Gerin, Godard, Gottlieb, Guillard, Gusdorf, Hartigan, He, Herbst, Hornekaer, Jäger, Janot-Pacheco, Joblin, Kaufman, Kemper, Kendrew, Kirsanova, Klaassen, Knight, Kwok, Álvaro Labiano, Lai, Lee, Lefloch, Petit, Li, Linz, Mackie, Madden, Mascetti, McGuire, Merino, Micelotta, Misselt, Morse,
  Mulas, Neelamkodan, Ohsawa, Omont, Paladini, Palumbo, Pathak, Pendleton, Petrignani, Pino, Puga, Rangwala, Rapacioli, Ricca, Roman-Duval, Roser, Roueff, Rouillé, Salama, Sales, Sandstrom, Sarre, Sciamma-O’Brien, Sellgren, Shannon, Shenoy, Teyssier, Thomas, Togi, Verstraete, Witt, Wootten, Ysard, Zettergren, Zhang, Zhang, \& Zhen}]{Berné_2022}
PDRs4AllTeam, Berné, O., Émilie Habart, {et~al.} 2022, Publications of the Astronomical Society of the Pacific, 134, 054301, \dodoi{10.1088/1538-3873/ac604c}

\bibitem[{{Peeters} {et~al.}(2024){Peeters}, {Habart}, {Bern{\'e}}, {Sidhu}, {Chown}, {Van De Putte}, {Trahin}, {Schroetter}, {Canin}, {Alarc{\'o}n}, {Schefter}, {Khan}, {Pasquini}, {Tielens}, {Wolfire}, {Dartois}, {Goicoechea}, {Maragkoudakis}, {Onaka}, {Pound}, {Vicente}, {Abergel}, {Bergin}, {Bernard-Salas}, {Boersma}, {Bron}, {Cami}, {Cuadrado}, {Dicken}, {Elyajouri}, {Fuente}, {Gordon}, {Issa}, {Joblin}, {Kannavou}, {Lacinbala}, {Languignon}, {Le Gal}, {Meshaka}, {Okada}, {Robberto}, {R{\"o}llig}, {Schirmer}, {Tabone}, {Zannese}, {Aleman}, {Allamandola}, {Auchettl}, {Baratta}, {Bejaoui}, {Bera}, {Black}, {Boulanger}, {Bouwman}, {Brandl}, {Brechignac}, {Br{\"u}nken}, {Buragohain}, {Burkhardt}, {Candian}, {Cazaux}, {Cernicharo}, {Chabot}, {Chakraborty}, {Champion}, {Colgan}, {Cooke}, {Coutens}, {Cox}, {Demyk}, {Meyer}, {Foschino}, {Garc{\'\i}a-Lario}, {Gerin}, {Gottlieb}, {Guillard}, {Gusdorf}, {Hartigan}, {He}, {Herbst}, {Hornekaer}, {J{\"a}ger}, {Janot-Pacheco}, {Kaufman}, {Kendrew}, {Kirsanova},
  {Klaassen}, {Kwok}, {Labiano}, {Lai}, {Lee}, {Lefloch}, {Le Petit}, {Li}, {Linz}, {Mackie}, {Madden}, {Mascetti}, {McGuire}, {Merino}, {Micelotta}, {Misselt}, {Morse}, {Mulas}, {Neelamkodan}, {Ohsawa}, {Paladini}, {Palumbo}, {Pathak}, {Pendleton}, {Petrignani}, {Pino}, {Puga}, {Rangwala}, {Rapacioli}, {Ricca}, {Roman-Duval}, {Roser}, {Roueff}, {Rouill{\'e}}, {Salama}, {Sales}, {Sandstrom}, {Sarre}, {Sciamma-O'Brien}, {Sellgren}, {Shenoy}, {Teyssier}, {Thomas}, {Togi}, {Verstraete}, {Witt}, {Wootten}, {Ysard}, {Zettergren}, {Zhang}, {Zhang}, \& {Zhen}}]{Peeters24}
{Peeters}, E., {Habart}, E., {Bern{\'e}}, O., {et~al.} 2024, \aap, 685, A74, \dodoi{10.1051/0004-6361/202348244}

\bibitem[{{Pineda} {et~al.}(2013){Pineda}, {Langer}, {Velusamy}, \& {Goldsmith}}]{Pineda13}
{Pineda}, J.~L., {Langer}, W.~D., {Velusamy}, T., \& {Goldsmith}, P.~F. 2013, \aap, 554, A103, \dodoi{10.1051/0004-6361/201321188}

\bibitem[{{Pineda} {et~al.}(2017){Pineda}, {Langer}, {Goldsmith}, {Horiuchi}, {Kuiper}, {Muller}, {Hughes}, {Ott}, {Requena-Torres}, {Velusamy}, \& {Wong}}]{Pineda17}
{Pineda}, J.~L., {Langer}, W.~D., {Goldsmith}, P.~F., {et~al.} 2017, \apj, 839, 107, \dodoi{10.3847/1538-4357/aa683a}

\bibitem[{{Ramachandran} {et~al.}(2019){Ramachandran}, {Hamann}, {Oskinova}, {Gallagher}, {Hainich}, {Shenar}, {Sander}, {Todt}, \& {Fulmer}}]{Ramachandran2019}
{Ramachandran}, V., {Hamann}, W.~R., {Oskinova}, L.~M., {et~al.} 2019, \aap, 625, A104, \dodoi{10.1051/0004-6361/201935365}

\bibitem[{{R{\'e}my-Ruyer} {et~al.}(2014){R{\'e}my-Ruyer}, {Madden}, {Galliano}, {Galametz}, {Takeuchi}, {Asano}, {Zhukovska}, {Lebouteiller}, {Cormier}, {Jones}, {Bocchio}, {Baes}, {Bendo}, {Boquien}, {Boselli}, {DeLooze}, {Doublier-Pritchard}, {Hughes}, {Karczewski}, \& {Spinoglio}}]{Remy-Ruyer2014}
{R{\'e}my-Ruyer}, A., {Madden}, S.~C., {Galliano}, F., {et~al.} 2014, \aap, 563, A31, \dodoi{10.1051/0004-6361/201322803}

\bibitem[{{R{\"o}llig} {et~al.}(2006){R{\"o}llig}, {Ossenkopf}, {Jeyakumar}, {Stutzki}, \& {Sternberg}}]{Rollig06}
{R{\"o}llig}, M., {Ossenkopf}, V., {Jeyakumar}, S., {Stutzki}, J., \& {Sternberg}, A. 2006, \aap, 451, 917, \dodoi{10.1051/0004-6361:20053845}

\bibitem[{{R{\"o}llig} \& {Ossenkopf-Okada}(2022)}]{Rollig2022}
{R{\"o}llig}, M., \& {Ossenkopf-Okada}, V. 2022, \aap, 664, A67, \dodoi{10.1051/0004-6361/202141854}

\bibitem[{{Roman-Duval} {et~al.}(2022){Roman-Duval}, {Jenkins}, {Tchernyshyov}, {Clark}, {De Cia}, {Gordon}, {Hamanowicz}, {Lebouteiller}, {Rafelski}, {Sandstrom}, {Werk}, \& {Yanchulova Merica-Jones}}]{Roman-Duval2022}
{Roman-Duval}, J., {Jenkins}, E.~B., {Tchernyshyov}, K., {et~al.} 2022, \apj, 928, 90, \dodoi{10.3847/1538-4357/ac5248}

\bibitem[{{Roueff} {et~al.}(2019){Roueff}, {Abgrall}, {Czachorowski}, {Pachucki}, {Puchalski}, \& {Komasa}}]{Roueff2019}
{Roueff}, E., {Abgrall}, H., {Czachorowski}, P., {et~al.} 2019, \aap, 630, A58, \dodoi{10.1051/0004-6361/201936249}

\bibitem[{{Salda{\~n}o} {et~al.}(2024){Salda{\~n}o}, {Rubio}, {Bolatto}, {Sandstrom}, {Swift}, {Verdugo}, {Jameson}, {Walker}, {Kulesa}, {Spilker}, {Bergman}, \& {Salazar}}]{Saldano2024}
{Salda{\~n}o}, H.~P., {Rubio}, M., {Bolatto}, A.~D., {et~al.} 2024, \aap, 687, A26, \dodoi{10.1051/0004-6361/202348436}

\bibitem[{{Salgado} {et~al.}(2016){Salgado}, {Bern{\'e}}, {Adams}, {Herter}, {Keller}, \& {Tielens}}]{Salgado16}
{Salgado}, F., {Bern{\'e}}, O., {Adams}, J.~D., {et~al.} 2016, \apj, 830, 118, \dodoi{10.3847/0004-637X/830/2/118}

\bibitem[{{Sandstrom} {et~al.}(2010){Sandstrom}, {Bolatto}, {Draine}, {Bot}, \& {Stanimirovi{\'c}}}]{Sandstrom10}
{Sandstrom}, K.~M., {Bolatto}, A.~D., {Draine}, B.~T., {Bot}, C., \& {Stanimirovi{\'c}}, S. 2010, \apj, 715, 701, \dodoi{10.1088/0004-637X/715/2/701}

\bibitem[{{Sandstrom} {et~al.}(2012){Sandstrom}, {Bolatto}, {Bot}, {Draine}, {Ingalls}, {Israel}, {Jackson}, {Leroy}, {Li}, {Rubio}, {Simon}, {Smith}, {Stanimirovi{\'c}}, {Tielens}, \& {van Loon}}]{Sandstrom12}
{Sandstrom}, K.~M., {Bolatto}, A.~D., {Bot}, C., {et~al.} 2012, \apj, 744, 20, \dodoi{10.1088/0004-637X/744/1/20}

\bibitem[{{Schneider} {et~al.}(2021){Schneider}, {R{\"o}llig}, {Polehampton}, {Comer{\'o}n}, {Djupvik}, {Makai}, {Buchbender}, {Simon}, {Bontemps}, {G{\"u}sten}, {White}, {Okada}, {Parikka}, \& {Rothbart}}]{Schneider2021}
{Schneider}, N., {R{\"o}llig}, M., {Polehampton}, E.~T., {et~al.} 2021, \aap, 653, A108, \dodoi{10.1051/0004-6361/202140824}

\bibitem[{{Schruba} {et~al.}(2012){Schruba}, {Leroy}, {Walter}, {Bigiel}, {Brinks}, {de Blok}, {Kramer}, {Rosolowsky}, {Sandstrom}, {Schuster}, {Usero}, {Weiss}, \& {Wiesemeyer}}]{Schruba12}
{Schruba}, A., {Leroy}, A.~K., {Walter}, F., {et~al.} 2012, \aj, 143, 138, \dodoi{10.1088/0004-6256/143/6/138}

\bibitem[{{Scowcroft} {et~al.}(2016){Scowcroft}, {Freedman}, {Madore}, {Monson}, {Persson}, {Rich}, {Seibert}, \& {Rigby}}]{Scowcroft16}
{Scowcroft}, V., {Freedman}, W.~L., {Madore}, B.~F., {et~al.} 2016, \apj, 816, 49, \dodoi{10.3847/0004-637X/816/2/49}

\bibitem[{{Seifried} {et~al.}(2020){Seifried}, {Haid}, {Walch}, {Borchert}, \& {Bisbas}}]{Seifried20}
{Seifried}, D., {Haid}, S., {Walch}, S., {Borchert}, E.~M.~A., \& {Bisbas}, T.~G. 2020, \mnras, 492, 1465, \dodoi{10.1093/mnras/stz3563}

\bibitem[{{Seifried} {et~al.}(2017){Seifried}, {Walch}, {Girichidis}, {Naab}, {W{\"u}nsch}, {Klessen}, {Glover}, {Peters}, \& {Clark}}]{Seifried17}
{Seifried}, D., {Walch}, S., {Girichidis}, P., {et~al.} 2017, \mnras, 472, 4797, \dodoi{10.1093/mnras/stx2343}

\bibitem[{{Seo} {et~al.}(2019){Seo}, {Goldsmith}, {Walker}, {Hollenbach}, {Wolfire}, {Kulesa}, {Tolls}, {Bernasconi}, {Kavak}, {van der Tak}, {Shipman}, {Gao}, {Tielens}, {Burton}, {Yorke}, {Young}, {Peters}, {Young}, {Groppi}, {Davis}, {Pineda}, {Langer}, {Kawamura}, {Stark}, {Melnick}, {Rebolledo}, {Wong}, {Horiuchi}, \& {Kuiper}}]{Seo2019}
{Seo}, Y.~M., {Goldsmith}, P.~F., {Walker}, C.~K., {et~al.} 2019, \apj, 878, 120, \dodoi{10.3847/1538-4357/ab2043}

\bibitem[{{Smith} {et~al.}(2007){Smith}, {Draine}, {Dale}, {Moustakas}, {Kennicutt}, {Helou}, {Armus}, {Roussel}, {Sheth}, {Bendo}, {Buckalew}, {Calzetti}, {Engelbracht}, {Gordon}, {Hollenbach}, {Li}, {Malhotra}, {Murphy}, \& {Walter}}]{Smith2007}
{Smith}, J.~D.~T., {Draine}, B.~T., {Dale}, D.~A., {et~al.} 2007, \apj, 656, 770, \dodoi{10.1086/510549}

\bibitem[{{Smith} {et~al.}(2014){Smith}, {Glover}, {Clark}, {Klessen}, \& {Springel}}]{Smith14}
{Smith}, R.~J., {Glover}, S. C.~O., {Clark}, P.~C., {Klessen}, R.~S., \& {Springel}, V. 2014, \mnras, 441, 1628, \dodoi{10.1093/mnras/stu616}

\bibitem[{{Sternberg} \& {Dalgarno}(1995)}]{Sternberg&Dalgarno95}
{Sternberg}, A., \& {Dalgarno}, A. 1995, \apjs, 99, 565, \dodoi{10.1086/192198}

\bibitem[{{St{\"o}rzer} \& {Hollenbach}(1998)}]{Storzer1998}
{St{\"o}rzer}, H., \& {Hollenbach}, D. 1998, \apj, 495, 853, \dodoi{10.1086/305315}

\bibitem[{{Tielens}(2010)}]{Tielens10}
{Tielens}, A.~G.~G.~M. 2010, {The Physics and Chemistry of the Interstellar Medium}

\bibitem[{{Tielens} \& {Hollenbach}(1985)}]{PDR1985}
{Tielens}, A.~G.~G.~M., \& {Hollenbach}, D. 1985, \apj, 291, 722, \dodoi{10.1086/163111}

\bibitem[{{Toribio San Cipriano} {et~al.}(2017){Toribio San Cipriano}, {Dom{\'\i}nguez-Guzm{\'a}n}, {Esteban}, {Garc{\'\i}a-Rojas}, {Mesa-Delgado}, {Bresolin}, {Rodr{\'\i}guez}, \& {Sim{\'o}n-D{\'\i}az}}]{Toribio17}
{Toribio San Cipriano}, L., {Dom{\'\i}nguez-Guzm{\'a}n}, G., {Esteban}, C., {et~al.} 2017, \mnras, 467, 3759, \dodoi{10.1093/mnras/stx328}

\bibitem[{{Van De Putte} {et~al.}(2024){Van De Putte}, {Meshaka}, {Trahin}, {Habart}, {Peeters}, {Bern{\'e}}, {Alarc{\'o}n}, {Canin}, {Chown}, {Schroetter}, {Sidhu}, {Boersma}, {Bron}, {Dartois}, {Goicoechea}, {Gordon}, {Onaka}, {Tielens}, {Verstraete}, {Wolfire}, {Abergel}, {Bergin}, {Bernard-Salas}, {Cami}, {Cuadrado}, {Dicken}, {Elyajouri}, {Fuente}, {Joblin}, {Khan}, {Lacinbala}, {Languignon}, {Le Gal}, {Maragkoudakis}, {Okada}, {Pasquini}, {Pound}, {Robberto}, {R{\"o}llig}, {Schefter}, {Schirmer}, {Tabone}, {Vicente}, {Zannese}, {Colgan}, {He}, {Rouill{\'e}}, {Togi}, {Aleman}, {Auchettl}, {Baratta}, {Bejaoui}, {Bera}, {Black}, {Boulanger}, {Bouwman}, {Brandl}, {Brechignac}, {Br{\"u}nken}, {Buragohain}, {Burkhardt}, {Candian}, {Cazaux}, {Cernicharo}, {Chabot}, {Chakraborty}, {Champion}, {Cooke}, {Coutens}, {Cox}, {Demyk}, {Meyer}, {Foschino}, {Garc{\'\i}a-Lario}, {Gerin}, {Gottlieb}, {Guillard}, {Gusdorf}, {Hartigan}, {Herbst}, {Hornekaer}, {Issa}, {J{\"a}ger}, {Janot-Pacheco}, {Kannavou}, {Kaufman},
  {Kemper}, {Kendrew}, {Kirsanova}, {Klaassen}, {Kwok}, {Labiano}, {Lai}, {Le Floch}, {Le Petit}, {Li}, {Linz}, {Mackie}, {Madden}, {Mascetti}, {McGuire}, {Merino}, {Micelotta}, {Morse}, {Mulas}, {Neelamkodan}, {Ohsawa}, {Omont}, {Paladini}, {Palumbo}, {Pathak}, {Pendleton}, {Petrignani}, {Pino}, {Puga}, {Rangwala}, {Rapacioli}, {Rho}, {Ricca}, {Roman-Duval}, {Roser}, {Roueff}, {Salama}, {Sales}, {Sandstrom}, {Sarre}, {Sciamma-O'Brien}, {Sellgren}, {Shenoy}, {Teyssier}, {Thomas}, {Witt}, {Wootten}, {Ysard}, {Zettergren}, {Zhang}, {Zhang}, \& {Zhen}}]{VanDePutte24}
{Van De Putte}, D., {Meshaka}, R., {Trahin}, B., {et~al.} 2024, \aap, 687, A86, \dodoi{10.1051/0004-6361/202449295}

\bibitem[{{van Dishoeck} \& {Black}(1988)}]{vanDishoeck&Black88}
{van Dishoeck}, E.~F., \& {Black}, J.~H. 1988, \apj, 334, 771, \dodoi{10.1086/166877}

\bibitem[{{Virtanen} {et~al.}(2020){Virtanen}, {Gommers}, {Oliphant}, {Haberland}, {Reddy}, {Cournapeau}, {Burovski}, {Peterson}, {Weckesser}, {Bright}, {van der Walt}, {Brett}, {Wilson}, {Jarrod Millman}, {Mayorov}, {Nelson}, {Jones}, {Kern}, {Larson}, {Carey}, {Polat}, {Feng}, {Moore}, {Vand erPlas}, {Laxalde}, {Perktold}, {Cimrman}, {Henriksen}, {Quintero}, {Harris}, {Archibald}, {Ribeiro}, {Pedregosa}, {van Mulbregt}, \& {Contributors}}]{Virtanen_2020}
{Virtanen}, P., {Gommers}, R., {Oliphant}, T.~E., {et~al.} 2020, Nature Methods, 17, 261, \dodoi{https://doi.org/10.1038/s41592-019-0686-2}

\bibitem[{{Wolfire} {et~al.}(2010){Wolfire}, {Hollenbach}, \& {McKee}}]{Wolfire10}
{Wolfire}, M.~G., {Hollenbach}, D., \& {McKee}, C.~F. 2010, \apj, 716, 1191, \dodoi{10.1088/0004-637X/716/2/1191}

\bibitem[{{Wolfire} {et~al.}(1995){Wolfire}, {Hollenbach}, {McKee}, {Tielens}, \& {Bakes}}]{Wolfire1995}
{Wolfire}, M.~G., {Hollenbach}, D., {McKee}, C.~F., {Tielens}, A.~G.~G.~M., \& {Bakes}, E.~L.~O. 1995, \apj, 443, 152, \dodoi{10.1086/175510}

\bibitem[{{Wolfire} {et~al.}(2022){Wolfire}, {Vallini}, \& {Chevance}}]{wolfire2022}
{Wolfire}, M.~G., {Vallini}, L., \& {Chevance}, M. 2022, \araa, 60, 247, \dodoi{10.1146/annurev-astro-052920-010254}

\bibitem[{{Wrathmall} {et~al.}(2007){Wrathmall}, {Gusdorf}, \& {Flower}}]{Wrathmall2007}
{Wrathmall}, S.~A., {Gusdorf}, A., \& {Flower}, D.~R. 2007, \mnras, 382, 133, \dodoi{10.1111/j.1365-2966.2007.12420.x}

\bibitem[{{Wu} {et~al.}(2018){Wu}, {Bron}, {Onaka}, {Le Petit}, {Galliano}, {Languignon}, {Nakamura}, \& {Okada}}]{Wu2018}
{Wu}, R., {Bron}, E., {Onaka}, T., {et~al.} 2018, \aap, 618, A53, \dodoi{10.1051/0004-6361/201832595}

\bibitem[{{Yanchulova Merica-Jones} {et~al.}(2017){Yanchulova Merica-Jones}, {Sandstrom}, {Johnson}, {Dalcanton}, {Dolphin}, {Gordon}, {Roman-Duval}, {Weisz}, \& {Williams}}]{Smidge}
{Yanchulova Merica-Jones}, P., {Sandstrom}, K.~M., {Johnson}, L.~C., {et~al.} 2017, \apj, 847, 102, \dodoi{10.3847/1538-4357/aa8a67}

\bibitem[{{Yang} {et~al.}(2010){Yang}, {Stancil}, {Balakrishnan}, \& {Forrey}}]{Yang2010}
{Yang}, B., {Stancil}, P.~C., {Balakrishnan}, N., \& {Forrey}, R.~C. 2010, \apj, 718, 1062, \dodoi{10.1088/0004-637X/718/2/1062}

\end{thebibliography}

\appendix
\section{PDR Models and Resulting Profiles}
\begin{figure*}[h]
    \centering
    \includegraphics[width=\textwidth, trim=0mm 0mm 0mm 0mm,clip]{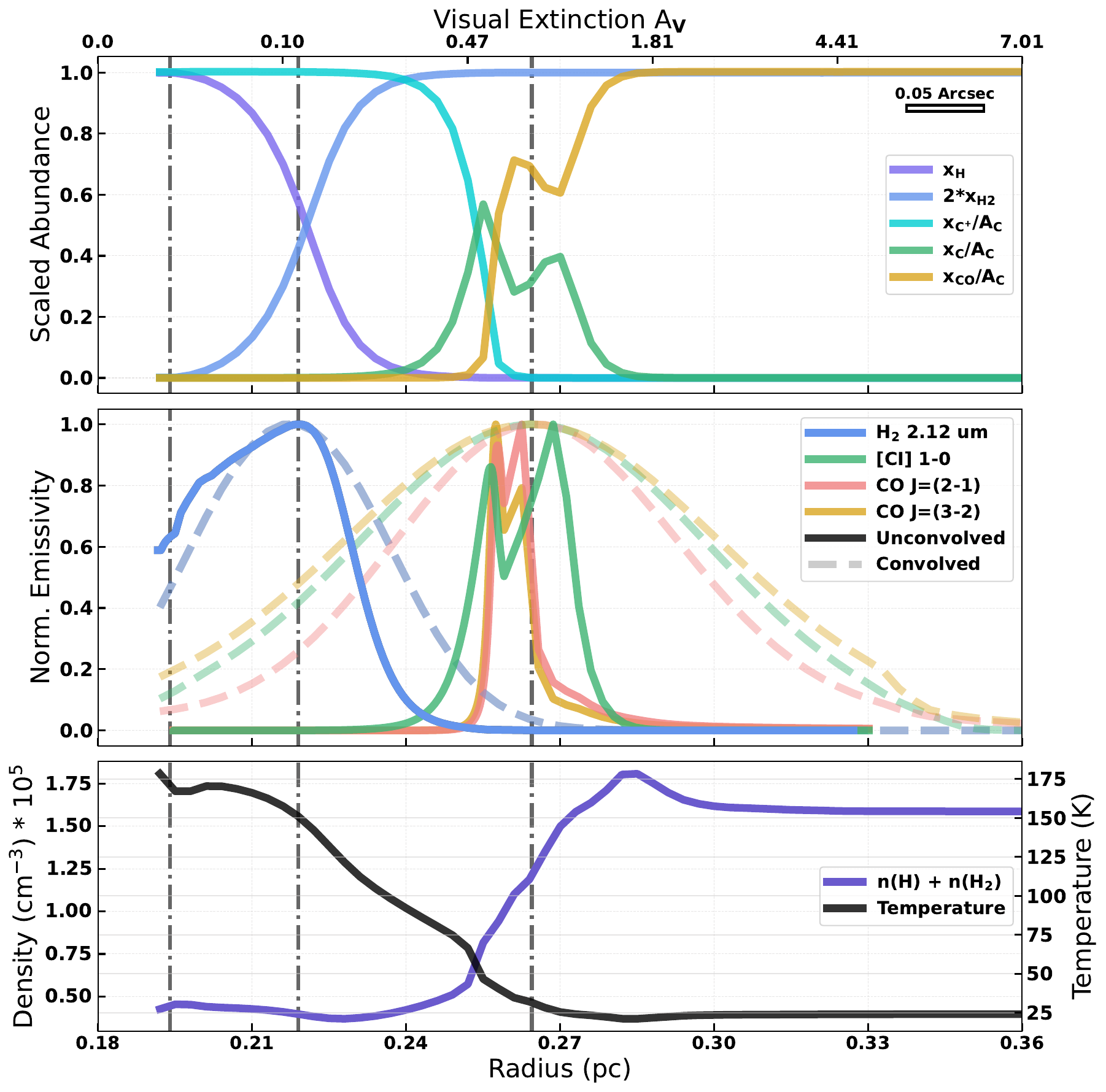}\\
    \caption{Top panel: We present the scaled radial abundance profiles of the $P_{\rm th}/k=7.6 \times 10^{6}$ K cm$^{-3}$ constant pressure model, where $x_i = n_i/n$ is the fractional abundance of species $i$ and A$_{\rm C} = 3.2 \times 10^{-5}$ is the gas phase abundance of Carbon per hydrogen nucleus. To show all profiles on the same plot while preserving their shapes, we scale the \htwo fractional abundance by 2 and the C$^+$/C/CO profiles by $1/A_{\rm C}$. Half of the gas is molecular ${\rm H_2}$ at $2x_{\rm H2} = 0.5$. Middle Panel: We present the unconvolved (solid) and convolved (dashed) emissivity profiles of the \htwo $2.12 \micron$, [CI] 1-0, ${}^{12}$CO J=2-1, ${}^{12}$CO J=3-2 lines from the constant pressure model mentioned in the top panel. Bottom panel: We show the modeled temperature profile along with the H$+$\htwo density profile across the PDR for the constant pressure case as presented in the top panel. We identify the ionization front (IF), the dissociation front (DF), and the C/CO transition from left to right as seen in the vertical dashed-dotted gray lines in each panel.}
    \label{fig:models_abund}
\end{figure*}

In the top panel of Figure \ref{fig:models_abund}, we present the abundance profiles for the constant pressure model that shows the best correspondence with the observations of the N13 PDR. In the figure $x_i = n_i/n$ is the fractional abundance of species $i$ and A$_{\rm C} = 3.2 \times 10^{-5}$ is the gas phase abundance of Carbon per hydrogen nucleus. The profiles are scaled by constant factors to be shown on the same plot (e.g., C$^+$/C/CO are scaled by $1/A_{\rm C}$). In the middle panel of this figure, we show the constant pressure model with unconvolved (solid) and convolved (dashed) emissivity profiles to match our resolution. In this panel, we present the emissivity profiles of the \htwo $2.12 \micron$, [CI] 1-0, ${}^{12}$CO J=2-1, ${}^{12}$CO J=3-2 lines. In the bottom panel of Figure \ref{fig:models_abund}, we present the temperature profile for the constant pressure model. Vertical dotted-dashed gray lines in both panels show the observed locations of the IF, DF, and C/CO transitions from left to right in both panels. We find the constant pressure model abundance profiles show a close agreement between the transition locations and the observed emission peaks, which we assign to the DF and C/CO transition. In the convolved constant pressure model, the separation between the peak of CO J=2-1 and [CI] 1-0 is 0.006 pc, corresponding to 0.4 A$_{\rm v}$. We note that the abundance profiles are not convolved to match the resolution of the observations. In addition, the constant pressure model also does a reasonable job of reproducing the IF location (the model curves end at the IF).

The models calculate the H$_2$ excitation and emission using 150 $vj$ levels. The radiative transition coefficients are taken from \cite{Roueff2019}. Collision coefficients for  collisions  with atomic H, o-H$_2$, and p-H$_2$ are taken from \cite{Wrathmall2007}, and \cite{Flower1998,Flower1999}. 
For CO, we use a maximum of 40 rotational levels with $A$ values from \cite{Chandra1996} and collision rates with o-H$_2$ and p-H$_2$ from \cite{Yang2010}. The line emissivity is computed following Equations B1, B2, and B3 in \citet{PDR1985}, with the line optical depth and escape probability as in B8 and B9. The model 1-0 S(1) emissivity peaks at the dissociation front, where the gas mass is half molecular, and is excited by FUV pumping. There, gas heating is dominated by photoelectric heating with some contribution from H$_2$ formation heating and balanced by mainly [OI] 63 $\mu$m fine-structure line cooling. The CO emission peaks at the edge of the C/CO transition, where the CO abundance starts to rise while the gas temperature is still sufficiently warm to excite the CO transitions. There, heating is dominated by photoelectric heating and balanced by mainly CO rotational line cooling.

We note that the observed 1-0 S(1) line emission exhibits a peak at the model DF. The observed 1-0 S(1)/2-1 S(1) ratio at the peak is $\sim 2.2$, confirming that the emission is due to FUV pumping. In steady state, the H$_2$ FUV pumping rate is proportional to the H$_2$ formation rate on grains $Rnn_{\rm HI}$, where $R$ is the formation rate coefficient, and $n_{\rm HI}$ is the atomic hydrogen density. One would expect, in the limit of constant $R$ and $n$, that the emission would simply follow the atomic hydrogen density and there would not be a peak in emission. However, there are several additional processes that affect the line profile.

At depths into the cloud before the peak, photo processes can suppress the abundance of vibrationally excited H$_2$ in $v=1$ $j=3$  whenever the photo rates are comparable to radiative de-excitation to ground \cite[e.g.,][]{BurtonHollenbachTielens90}. With increasing column density, H$_2$ self-shielding reduces the photo depopulation of excited H$_2$ in  $v=1$, $j=3$, causing a rise in H$_2$ 1-0 S(1) line emission.  Increasing self-shielding in the lower H$_2$ rotational levels reduces the FUV pumping and causes a drop on the far side of the peak. The rise in H$_2$ abundance is a result of the same H$_{\rm 2}$ self-shielding. As a result, the 1-0 S(1) line appears as a peak in emission closely associated with the H$_2$ dissociation front. In addition, in constant pressure models, a rapidly rising gas temperature towards the cloud surface reduces both $n$ and $n_{\rm H}$, causing a rapid drop in the H$_2$ formation rate and, in steady state, a drop in the FUV pumping rate.

Models also predict a peak in the 1-0 S(1) line at the DF when collisional excitation dominates the 1-0 S(1) line emission (see for e.g., \citealt{BurtonHollenbachTielens90}, Fig.\ 1; \citealt{Allers05}, Fig.\ 6; \citealt{Goicoechea19}, Fig.\ 10). The high gas temperature, combined with the increasing H$_2$ fraction, produces a peak that falls off when the gas temperature drops. We suggest the 1-0 S(1) emission peak at the DF occurs for a wide range of conditions for both collisionally excited and FUV pumped excitation, however for $G_0$ much less than $\sim 1000$, and for FUV pumping, the peak will likely to be minor since photo depopulation will no longer compete with downward radiative de-excitation.

More recent updates to PDR models include the photodissociation and photoionization rates from \cite{Heays17}, $^{13}$C chemistry, measured dissociative recombination rates of OH$^+$ \citep{Kalosi23} and CH$^+$ \citep{Paul2022},  and collisional excitation of C by H$_2$ \citep[][]{Klos21,Goicoechea25}.

\begin{deluxetable*}{c|c|c|c|c|c|c}
\tablecaption{Radial Profile Data with Coordinates\label{tab:radial slice h2 co}}
\tablewidth{0pt}
\tablehead{
\colhead{RA} & \colhead{Dec} & \colhead{Radii} & \colhead{\htwo 2.12 $\micron$}  & \colhead{CO J=2-1}  & \colhead{CO J=3-2}  & \colhead{[CI] 1-0} \\
\colhead{H:M:S} & \colhead{D:M:S} & \colhead{$''$} & \colhead{erg s$^{-1}$ cm$^{-2}$ sr$^{-1}$}  &\colhead{K km/s}  & \colhead{K km/s}  & \colhead{K km/s} \\
\colhead{} & \colhead{} & \colhead{} & \colhead{$\times$ 10$^{-5}$}  &\colhead{}  & \colhead{}  & \colhead{} 
}
\startdata
$0:45:26.913$ & $-73:22:55.129$ & $1.57$ & $1.40 \pm 0.03$ & $38.45 \pm 0.55$ & $42.18 \pm 0.36$ & $4.85 \pm 0.02$ \\ 
$0:45:26.915$ & $-73:22:55.122$ & $1.58$ & $1.36 \pm 0.04$ & $37.81 \pm 0.55$ & $41.58 \pm 0.43$ & $4.93 \pm 0.02$ \\ 
$0:45:26.917$ & $-73:22:55.116$ & $1.59$ & $1.32 \pm 0.04$ & $37.18 \pm 0.57$ & $40.98 \pm 0.49$ & $5.00 \pm 0.02$ \\ 
$0:45:26.918$ & $-73:22:55.110$ & $1.60$ & $1.30 \pm 0.04$ & $36.53 \pm 0.59$ & $40.36 \pm 0.55$ & $5.05 \pm 0.02$ \\ 
$0:45:26.920$ & $-73:22:55.103$ & $1.61$ & $1.30 \pm 0.04$ & $35.88 \pm 0.60$ & $39.72 \pm 0.60$ & $5.08 \pm 0.02$ \\ 
$0:45:26.922$ & $-73:22:55.097$ & $1.62$ & $1.31 \pm 0.05$ & $35.23 \pm 0.60$ & $39.06 \pm 0.64$ & $5.09 \pm 0.04$ \\ 
$0:45:26.924$ & $-73:22:55.091$ & $1.63$ & $1.34 \pm 0.05$ & $34.59 \pm 0.58$ & $38.39 \pm 0.68$ & $5.07 \pm 0.06$ \\ 
$0:45:26.926$ & $-73:22:55.084$ & $1.64$ & $1.38 \pm 0.05$ & $33.93 \pm 0.56$ & $37.70 \pm 0.71$ & $5.02 \pm 0.09$ \\ 
$0:45:26.927$ & $-73:22:55.078$ & $1.65$ & $1.43 \pm 0.04$ & $33.23 \pm 0.54$ & $37.00 \pm 0.75$ & $4.93 \pm 0.12$ \\ 
$0:45:26.929$ & $-73:22:55.072$ & $1.66$ & $1.48 \pm 0.04$ & $32.52 \pm 0.53$ & $36.28 \pm 0.78$ & $4.82 \pm 0.12$ \\ 
$0:45:26.931$ & $-73:22:55.065$ & $1.67$ & $1.52 \pm 0.04$ & $31.80 \pm 0.52$ & $35.55 \pm 0.81$ & $4.70 \pm 0.11$ \\ 
$0:45:26.933$ & $-73:22:55.059$ & $1.68$ & $1.56 \pm 0.04$ & $31.03 \pm 0.54$ & $34.82 \pm 0.84$ & $4.58 \pm 0.10$ \\ 
$0:45:26.935$ & $-73:22:55.053$ & $1.69$ & $1.57 \pm 0.03$ & $30.28 \pm 0.57$ & $34.10 \pm 0.86$ & $4.45 \pm 0.10$ \\ 
$0:45:26.937$ & $-73:22:55.046$ & $1.70$ & $1.57 \pm 0.02$ & $29.76 \pm 0.53$ & $33.37 \pm 0.86$ & $4.31 \pm 0.10$ \\ 
$0:45:26.938$ & $-73:22:55.040$ & $1.71$ & $1.52 \pm 0.01$ & $29.37 \pm 0.48$ & $32.65 \pm 0.86$ & $4.18 \pm 0.11$ \\ 
$0:45:26.940$ & $-73:22:55.034$ & $1.72$ & $1.45 \pm 0.00$ & $29.01 \pm 0.44$ & $31.93 \pm 0.86$ & $4.06 \pm 0.12$ \\ 
$0:45:26.942$ & $-73:22:55.027$ & $1.73$ & $1.39 \pm 0.00$ & $28.66 \pm 0.42$ & $31.21 \pm 0.84$ & $3.93 \pm 0.13$ \\ 
$0:45:26.944$ & $-73:22:55.021$ & $1.74$ & $1.35 \pm 0.01$ & $28.30 \pm 0.41$ & $30.50 \pm 0.83$ & $3.82 \pm 0.15$ \\ 
$0:45:26.946$ & $-73:22:55.015$ & $1.75$ & $1.32 \pm 0.01$ & $27.89 \pm 0.42$ & $29.79 \pm 0.81$ & $3.72 \pm 0.17$ \\ 
$0:45:26.947$ & $-73:22:55.008$ & $1.76$ & $1.30 \pm 0.01$ & $27.43 \pm 0.43$ & $29.09 \pm 0.80$ & $3.64 \pm 0.18$ \\ 
$0:45:26.949$ & $-73:22:55.002$ & $1.77$ & $1.31 \pm 0.01$ & $26.93 \pm 0.43$ & $28.40 \pm 0.78$ & $3.57 \pm 0.17$ \\ 
$0:45:26.951$ & $-73:22:54.996$ & $1.78$ & $1.35 \pm 0.01$ & $26.39 \pm 0.42$ & $27.73 \pm 0.77$ & $3.49 \pm 0.16$ \\ 
$0:45:26.953$ & $-73:22:54.990$ & $1.79$ & $1.38 \pm 0.00$ & $25.81 \pm 0.40$ & $27.07 \pm 0.76$ & $3.39 \pm 0.16$ \\ 
$0:45:26.955$ & $-73:22:54.983$ & $1.80$ & $1.40 \pm 0.00$ & $25.19 \pm 0.39$ & $26.42 \pm 0.76$ & $3.30 \pm 0.15$ \\ 
$0:45:26.957$ & $-73:22:54.977$ & $1.81$ & $1.43 \pm 0.00$ & $24.51 \pm 0.36$ & $25.77 \pm 0.75$ & $3.19 \pm 0.15$ \\ 
$0:45:26.958$ & $-73:22:54.971$ & $1.82$ & $1.44 \pm 0.01$ & $23.80 \pm 0.33$ & $25.14 \pm 0.74$ & $3.07 \pm 0.14$ \\ 
$0:45:26.960$ & $-73:22:54.964$ & $1.83$ & $1.45 \pm 0.02$ & $23.04 \pm 0.32$ & $24.52 \pm 0.74$ & $2.95 \pm 0.14$ \\ 
$0:45:26.962$ & $-73:22:54.958$ & $1.84$ & $1.45 \pm 0.02$ & $22.24 \pm 0.32$ & $23.91 \pm 0.72$ & $2.81 \pm 0.13$ \\ 
$0:45:26.964$ & $-73:22:54.952$ & $1.85$ & $1.44 \pm 0.03$ & $21.43 \pm 0.30$ & $23.35 \pm 0.70$ & $2.67 \pm 0.13$ \\ 
$0:45:26.966$ & $-73:22:54.945$ & $1.86$ & $1.43 \pm 0.03$ & $20.72 \pm 0.26$ & $22.78 \pm 0.67$ & $2.51 \pm 0.13$ \\ 
$0:45:26.967$ & $-73:22:54.939$ & $1.87$ & $1.42 \pm 0.04$ & $20.07 \pm 0.24$ & $22.22 \pm 0.65$ & $2.35 \pm 0.12$ \\ 
$0:45:26.969$ & $-73:22:54.933$ & $1.88$ & $1.43 \pm 0.05$ & $19.40 \pm 0.23$ & $21.65 \pm 0.62$ & $2.18 \pm 0.11$ \\ 
\enddata
\tablecomments{We present the last 30 rows of our measured radial profiles along Slice 1 for \htwo 2.12 $\micron$, CO J=2-1, CO J=3-2, [CI] 1-0. The second row of the table header indicates the units and the third column indicates the scaling factor for the number in that column. The \htwo 2.12 $\micron$ is scaled as $10^{-5}$ for readability, and is in units of erg s$^{-1}$ cm$^{-2}$ sr$^{-1}$, while the [CI] 1-0 and CO lines are not scaled and in units of K Km/s. The full version is available for download.
}
\end{deluxetable*}

\begin{deluxetable*}{c|c|c|c|c|c}
\caption{Radial Profile Data: PAH 3.3 $\micron$, HI Paschen 4-3, HI Brackett 5-4 \label{tab:radial slice hi pah}}
\tablewidth{0pt}
\tablehead{
\colhead{RA} & \colhead{Dec} & \colhead{Radii} & \colhead{PAH 3.3$\micron$}  & \colhead{Paschen 4-3}  & \colhead{Brackett 5-4}  \\
\colhead{H:M:S} &\colhead{D:M:S} & \colhead{$''$} & \colhead{erg s$^{-1}$ cm$^{-2}$ sr$^{-1}$}  &\colhead{erg s$^{-1}$ cm$^{-2}$ sr$^{-1}$}  & \colhead{erg s$^{-1}$ cm$^{-2}$ sr$^{-1}$} \\
\colhead{} & \colhead{} & \colhead{} & \colhead{$\times$ 10$^{-5}$}  &\colhead{$\times$ 10$^{-5}$}  & \colhead{$\times$ 10$^{-5}$} 
}
\startdata
$0:45:26.915$ & $-73:22:55.122$ & $1.57$ & $11.70 \pm 0.05$ & $30.33 \pm 0.20$ & $8.27 \pm 0.03$ \\ 
$0:45:26.917$ & $-73:22:55.116$ & $1.58$ & $11.64 \pm 0.07$ & $30.17 \pm 0.19$ & $8.24 \pm 0.03$ \\ 
$0:45:26.918$ & $-73:22:55.110$ & $1.59$ & $11.57 \pm 0.05$ & $29.94 \pm 0.14$ & $8.18 \pm 0.03$ \\ 
$0:45:26.920$ & $-73:22:55.103$ & $1.60$ & $11.44 \pm 0.03$ & $29.55 \pm 0.09$ & $8.12 \pm 0.03$ \\ 
$0:45:26.922$ & $-73:22:55.097$ & $1.61$ & $11.39 \pm 0.03$ & $29.20 \pm 0.07$ & $8.04 \pm 0.03$ \\ 
$0:45:26.924$ & $-73:22:55.091$ & $1.62$ & $11.35 \pm 0.04$ & $28.71 \pm 0.02$ & $7.95 \pm 0.02$ \\ 
$0:45:26.926$ & $-73:22:55.084$ & $1.63$ & $11.28 \pm 0.06$ & $28.18 \pm 0.02$ & $7.87 \pm 0.03$ \\ 
$0:45:26.927$ & $-73:22:55.078$ & $1.64$ & $11.22 \pm 0.10$ & $27.72 \pm 0.05$ & $7.81 \pm 0.03$ \\ 
$0:45:26.929$ & $-73:22:55.072$ & $1.65$ & $11.20 \pm 0.15$ & $27.31 \pm 0.08$ & $7.77 \pm 0.03$ \\ 
$0:45:26.931$ & $-73:22:55.065$ & $1.66$ & $11.23 \pm 0.21$ & $26.94 \pm 0.10$ & $7.73 \pm 0.02$ \\ 
$0:45:26.933$ & $-73:22:55.059$ & $1.67$ & $11.34 \pm 0.27$ & $26.63 \pm 0.11$ & $7.70 \pm 0.02$ \\ 
$0:45:26.935$ & $-73:22:55.053$ & $1.68$ & $11.58 \pm 0.34$ & $26.40 \pm 0.10$ & $7.66 \pm 0.01$ \\ 
$0:45:26.937$ & $-73:22:55.046$ & $1.69$ & $12.06 \pm 0.41$ & $26.31 \pm 0.06$ & $7.62 \pm 0.00$ \\ 
$0:45:26.938$ & $-73:22:55.040$ & $1.70$ & $12.71 \pm 0.41$ & $26.52 \pm 0.02$ & $7.59 \pm 0.01$ \\ 
$0:45:26.940$ & $-73:22:55.034$ & $1.71$ & $13.39 \pm 0.40$ & $26.96 \pm 0.01$ & $7.54 \pm 0.02$ \\ 
$0:45:26.942$ & $-73:22:55.027$ & $1.72$ & $13.71 \pm 0.39$ & $27.31 \pm 0.01$ & $7.50 \pm 0.02$ \\ 
$0:45:26.944$ & $-73:22:55.021$ & $1.73$ & $13.75 \pm 0.36$ & $27.58 \pm 0.01$ & $7.48 \pm 0.03$ \\ 
$0:45:26.946$ & $-73:22:55.015$ & $1.74$ & $13.77 \pm 0.34$ & $27.86 \pm 0.01$ & $7.47 \pm 0.03$ \\ 
$0:45:26.947$ & $-73:22:55.008$ & $1.75$ & $14.17 \pm 0.34$ & $28.05 \pm 0.03$ & $7.45 \pm 0.04$ \\ 
$0:45:26.949$ & $-73:22:55.002$ & $1.76$ & $14.87 \pm 0.35$ & $28.20 \pm 0.04$ & $7.42 \pm 0.04$ \\ 
$0:45:26.951$ & $-73:22:54.996$ & $1.77$ & $15.48 \pm 0.33$ & $28.36 \pm 0.04$ & $7.41 \pm 0.04$ \\ 
$0:45:26.953$ & $-73:22:54.990$ & $1.78$ & $15.86 \pm 0.27$ & $28.46 \pm 0.05$ & $7.39 \pm 0.04$ \\ 
$0:45:26.955$ & $-73:22:54.983$ & $1.79$ & $16.01 \pm 0.24$ & $28.47 \pm 0.07$ & $7.39 \pm 0.04$ \\ 
$0:45:26.957$ & $-73:22:54.977$ & $1.80$ & $16.07 \pm 0.24$ & $28.40 \pm 0.07$ & $7.38 \pm 0.04$ \\ 
$0:45:26.958$ & $-73:22:54.971$ & $1.81$ & $16.07 \pm 0.26$ & $28.27 \pm 0.07$ & $7.38 \pm 0.04$ \\ 
$0:45:26.960$ & $-73:22:54.964$ & $1.82$ & $16.04 \pm 0.29$ & $28.10 \pm 0.06$ & $7.38 \pm 0.04$ \\ 
$0:45:26.962$ & $-73:22:54.958$ & $1.83$ & $16.25 \pm 0.20$ & $27.91 \pm 0.04$ & $7.37 \pm 0.04$ \\ 
$0:45:26.964$ & $-73:22:54.952$ & $1.84$ & $16.38 \pm 0.16$ & $27.73 \pm 0.02$ & $7.36 \pm 0.04$ \\ 
$0:45:26.966$ & $-73:22:54.945$ & $1.85$ & $16.50 \pm 0.14$ & $27.59 \pm 0.01$ & $7.35 \pm 0.04$ \\ 
$0:45:26.967$ & $-73:22:54.939$ & $1.86$ & $16.55 \pm 0.12$ & $27.48 \pm 0.01$ & $7.33 \pm 0.03$ \\ 
$0:45:26.969$ & $-73:22:54.933$ & $1.87$ & $16.46 \pm 0.12$ & $27.45 \pm 0.01$ & $7.34 \pm 0.03$ \\ 
\enddata
\tablecomments{Continuation of Table \ref{tab:radial slice h2 co}. We present the last 30 rows of the measured radial profiles along Slice 1 for PAH 3.3 $\micron$, HI Paschen 4-3, HI Brackett 5-4. The second row of the table header indicates the units and the third column indicates the scaling factor for the data in that column. The full version is available for download.}
\end{deluxetable*}

\end{document}